\documentclass[12pt]{article}
\def\lsim{\mathrel{\rlap {\raise.5ex\hbox{$ < $}}
{\lower.5ex\hbox{$\sim$}}}}
\def\gsim{\mathrel{\rlap {\raise.5ex\hbox{$ > $}}
{\lower.5ex\hbox{$\sim$}}}}
\def\sqr#1#2{{\vcenter{\vbox{\hrule height.#2pt
        \hbox{\vrule width.#2pt height#1pt \kern#1pt
           \vrule width.#2pt}
        \hrule height.#2pt}}}}

\def\lsim{{\displaystyle
{{\raise-8pt\hbox{$ <$}}
\atop{\raise5pt\hbox{$\sim$}}}}}
\def\gsim{{\displaystyle
{{\raise-8pt\hbox{$ >$}}
\atop{\raise5pt\hbox{$\sim$}}}}}
\def\slsim{{\displaystyle
{{\raise-8pt\hbox{$\scriptstyle <$}}
\atop{\raise5pt\hbox{$\scriptstyle \sim$}}}}}
\def\sgsim{{\displaystyle
{{\raise-8pt\hbox{$\scriptstyle  >$}}
\atop{\raise5pt\hbox{$\scriptstyle \sim$}}}}}
\newskip\humongous \humongous=0pt plus 1000pt minus 1000pt
\def\caja{\mathsurround=0pt}
\def\eqalign#1{\,\vcenter{\openup1\jot \caja
        \ialign{\strut \hfil$\displaystyle{##}$&$
        \displaystyle{{}##}$\hfil\crcr#1\crcr}}\,}
\newcommand{\oao}[2]{{#1\atopwithdelims[]#2}}

\newcommand{\sump}[0]{\sum_{(h,g)}\!{\raise 4pt \hbox{$'$}}\,}
\newcommand{\dv}[0]{\delta_v}
\newcommand{\dhh}[0]{\delta_h}
\newcommand{\dvp}[0]{\delta_v^{\phantom{i}}}
\newcommand{\dhhp}[0]{\delta_h^{\phantom{i}}}
\newcommand{\fvw}[0]{f_v^w}
\catcode`\@=11
\newcount\hour
\newcount\minute
\newtoks\amorpm
\hour=\time\divide\hour by60
\minute=\time{\multiply\hour by60 \global\advance\minute by-\hour}
\edef\standardtime{{\ifnum\hour<12 \global\amorpm={am}%
        \else\global\amorpm={pm}\advance\hour by-12 \fi
        \ifnum\hour=0 \hour=12 \fi
        \number\hour:\ifnum\minute<10 0\fi\number\minute\the\amorpm}}
\edef\militarytime{\number\hour:\ifnum\minute<10 0\fi\number\minute}
\def\draftlabel#1{{\@bsphack\if@filesw {\let\thepage\relax
   \xdef\@gtempa{\write\@auxout{\string
      \newlabel{#1}{{\@currentlabel}{\thepage}}}}}\@gtempa
   \if@nobreak \ifvmode\nobreak\fi\fi\fi\@esphack}
        \gdef\@eqnlabel{#1}}
\def\@eqnlabel{}
\def\@vacuum{}
\def\draftmarginnote#1{\marginpar{\raggedright\scriptsize\tt#1}}
\def\draft{\oddsidemargin -.2truein
        \def\@oddfoot{\sl preliminary draft \hfil
        \rm\thepage\hfil\sl\today\quad\militarytime}
        \let\@evenfoot\@oddfoot \overfullrule 3pt
        \let\label=\draftlabel
        \let\marginnote=\draftmarginnote
   \def\@eqnnum{(\theequation)\rlap{\kern\marginparsep\tt\@eqnlabel}%
\global\let\@eqnlabel\@vacuum}  }
\relax
\def\Tr{\,{\rm Tr}\, }
\def\Trp{\,{\rm Tr'}\, }

\def\Im{\,{\rm Im}\, }
\def\Re{\,{\rm Re}\, }
\def\bJ{\overline{J}}
\def\bj{\bar j}

\def\bP{\overline{P}}

\def\bE{\overline{E}}

\def\bT{\overline{T}}
\def\bU{\overline{U}}
\def\bOmega{\overline{\Omega}}
\def\bLambda{\overline{\Lambda}}

\def\thefootnote{\fnsymbol{footnote}}
\def\be{\begin{equation}}
\def\ee{\end{equation}}
\def\ba{\begin{eqnarray}}
\def\ea{\end{eqnarray}}
\def\bs{\begin{subequations}}
\def\es{\end{subequations}}

\def\th{\vartheta}

\def\r{\rho}
\def\l{\lambda}

\def\t{\tau}

\def\gs{g_{\rm s}}

\def\t{\tau}
\def\im{\tau_2}
\def\iT{T_2}
\def\iU{U_2}
\def\I{\rm I}
\def\II{\rm II}

\def\X{\rm X}

\def\sp{\ , \ \ }
\def\ifd{\int_{\cal F}\frac{{\rm d}^2\tau}{\im}}
\def\bv{\bar v}
\def\fh{f_h^w}
\def\fv{f_v^w}
\def\bgra{b_{\rm grav}}
\def\bgrav{b_{\rm grav}}

\def\ee{\end{equation}}
\def\bea{\begin{eqnarray}}
\def\eea{\end{eqnarray}}
\def\nn{\nonumber}

\def\np#1#2#3{Nucl. Phys. {\bf{B#1}} (#2) #3}
\def\pl#1#2#3{Phys. Lett. {\bf{B#1}} (#2) #3}
\def\plo#1#2#3{Phys. Lett. {\bf{#1B}} (#2) #3}

\def\mpl#1#2#3{Mod. Phys. Lett. {\bf{A#1}} (#2) #3}
\newcommand{\ihd}[0]{\int_{\cal H}\frac{{\rm d}^2\tau}{\tau_2^2}}
\newcommand{\ijd}[1]{\int_{\cal S}\frac{{\rm d}^2\tau}{\tau_2^#1}}
\newcommand{\Li}[0]{{\cal L}i}
\newcommand{\Lii}[1]{{{\cal L}i}_1\left(e^{2\pi
i\left(#1\right)}\right)}
\newcommand{\PP}[1]{{\cal P}\left(#1\right)}
\newcommand{\TTH}[1]{{\Theta}\left(#1\right)}
\newcommand{\ABS}[1]{{\left|#1\right|}}
\newcommand{\uarrw}[0]{\mathrel{
{\raise.5ex\vbox{\hrule width 1cm}\hskip-6pt\rightarrow}}}
\newcommand{\underarrow}[1]{\mathop{\uarrw}_{#1}}
\def\thebibliography#1{%
\vskip 0.5cm \centerline{\bf References}
\list{%
[\arabic{enumi}]}{\settowidth\labelwidth{[#1]}
\leftmargin\labelwidth
\advance\leftmargin\labelsep
\usecounter{enumi}}
\def\newblock{\hskip .11em plus .33em minus .07em}
\sloppy\clubpenalty4000\widowpenalty4000
\sfcode`\.=1000\relax}

\renewcommand{\theequation}{\arabic{section}.\arabic{equation}}

\renewcommand{\section}{\setcounter{equation}{0}\@startsection%
{section}{1}{0mm}{-\baselineskip}{0.5\baselineskip}%
{\normalfont\normalsize\bfseries}}

\renewcommand{\subsection}{\@startsection%
{subsection}{2}{0mm}{-\baselineskip}{0.5\baselineskip}%
{\normalfont\normalsize\slshape}}
\begin{document}
\renewcommand{\theequation}{\arabic{section}.\arabic{equation}}
\begin{titlepage}
\begin{flushright}
CERN-TH/97-44, NEIP-97-007, IOA-97-08 \\
LPTENS/97/11, CPTH-S499.0397 \\
hep-th/9807067 \\
\end{flushright}
\begin{centering}
%
{\bf STRING THRESHOLD CORRECTIONS IN MODELS WITH SPONTANEOUSLY BROKEN
SUPERSYMMETRY}$^{\ \ast}$\\
\vspace{5pt}
{E. KIRITSIS$^{\ 1}$, C. KOUNNAS$^{\ 1,\, \ddagger}$,
P.M. PETROPOULOS$^{\ 1,\, 2,\, \diamond}$ \\
and \\
J. RIZOS$^{\ 1,\, 3}$}\\
\vspace{3pt}
{\it $^1 $ Theory Division, CERN}\\
{\it 1211 Geneva 23, Switzerland}\\[2pt]
{\it $^2 $ Institut de Physique Th\'eorique, Universit\'e de
Neuch\^atel}\\
{\it 2000 Neuch\^atel, Switzerland}\\
{\it and}\\
{\it $^3 $ Division of Theoretical Physics, Physics Department,
University of Ioannina}\\
{\it 45110 Ioannina, Greece}\\
\vspace{.15in}
{\bf Abstract}\\
\end{centering}
We analyse a class of four-dimensional heterotic ground states with
$N=2$ space-time supersymmetry. From the ten-dimensional perspective,
such models can be viewed as
compactifications  on a six-dimensional manifold with $SU(2)$
holonomy,
which is locally but {\it
not globally} $K3 \times T^2$.
The maximal $N=4$ supersymmetry is {\it spontaneously} broken to
$N=2$. The masses of the two massive gravitinos depend on the ($T,U$)
moduli of $T^2$.
We evaluate the one-loop threshold corrections of gauge and $R^2$
couplings and we show that they
fall in several universality
classes, in contrast to what happens in usual $K3 \times T^2$
compactifications,
where the
$N=4$ supersymmetry is explicitly broken to $N=2$, and where
a single universality class appears. These universality properties
follow  from
the structure of the elliptic genus. The behaviour of the threshold
corrections
as functions of the moduli is analysed in detail: it is singular
across
several rational lines of the $T^2$ moduli because
of the appearance of extra massless states, and suffers only from
logarithmic singularities at large radii. These features differ
substantially
from the ordinary $K3 \times T^2$ compactifications,
thereby reflecting the existence of spontaneously-broken
$N=4$ supersymmetry. Although our results are valid in the general
framework defined above, we also point out several properties,
specific
to orbifold constructions, which might be of phenomenological
relevance.
\begin{flushleft}
CERN-TH/97-44, NEIP-97-007, IOA-97-08, LPTENS/97/11, CPTH-S499.0397
\\
June 1998 \\
\end{flushleft}
\hrule width 6.7cm \vskip.1mm{\small \small \small
$^\ast$\ Research partially supported by the EEC under the contracts
TMR-ERBFMRX-CT96-0090, TMR-ERB-4061-PL95-0789, TMR-ERBFMRX-CT96-0045
and ERBCHBG-CT94-0634.
\\
$^\ddagger$\ On leave from {\it Laboratoire de Physique Th\'eorique de
l'Ecole Normale Sup\'erieure,
CNRS,} 24 rue Lhomond, 75231 Paris Cedex 05, France.\\
$^\diamond$ On leave from {\it Centre de Physique Th\'eorique,
Ecole Polytechnique,
CNRS,} 91128 Palaiseau Cedex, France.}
\end{titlepage}
\newpage
\setcounter{footnote}{0}
\renewcommand{\thefootnote}{\arabic{footnote}}

\setcounter{section}{0}
\section{Introduction}

In four dimensions the maximal number of possible space-time
supersymmetries is $N=8$. This upper limit on $N$ follows from the
requirement that no massless states with spin greater than 2 exist
in the theory. In a realistic world, and for energies
above the electroweak scale, $E>M_Z$, we need chiral matter, and among
supersymmetric theories
only the $N=1$ possess chiral representations.
There is a general belief that in field theory spontaneous breaking of
an
$N>1$ supersymmetric
theory necessarily produces a non-chiral spectrum.
This impeded attempts \cite{fayet} to use $N>1$ supersymmetric theories
in order to describe physics beyond the electroweak scale.
In string theory, this question does not apply.
In Ref. \cite{tobe} it was shown that there are perturbative heterotic
ground states where supersymmetry is spontaneously broken from $N=2$
down to
$N=1$, which possess a chiral four-dimensional spectrum.
This opens more possibilities in string model-building, and  obviously
a more careful investigation is required when  $N>1$ is spontaneously
broken
to chiral $N=1$ models.

In general, we can assume that there might be a sequence of
supersymmetry-breaking
transitions,
$N=8\to 4 \to 2 \to 1$, that occur at intermediate-energy scales,
$\Lambda_{N}$.
We can also assume that the final scale,
corresponding to the $N=1\to 0$ supersymmetry breaking,
is relatively low, $\Lambda_{N=1}\sim O(1)$ TeV,
while  $M_{\rm s}\sim O\left(10^{17}\right){\rm \ TeV} >
\Lambda_{N>1}>O(1)$ TeV.
This scenario provides a
solution to the
hierarchy problem, and, depending on the value of the intermediate
scales
$\Lambda_N$,
$\Lambda_{N=1}$ can be pushed higher by no more than a few orders of
magnitude.
In this framework, it is important to estimate the physical
consequences of the existence of other supersymmetry-breaking scales,
$\Lambda_{N=8}$, $\Lambda_{N=4}$ and $\Lambda_{N=2}$.
To do this, we need to analyse the behaviour of the couplings in
string theory, and in particular their threshold corrections as a
function of the compactification moduli and the supersymmetry-breaking
scales $\Lambda_{N}$.

The origin of $\Lambda_{N}$ (including $\Lambda_{N=1}$)
can be either
perturbative or non-perturbative \cite{tobe}. The recent remarkable
progress in
understanding the
non-perturbative structure of string theories gives us the possibility
to study
some of the
non-perturbative aspects of partial breaking at scales $\Lambda_{N}$,
by
performing
perturbative calculations in dual string theories. We are thus led to
reconsider ``spontaneous"
versus ``explicit" supersymmetry breaking beyond perturbation theory.
Indeed,
in perturbative
string theory there exist two qualitatively different ways of reducing
the number
of
supersymmetries. In the language of orbifold compactification, some of
the
original gravitinos
are projected out from the spectrum.  We would like to distinguish the
freely-acting orbifolds
(spontaneous breaking), from the non-freely-acting ones (explicit
breaking).
A free
orbifold action is
the one that has no fixed points (strictly speaking, it should not be
called
orbifold), whereas
in a non-free action there are fixed points and some extra {\it twisted
states}
are added in the
theory. Such a definition relies on a geometrical interpretation of a
given
ground state. It
can, however, be extended to non-geometric ground states.  If the
orbifold action
that breaks
supersymmetry is free, then the associated non-invariant gravitinos
are not
projected out but
become massive. This is a stringy generalization \cite{ss,ant} of the
Scherck--Schwarz
idea \cite{ssw} of
breaking supersymmetry\footnote{There are two other mechanisms for
breaking the supersymmetry. The first is gaugino condensation
\cite{gc}, while
the second uses internal magnetic fields \cite{b}. None of them seems
to fit in the stringy
Scherk--Schwarz context.} in the context of Kaluza--Klein theories. The
low-energy behaviour of
the couplings of these two classes is markedly different.
In the non-freely-acting case
(explicit breaking) \cite{ka}--\cite{kkpra}, the low-energy theory has
no
memory of the original supersymmetry, while in
the freely-acting case (spontaneous breaking) it does \cite{tobe}.
This was verified by explicit
calculation in several classes of ground states with
spontaneously-broken
supersymmetry: $N=4$ to $N=2$
\cite{decoa}, $N=8$ to $N=6$ \cite{fkda} as well as
$N=8$ to $N=3$ \cite{kku};
furthermore the non-perturbative aspects of this problem
have been  studied utilizing
the heterotic/type~II duality \cite{tobe, GKKOPP}.

String ground states with spontaneously-broken supersymmetry
have very peculiar high-energy properties that might be desirable:
a logarithmically growing gauge
and gravitational thresholds at large moduli despite the existence of
towers of charged
states below the Planck mass, and a possibly special behaviour of the
vacuum energy. Obviously, this issue is of crucial importance in
choosing
string models that should represent the real low-energy world.

In this paper our approach is more modest. One of our goals will be
to understand in more detail the generic properties
of low-energy couplings in heterotic models, where supersymmetry
is spontaneously broken from $N=4$ to $N=2$, relevant for the physics
at
energy scales $E \sim \Lambda_{N=4}$.
In situations where
supersymmetry is further reduced, it turns out that the dependence of
the low-energy
couplings on the volumes of the internal manifold can also be obtained
from
the calculations presented here.
In a sense this paper is a generalization of \cite{decoa} to a much
wider class of heterotic ground states with spontaneously-broken
$N=4$ to $N=2$ supersymmetry.

The heterotic ground states that will be considered in the following
have
8 unbroken supercharges.
These can be thought of as compactifications of the ten-dimensional
heterotic string on a six-dimensional manifold with $SU(2)$ holonomy,
which is locally but not globally of the $K3 \times T^2$ type.
They are characterized by a set of shift vectors $w$ that act on the
two-torus. Some of these models can be constructed  starting
from the heterotic string on $T^6=T^4\times T^2$ and orbifolding by a
symmetry that involves translations on $T^2$ and non-freely-acting
transformations on $T^4$.
Another orbifold construction that belongs to the above class is the
following: orbifold a standard $K3\times T^2$
compactification by using a symmetry that is non-freely-acting on $K3$,
but preserves the hyper-K\"ahler structure and acts as a translation on
the
two-torus. It is important to stress, however, that these examples do
not exhaust all the possibilities: our analysis is valid beyond any
orbifold construction.

We will restrict ourselves to the simplest translation on the
two-torus, namely $Z_2$. In this case the shift is a half-lattice
vector of the $T^2$ Narain lattice. We will be quite general, making
no detailed
assumptions on the structure of internal (4,0) superconformal
theory.
As will be clear from our discussion below, our techniques are
directly applicable to a general translation  group on the two-torus.
In the entire class of models we will be dealing with, the original
$N=4$ supersymmetry is
spontaneously broken to $N=2$, and the two massive gravitinos have
masses
that depend on the two-torus moduli.

We will focus on threshold corrections to the gauge and $R^2$
couplings. In the presence of $N=2$ space-time supersymmetry, it is
known that the only perturbative corrections to such couplings come
from
one loop. Moreover, the only massive states that contribute at one loop
are
BPS multiplets, since the threshold is proportional to the supertrace
of
the helicity squared.
Thus, such thresholds depend on the elliptic genus of the internal
(4,0) superconformal field theory. This property is essential and it
implies, along with modular invariance, that the thresholds possess
certain
universality properties and depend only on some low-energy data.
This was already demonstrated for standard $K3\times T^2$
compactifications in \cite{kkpr, kkpra}.
Here, the thresholds will be shown to depend on the
lattice shift vector $w$, the beta-function coefficients $b_i$, the
levels $k_i$
of the associated current algebras, and the jumps of the
beta functions $\delta_h b_i$ and $\delta_v b_i$ at special
submanifolds of the moduli space of the two-torus,
where extra massless states appear
(but where gauge symmetry is not necessarily enhanced).
We will show in particular that the usual decomposition for the gauge
threshold corrections, which holds, for instance, in the standard
$K3\times T^2$
compactification, is not valid any longer and must be replaced by
$$
\Delta_i^w= b_i^{\phantom{i}}\,  \Delta^w_{\phantom{i}}(T,U) +
\dhhp b_i^{\phantom{i}} \, H^w_{\phantom{i}}(T,U) +
\dvp b_i^{\phantom{i}}\, V^w_{\phantom{i}}(T,U) +
k_i^{\phantom{i}}\, Y^w_{\phantom{i}}(T,U) \, .
$$
The moduli-dependent functions $\Delta^w$, $H^w$ and $V^w$ are
{\it universal}: they only depend on the shift vector $w$. 
On the other hand, $Y^w$ depends also on the gravitational anomaly
of the model: $Y^w_{\vphantom 1}=Y^w_1 + \bgra\,  Y^w_2$, with 
$Y^w_{1,2}$ universal.

The models under consideration in this paper have type II duals.
The easiest way to see this is to employ the construction of the
heterotic
ones as freely-acting orbifolds of the
heterotic string on $T^6$. Since the heterotic string on $T^6$ is dual
(via $S\leftrightarrow T$ interchange) to the type II string on
$K3\times
T^2$, freely orbifolding both sides will produce a new dual pair.
In \cite{tobe} heterotic/type II duality was utilized to study some
aspects of this problem.
In particular, it was shown that heterotic ground states with $N=4$
supersymmetry
spontaneously broken to $N=2$ and massive gravitinos in the
perturbative spectrum are sometimes dual to type II models
without massive gravitinos in the perturbative spectrum.
Thus, at the perturbative level, supersymmetry in the type II context
seems explicitly broken. The massive gravitinos are
BPS multiplets of the unbroken $N=2$ supersymmetry.
They  can therefore be
identified in the type II description to monopoles whose mass
is of order $1/g^2_{\II}$,
where $g_{\II}$ is the type II string coupling.
In particular they become very light at strong type II coupling,
thereby enhancing the supersymmetry.
This might indicate the possibility that in string
theory
supersymmetry is always spontaneously broken, either perturbatively, or
non-perturbatively.
There is another possibility, though. In both theories of the dual pair
there
are
potential non-perturbative corrections. It is thus possible that an
analogue
of the Seiberg--Witten non-restoration of gauge symmetry is happening
here:
non-perturbative effects
do not enable supersymmetry restoration at strong coupling.
A more careful study of this problem is necessary, which we leave for
the
future.

As we pointed out previously, the appearance of non-perturbative
corrections to the gauge
and $R^2$ couplings cannot be excluded in general.
In $N=2$ ground states, we always have the appropriate Higgs
expectation
value
that cuts off the infra-red. Thus, all non-perturbative effects are
expected to be due to instantons.
Moreover, since the $F^2$ and $R^2$ couplings are of the BPS-saturated
type
\cite{hm,bk,bk1,KO} only supersymmetric instantons (that preserve one
out
of
the two supersymmetries) are expected to contribute.
Thus, in  the four-dimensional heterotic ground states we expect
instanton corrections due to the heterotic five-brane wrapped around
the
compact internal six-dimensional manifold.

More information about the perturbative and non-perturbative
contributions is reached by
decompactifying one of the directions of the two-torus to obtain a
five-dimensional theory. Here, there are two possibilities:
(\romannumeral1)
the five-dimensional model has 16 supercharges, and in this case the
five-dimensional perturbative
thresholds are zero, as implied by the extended supersymmetry;
(\romannumeral2)
the five-dimensional model has only 8
supercharges and now the perturbative thresholds are non-zero.
In both cases, however, there are no non-perturbative instanton
corrections:
the six-dimensional world-volume of the Euclidean heterotic five-brane
cannot be wrapped around the internal space and have finite action.

The structure of this paper is as follows. In Section \ref{N2} we
present
a general description of $N=2$ heterotic ground states in terms of
their helicity-generating partition function.
The latter is expressed in terms of the elliptic genus corresponding to
the internal manifold. This is very useful for the determination of
threshold corrections. Moreover, it allows us to define the class of
models that we will be analysing throughout the paper, by giving the
generic form of their elliptic genus.

In Section \ref{N2orb} we briefly recall the general procedure that is
used for computing gauge and gravitational threshold corrections in
supersymmetric string vacua. We also present the basic properties of
these corrections in $N=2$ heterotic compactifications, where a
two-torus is factorized. These models play an important role in our
subsequent analysis, because they turn out to share some
decompactification limits with the models where the two-torus undergoes
a shift and where supersymmetry is promoted to spontaneously-broken
$N=4$.

Section \ref{spbr} is devoted to the description of the class of models
where the two-torus is not factorized. Here we stress the role of the
shift on the $T^2$, which is interpreted as a stringy Sherck--Schwarz
mechanism. Depending on the kind of shift vector, several
decompactification scenarios appear. In models where the norm $\l$ of
the shift vector vanishes, two possible decompactification limits exist
in the $(T,U)$ plane: with and without restoration of $N=4$
supersymmetry. When $\l = 1$ (the only relevant alternative), $N=4$
supersymmetry is always restored. This is in agreement with the partial
breaking of the target-space duality group, which makes several
directions in the moduli space inequivalent.

In Section \ref{thrN4} we proceed to the computation of threshold
corrections.
This is achieved by advocating general holomorphicity and
modular-covariance properties. Most of the model- and moduli-dependence
is lost at the level of the thresholds, which turn out to depend only
on the two-torus moduli $(T,U)$ as well as on several rational
parameters (discrete Wilson lines) related to some low-energy data of
the model. These are $b_i$ and $k_i$ but also $\delta b_i$, the
discontinuities of the beta-function coefficients along some specific
lines in the two-torus moduli space, where additional vector multiplets
and/or hypermultiplets become massless.
Across these lines, the thresholds diverge logarithmically. We also
observe that in the class of models under consideration, the above
low-energy parameters are in fact related in a very specific way. This
leaves some arbitrariness in the splitting of the gauge threshold
corrections into gauge-factor-dependent and gauge-factor-independent
pieces, even though we demand the latter contribution to be regular in
the $(T,U)$ space.
Moreover, some model-dependence survives in the
group-factor-independent term
$Y^w_{\phantom{i}}(T,U)$, which is not fully universal. A similar model-dependence appears in the gravitational thresholds.
These features are to be contrasted to what happens in models in which
a two-torus is factorized: the gauge threshold is uniquely
defined as a sum of two terms, one being universal and the other
group-factor-dependent, and both regular; the gravitational threshold corrections are model-independent. Finally, the behaviour of the
thresholds at various decompactification limits is analysed, and turns
out to agree with what is expected on general grounds based on the
restoration of $N=4$ supersymmetry. Again the results strongly depend
on whether the norm $\l$ of the shift vector  equals 0 or 1. The
existence of a common decompactification limit in these models and in
models with a factorized two-torus is also observed in the behaviour of
the thresholds.

As an application, we examine in  Section \ref{orb} the subclass of
$Z_2$ orbifolds. In this case, more can be said about the nature of the
extra massless states appearing along the rational lines of the $T^2$
moduli space.
In fact, a priori, these can be either vector multiplets or
hypermultiplets depending on the specific model at hand and on the
shift vector acting on the two-torus. In the case of orbifolds, {\it
only extra hypermultiplets} become massless, except for the lines $T=U$
and $T=-1/U$,
present systematically, where either hypermultiplets or vector
multiplets may appear, depending on the shift vector. This information
might be of some phenomenological relevance. The last part of Section
\ref{orb} is devoted to some specific orbifold examples for both the
situations $\l = 0$ and $\l = 1$. We construct in particular
four-dimensional ground states whose gauge group contains factors such
as $E_8 \times E_8$, $SO(40)$ or even $E_8$ realized at level 2.

Most of the technicalities are presented in appendices. In Appendix A,
we give an overview  of $Z_2$-shifted $(2,2)$ lattice sums. Rational
lines and asymptotic behaviours of the latter are also analysed there.
Appendices B and C contain the machinery used for the determination of
gauge and gravitational corrections. Finally, in Appendix D, we perform
explicitly the general
integrals over the fundamental domain, which are involved in our
expressions for the thresholds. We also analyse their singularities and
asymptotic behaviours.

\boldmath
\section{General description of $N=2$ heterotic ground
states}\label{N2}
\unboldmath

In this section we will give a brief description of the heterotic
ground states that we will be studying in the following.
They will be best described by writing their
(four-dimensional) helicity-generating partition functions.
Indeed, our motivation is eventually to compute couplings associated
with interactions such
as $F^n R^m$. Therefore,  we need in general to evaluate amplitudes
involving
operators like $i \left( x^{\mu}
\buildrel{\leftrightarrow}\over{\partial}
x^{\nu} +2\,  \psi^{\mu}
\psi^{\nu}\right)\bJ^k$,  where $\bJ^k$ is an appropriate right-moving
current and the left-moving factor corresponds to the left-helicity
operator.
We will not expand here on the various procedures that have been used
in order to
calculate exactly (i.e. to all orders in $\alpha '$ by properly taking
into account corrections due to the gravitational
back-reaction, and without infra-red ambiguities) these correlation
functions;
details on the determination of the amplitude-generating
functions relevant for gauge and gravitational
couplings can be found in Refs. \cite{kk}--\cite{kkpra}. We will
restrict ourselves to the
helicity-generating partition functions (which are also very useful
in the analysis of $S$-duality issues), defined as:
\be
Z(v,\bar v)=\Trp
q^{L_0-{c\over 24}}\,
\bar q^{\bar L_0 - {\bar c \over 24}}\,
e^{2\pi i \left(v \l-\bar v \bar \l \right)}
\, ,
\label{hel}
\ee
where the prime over the trace excludes the zero-modes related to the
space-time coordinates (consequently
$Z(v,\bar v)\vert_{v=\bar v=0}= \t_2 Z$, where $Z$ is the vacuum
amplitude), and $\l,\bar \l$
stand for the left- and right-helicity contributions to
the four-dimensional physical helicity. Various helicity
supertraces are finally obtained by taking appropriate
derivatives of (\ref{hel}).
More on these issues can be found in
Appendix G of \cite{book}.

For four-dimensional heterotic $N=4$
solutions
with maximal-rank gauge group ($r=22$) (\ref{hel}) reads\footnote{We
use the short-hand
notation $\vartheta \oao{a}{b} (v)$
for $\vartheta \oao{a}{b} (v\vert \t)$, and $\vartheta \oao{a}{b}$
for $\vartheta \oao{a}{b} (0\vert \t)$.}:
\ba
Z_{N=4}(v,\bar v)&=&{1 \over \vert\eta \vert^4}\,
{1\over 2}\sum_{a,b=0}^1 (-1)^{a+b+ab}\,
{\vartheta\oao{a}{b}(v)\over \eta}
\left({\vartheta{a\atopwithdelims[]b}\over \eta}\right)^3
\xi(v)\, \bar \xi(\bv)\,
Z_{6,22} \nn \\
&=&{1 \over \vert\eta \vert^4}
\left({\vartheta\oao{1}{1}\left({v\over 2}\right)\over \eta}\right)^4
\xi(v)\,\bar \xi(\bv)\,
Z_{6,22} \, ,
\label{het4helj}
\ea
where
$$
\xi(v)=\prod_{n=1}^{\infty}{(1-q^n)^2\over \left(1-q^n e^{2\pi
iv}\right)\left(1-q^n e^{-2\pi iv}\right)}=
{\sin\pi v\over \pi}{\vartheta_1'(0)\over
\vartheta_1(v)}
$$
counts the helicity contributions of the space-time bosonic
oscillators,
and $Z_{6,22}\equiv \Gamma_{6,22}/\eta^6 \bar  \eta^{22} $ denotes the
partition function of
six compactified coordinates as well as of sixteen right-moving
currents; it depends
generically on 132 moduli, namely 36 internal background metric and
antisymmetric tensor fields, and 96 internal background gauge fields
(Wilson lines). It is possible to continuously connect several
extended-symmetry
points such as $U(1)^6 \times E_8  \times E_8$, $U(1)^6 \times SO(32)$
or $SO(44)$. The $(4,0)$ supersymmetry is read off automatically from
expression
(\ref{het4helj}),
which has a fourth-order zero at $v=0$. Theories with lower-rank gauge
group and the same
supersymmetry can be easily
constructed by modding out discrete symmetries, which correspond
to outer automorphisms and act without fixed points on the lattice
\cite{GKKOPP,CHL}.

Supersymmetry can be reduced to $N=2$ in various ways. Generically,
the helicity-generating function reads:
\be
Z^{\rm generic}_{N=2}(v,\bar v)={1 \over \vert\eta \vert^4}\,
{1\over 2}\sum_{a,b=0}^1 (-1)^{a+b+ab}\,
{\vartheta\oao{a}{b}(v)\over \eta}
{\vartheta{a\atopwithdelims[]b}\over \eta}
\xi(v)\, \bar \xi(\bv)\,
C_{6,22}\oao{a}{b}(v)\, ,
\label{het2hel}
\ee
where $C_{6,22}\oao{a}{b}(v)$ are traces in the internal superconformal
field
theory.
We have kept explicit the $\vartheta$-function contribution of
the left-moving fermions of the two transverse space-time
coordinates as well as those of the
internal two-dimensional free theory.
The $(6,22)$ internal field
theory has $c_{\rm R}=22$ on the right sector, while on the left sector
$c_{\rm L}=8$. This sector has
a two plus four split: the two-dimensional
part of it is
free, with central charge $c = 2$, while the four-dimensional one is
$N=4$ superconformal with $\hat c = 4$ \cite{bd}.
Space-time supersymmetry can be
used again to write (\ref{het2hel}) as:
\be
Z^{\rm generic}_{N=2}(v,\bv)={1 \over \vert\eta \vert^4}
 \left({\vartheta{1\atopwithdelims[]1}
\left({v\over 2}\right)\over \eta}\right)^2
\xi(v)\,\bar \xi(\bv)\,C_{6,22}{1\atopwithdelims[]1}\left({v\over
2}\right)\, ,
\label{het2helj}
\ee
where
$$
C_{6,22}{1\atopwithdelims[]1}\left({v\over 2}\right)=\Tr ^{\rm
int}_{\rm R}
(-1)^{F^{\rm int}}q^{L^{\rm int}_0-{1\over 3}} \,
\bar q^{\bar L^{\rm int}_0 - {11 \over 12}} \, e^{2\pi i v J_0}
$$
is the (generalized) elliptic genus of the internal conformal field
theory. The standard elliptic genus \cite{ellwit,Windey}, relevant for
the
gravitational
threshold corrections is obtained at $v=0$. The charge
$J_0$ is the sum of the internal $U(1)$-current zero-modes of the $N=2$
and
$N=4$ internal superconformal algebras.

An interesting class of models is provided when the ten-dimensional
theory is
generically compactified to six dimensions in a way that preserves 8
supercharges out of 16, and is toroidally compactified further down to
four dimensions. In that case, the $T^2$ contribution factorizes
completely.
We obtain:
\be
C_{6,22}{1\atopwithdelims[]1}\left({v\over 2}\right)=
C_{4,20}\left({v\over 2}\right) Z_{2,2}\, ,
\label{g2}
\ee
where $C_{4,20}(v/2)$ is
a (left-helicity-generating)
conformal block with two left and no right world-sheet supersymmetries,
and central charges $\hat c_{\rm L}=4$, $c_{\rm R}=20$. It accounts for
the compactification from ten to six
dimensions, and actually defines the generalized elliptic genus
for the four-dimensional compact manifold plus a gauge bundle on it
with instanton number 24.
This conformal block depends in particular on several moduli (other
than
the two-torus ones). On the other hand,
$Z_{2,2}\equiv \Gamma_{2,2}/ \vert\eta \vert^4$ is the
partition function of the two-torus (\ref{z22}),
whose complex moduli are $T$ and
$U$. It is invariant under the full
target-space duality group $SL(2,Z)_T^{\vphantom U} \times
SL(2,Z)^{\vphantom T}_U \times Z_2^{T \leftrightarrow U}$. More
details about lattice sums can be found in Appendix A.

The above $N=2$ construction
(\ref{g2})
captures many compactifications such as
$K3$ or orbifold models.
For example, (symmetric or asymmetric) $Z_2$ twists acting at the level
of the $N=4$ model
(\ref{het4helj}) leave invariant a single complex
plane corresponding to a $T^2$ compactification from six to four
dimensions. They therefore belong to the above class of $N=2$ ground
states.
Their helicity-generating
function is given by Eqs. (\ref{het2helj}) and (\ref{g2}) with
\be
C^{\rm orb}_{4,20}\left({v\over 2}\right)=
{1\over 2}\sum_{h,g=0}^1
{\vartheta{1+h\atopwithdelims[]1+g}
\left({v\over 2}\right)\over \eta}
{\vartheta{1-h\atopwithdelims[]1-g}
\left({v\over 2}\right)\over \eta}\,
 Z_{4,20}^{\vphantom{o}} {h\atopwithdelims[]g}\, ,
\label{corb}
\ee
where
$Z_{4,20} {h\atopwithdelims[]g}\equiv
{\Gamma_{4,20}{h\atopwithdelims[]g}\Big/ \eta ^4 \,
\bar \eta^{20}}$ summarize the bosonic orbifold blocks; in particular
the untwisted partition function that describes the right-moving
currents and the four compactified coordinates
is $Z_{4,20} {0\atopwithdelims[]0}\equiv Z_{4,20}$; it
depends on 80 moduli.
The  $Z_2$-twisted contributions
are  moduli-independent. They  can be
constructed in many consistent ways, provided they satisfy the
periodicity and
modular-invariance requirements (see Eqs. (\ref{255a}), (\ref{256a})
with $\l=0$).

The best-known example of the orbifold construction is the
symmetric $Z_2$ orbifold, which turns out to belong also to the $K3$
moduli space.
This model is achieved by going to a
point of the $(4,20)$ moduli space, where a four-torus is
factorized\footnote{Its $(4,20)$
lattice sum is actually given in (\ref{420O1r}), which provides also a
relevant example in
the framework of heterotic ground states where the $T^2$ is not
factorized.}.
The gauge group
is $E_{8}\times
E_7\times SU(2)\times{U(1)}^2$. Thus $N_V=386$, while
$N_H=628$
($N_V$ and $N_H$ are the number of
vector multiplets and hypermultiplets, respectively).

The gravitational and gauge couplings of the $N=2$ heterotic ground
states
with a factorized two-torus
have been studied
extensively. In the present paper, our goal is to
analyse more
general situations, where Eq. (\ref{g2}) does not hold any longer and
is replaced by
\be
C_{6,22}{1\atopwithdelims[]1}\left({v\over 2}\right)=
{1\over 2}\sum_{h,g=0}^1
C_{4,20}^{\l}\oao{h}{g}\left({v\over 2}\right)
\,
Z_{2,2}^{w}{h\atopwithdelims[]g} \, ,
\label{helsb}
\ee
where
$Z_{2,2}^{w}{h\atopwithdelims[]g}$
stands for the shifted partition function of the two-torus.
Such a structure appears, for instance, in freely-acting orbifolds that
reduce $N=4$ supersymmetry to $N=2$,
and act with a lattice shift on the two-torus.
Equations (\ref{het2helj}) and (\ref{helsb}) describe more general
$N=2$ ground states, which have always a
$U(1)^2$
right-moving gauge group coming from the two-torus. They are obviously
not of
the factorized form
$K3\times T^2$, but they correspond to compactifications on
six-dimensional
manifolds of $SU(2)$ holonomy.
In  Section
\ref{spbr},
where these models are described in detail,
we will argue that $N=2$ supersymmetry is promoted to {\it
spontaneously-broken} $N=4$. The corresponding threshold corrections
will be
computed in Section \ref{thrN4}.

\boldmath
\section{Thresholds in general and toroidal compactifications
from six dimensions}\label{N2orb}
\unboldmath

\subsection{Computation in generic heterotic
supersymmetric string vacua}\label{thrgen}

Threshold corrections appear in the relation between the
running gauge coupling $g_i(\mu)$
of the  low-energy effective field theory
and the string coupling constant $\gs$, which is associated with the
expectation value
of the dilaton field. For supersymmetric ground states
and non-anomalous $U(1)$'s, one can introduce, at the string level, an
infra-red
regularization prescription such that it becomes possible to match
unambiguously string theory and low-energy effective-field-theory
amplitudes
\cite{kk}--\cite{kkpra}, thus leading to the relation:
\be
{16\, \pi^2\over g_{i}^2(\mu)} = k_{i}{16\, \pi^2\over \gs^2} +
 b_{i}\log {M_{\rm s}^2\over \mu^2} +\Delta_{i}\, ,
\label{1}
\ee
which is actually expected if one assumes the decoupling of massive
modes
\cite{ka}--\cite{agnt}.
In this expression, $\mu$ is the infra-red scale, while
$M_{\rm s}=1/\sqrt{\alpha'}$ is the string scale. String unification
relates
the
latter to the Planck scale
$M_{\rm P}={1/\sqrt{32\, \pi G_{\rm N}}}$
and to the
string
coupling constant. At the tree level this relation reads:
\begin{equation}
M_{\rm s} = \gs\,  M_{\rm P}\label{msmp}\, ;
\label{three}
\end{equation}
notice that for supersymmetric vacua (\ref{three}) does not receive
any
perturbative correction \cite{kkpr}.

In the $\overline{DR}$ scheme for the
effective field theory,
the thresholds read \cite{kkpr, kkpra}:
\be
\Delta_{i}=\int_{\cal F}{{\rm d}^2\t\over \im}
\left(
F_i-
b_{i}
\right)
+b_{i}\log {2\,e^{1-\gamma}\over \pi\sqrt{27}} \, ,
\label{2}
\ee
where, in presence of supersymmetry, the function $F_i$ is defined by
the following genus-one string amplitude:
\be
F_i=\left\langle
-{\l}^2\left(
\bP_i^2 - {k_{i}\over 4\pi{\im}}
\right)\right\rangle_{\rm genus-one}\, .
\label{Fi}
\ee
Here $\l$ is the left-helicity operator introduced above,
$\bP_i$ is the charge operator of the gauge
group $G_i$ (for conventions see \cite{kk,pr}),
$k_i$ is the level of the $i$th gauge group factor, and $b_i$ are
the full beta-function coefficients,
\be
b_{i}=
\lim_{\im\to\infty}
F_i \, .
\label{2a}
\ee
Generically, we can express any $N=1$ heterotic vacuum
amplitude in the canonical form
\be
Z={1\over \im |\eta|^4}\, {1\over 2}\sum_{a,b=0}^1
{\vartheta{a\atopwithdelims[]b}\over \eta}\,
C{a\atopwithdelims[]b}\, ,
\label{34a}
\ee
where $C{a\atopwithdelims[]b}$ are related to the various sectors of
the internal six-dimensional
partition function. By using this form, we can recast Eq. (\ref{Fi})
as follows:
\be
F_i =
{i\over 2\pi}\, {1 \over|\eta|^4}
\sum_{a,b=0}^1
{\partial_{\t}
\vartheta{a\atopwithdelims[]b}\over\eta}
\left(
\bP_i^2 - {k_{i}\over 4\pi{\im}}\right) C{a\atopwithdelims[]b}\, .
\label{2gau}
\ee

One can similarly introduce the function
\be
F_{\rm grav} =
{i\over 24\, \pi}\, {1 \over|\eta|^4}
\sum_{a,b=0}^1
{\partial_{\t}
\vartheta{a\atopwithdelims[]b}\over\eta}
\left(
\bE_2 - {3\over \pi{\im}}\right) C{a\atopwithdelims[]b}\, ,
\label{2grav}
\ee
where $E_{2n}$ is the $n$th Eisenstein series (see
Appendix B).
This function plays a similar role in the determination of
the gravitational threshold corrections, which appear in the
renormalization of the $R^2$ term \cite{kkpr}:
\be
\Delta_{\rm grav}=\int_{\cal F}{{\rm d}^2\t\over \im}
\left(F_{\rm grav}-
b_{\rm grav}
\right)
\label{gravth}
\ee
(up to a constant term),
where
\be
b_{\rm grav} =  \lim_{\im \to\infty}\left(F_{\rm grav}
+{1 \over 12\, \bar q}\right)
\label{2ag}
\ee
is the gravitational anomaly in units where  a hypermultiplet
contributes
$1/12$ \cite{agnt}.

\subsection{The case of the $N=2$ ground states with a
factorized $T^2$}\label{thrN2}

Let us now concentrate on $N=2$ ground states that come from toroidal
compactification of generic
six-dimensional $N=1$ string  theories.
We recall here the determination of the gravitational and gauge
couplings for these models \cite{kkpra}, because it plays a significant
role in the analysis
of the more general constructions presented in Section \ref{spbr}:
those
two classes of ground states turn out to share large-moduli
limits (see Section \ref{spbr2}).
We will focus in particular on the couplings of group factors
corresponding to the rank-20 part\footnote{Actually, this part of the
gauge group is at most of rank 20.
For convenience, we will, however, keep on referring to it as the
``rank-20 component", in order to distinguish it from the two-torus
contribution.}
 of the gauge group, which were already present in
the six-dimensional theory and not on the corrections to the couplings
of the $U(1)$'s
originated from the two-torus (or the $SU(2)$'s or $SU(3)$ appearing at
extended-symmetry points of the $T$, $U$ moduli). In other words, the
charge
operator
$\bP_{i}^2$ will not act on the lattice sum $\Gamma_{2,2}$.
For the models at hand the
helicity-generating
function is given by (\ref{het2helj}) with (\ref{g2}).

By comparing the latter
(at $v=0$) to
(\ref{34a}), and using (\ref{2gau}) and (\ref{2grav}), we obtain:
\be
F_{i}=-{\Gamma_{2,2}\over \bar \eta^{24}}
\left(\bP_{i}^2-{k_{i}\over 4\pi\im}
\right)
\overline{\Omega}\ , \ \
F_{\rm grav}=-{\Gamma_{2,2}\over \bar \eta^{24}}
\left({\bE_2\over 12}-{1\over 4\pi\im}
\right)
\overline{\Omega}
\, ,
\label{57}
\ee
where
\be
\overline{\Omega}\equiv -{1\over 2}\,
\bar \eta^{20} \left. C_{4,20}\right\vert_{v=0}\, .
\label{58}
\ee

A few remarks are in order here. The first one concerns the
antiholomorphicity and universality properties
of the function $\overline{\Omega}$ defined in Eq. (\ref{58}). Indeed,
by analysing the relevant two-point amplitudes in six dimensions, it
was shown in \cite{ler} that, for vanishing $v$, $C_{4,20}$ is a purely
antiholomorphic function, the elliptic genus.
Put another way, $C_{4,20}\vert_{v=0}$
is essentially the supertrace of the
left-helicity squared (see Eq. (\ref{Fi})), and thus it
receives contributions from massless and massive BPS states only.
Such states are necessarily of the form
left-moving vacuum times right-moving excitations.

Six-dimensional anomaly cancellation\footnote{At the level of the
four-dimensional spectrum, this constraint reads
$N_H - N_V =242$, at generic points of the two-torus moduli space.
It translates into
$b_{\rm grav}= 22$, since in our normalizations
$b_{\rm grav} = {22 - N_V +N_H\over 12}$.
Along the enhanced-symmetry line $T=U$, $N_H - N_V =240$; it can even
reach
the values $238$ or $236$ when $T=U=i$ or $T=U=\rho$ respectively.}
forces the function $C_{4,20}\vert_{v=0}$ to be independent of the
kind of compactification that has been used to go from ten to six
dimensions.
Following \cite{kkpr, kkpra},
we therefore conclude that for the models under
consideration\footnote{Notice that in the case of orbifolds, Eqs.
(\ref{corb}) and (\ref{58}) lead to
the result (\ref{Omb}) by direct calculation.}
\be
\overline{\Omega}=\bE_4 \, \bE_6\, .
\label{Omb}
\ee

It is important to stress here that the above universality property
applies
exclusively to the elliptic genus and could not be promoted at the
level of the full model. In other words, the data
$C_{4,20}\vert_{v=0}=-{2 \bE_4  \bE_6/ \bar \eta^{20}}$
do not enable us to reconstruct the full function $C_{4,20}(v/2)$,
which is in general model- (and moduli-) dependent.

By using the above result (\ref{Omb}), we can go further: if we
advocate again
holomorphicity properties and demand (\ref{2a}) as well as the
absence of tachyon contribution in $F_i$, we determine the
action of the charge operator $\bP_{i}^2$ with the result
\cite{hm, pr, kkpr, kkpra}:
\ba
F_{i}&=&k_{i}\left( F_{\rm grav}  +  \Gamma_{2,2}
\left(
{\bar j\over 12}-84
\right)
\right)+b_i\, \Gamma_{2,2}\cr
&=&-{k_i \over 12}\, \Gamma_{2,2}\left(
{\widehat E_2 \, \bE_{4}\, \bE_{6}\over \bar
\eta^{24}}
-\bj+1008\right)+b_i\, \Gamma_{2,2}
\, ,
\label{59}
\ea
where $\widehat E_2$ stands for the modular covariant combination
$\bE_{2}-{3\over \pi\im}$ and $j(\t)={1\over q}+744+O(q)$,
$q=\exp 2\pi i\t$, is the standard $j$-function.
Therefore one can write
\be
\Delta_{i}=b_{i}^{\vphantom N}\,
\Delta-k_{i}^{\vphantom N} \, Y\, ,
\label{66a}
\ee
with
\ba
\Delta &=&
\int_{\cal F}{{\rm d}^2\t\over \im}
\Big(\Gamma_{2,2}\left(T,U,\bT,\bU\right)-1\Big)
+\log {2\,{\rm e}^{1-\gamma}\over \pi\sqrt{27}}\cr
&=&-\log\left(4\pi^2 \,\big|\eta(T)\big|^4 \, \big|\eta(U)\big|^4\,
T_2 \, U_2\right)
\label{67}
\ea
and
\be
Y={1 \over 12}\int_{\cal F}{{\rm d}^2\t\over \im}\,
\Gamma_{2,2}\left(T,U,\bT,\bU\right) \left(
{\widehat E_2\, \bE_{4}\, \bE_{6}\over \bar
\eta^{24}}-\bj+1008
\right) .
\label{68}
\ee
A further analysis of this gauge-factor-independent threshold can be
found in \cite{kkpra}.

Our second comment is related to the absence of ${1\over \bar q}$-pole
in
expression (\ref{59}). If such a pole were present, combined with the
lattice sum $\Gamma_{2,2}$, it would generate an extra constant term in
$F_i$, at some special line of the two-torus moduli space (such as
$T=U$, $T=U=i$ or $T=U=\rho\equiv\exp {2\pi i \over 3}$). This would
lead to
a jump in the
beta-function coefficients $b_i$, proportional to $k_i$ and due to
extra massless states charged under the $i$th gauge-group factor.
It is clear from the above analysis that this phenomenon does not
occur. In other words the extra massless states that do appear at
extended-symmetry points carry no charge with respect to the rank-20
component of the gauge group that is considered here. A
straightforward consequence of this situation is
the regularity of universal contributions $Y$ (see Eq. (\ref{68})) all
over
the $T^2$ moduli space. As we will see in the following, this picture
will change drastically in ground states where $N=2$ supersymmetry is
realized as a spontaneous breaking of $N=4$. Notice
that, already in the case at hand, the gravitational anomaly receives
an extra
contribution
\be
\delta_t\bgrav=-{\delta \over 6}\ , \ \ \delta = 1, 2 {\rm \ or \ } 3,
\label{d69}
\ee
when $T=U$, $T=U=i$, or $T=U=\r$,
respectively. This is due to the appearance of
$\delta_t N_V = 2 \delta$
extra vector multiplets, while $\delta_t N_H=0$.

The gravitational
thresholds
are given by (see (\ref{gravth}), (\ref{57}) and (\ref{Omb}))
\be
\Delta_{\rm grav}^{\rm gen}=-{1 \over 12}\int_{\cal F}{{\rm d}^2\t\over
\im}\, \left(
\Gamma_{2,2}
{\widehat E_2\, \bE_{4}\, \bE_{6}\over \bar
\eta^{24}}+264
\right)
\label{69}
\ee
at generic points of the $T^2$ moduli space and have a singular
behaviour along the line $T=U$. For these values of the moduli, it is
necessary to properly subtract the full $\bgrav=22+\delta_t \bgrav$
so as to avoid
logarithmic divergences in the integral (\ref{69}). Notice finally
that the gravitational thresholds are identical for all $N=2$ models
with a factorized two-torus (as is the gauge-factor-independent term of
the gauge thresholds).

The last observation we would like to make here concerns the
determination
of ${\overline{\Omega}}$ defined in (\ref{58}). Although the solution
given in
(\ref{Omb}) is the one that satisfies all the requirements
(antiholomorphicity, regularity in the $\tau$-plane, \dots), there is
another
possibility
that one should not disregard, namely $\overline{\Omega}=0$.
In that case $F_i=F_{\rm grav}=0$ and, as a consequence, all
beta-function coefficients
and the gravitational anomaly vanish: the ground state at hand
actually possesses
$N=4$
supersymmetry. In that case, the two extra space-time supersymmetries
appear as a
 conspiracy of left-moving zero-modes originated from the $(\hat
c_L=4,c_R=20)$
conformal block (this can happen, e.g. in orbifolds, since in some cases
two extra
massless gravitinos can appear from the twisted sector). Such models
will appear in Section~\ref{thrN4}, sharing
decompactification limits with $N=2$ models where $N=4$
supersymmetry
is broken spontaneously.

\boldmath
\section{Models with spontaneously-broken $N=4$ to $N=2$
supersymmetry}\label{spbr}
\unboldmath

\subsection{General models and helicity-generating
function}\label{spbr1}
As was announced at the end of Section \ref{N2},
we will now construct different $N=2$ models in four dimensions,
which can be
represented as ground states where $N=4$ supersymmetry is
{\it spontaneously broken} to $N=2$.

We will describe a representative orbifold construction \cite{decoa},
which we will
then generalize beyond orbifolds.
Orbifolding consists in performing a $Z_2$ rotation in the $(4,20)$
part
of the original $N=4$ model (and which would project out two
of the four gravitinos) together with a $Z_2$ lattice shift on the
$T^2$.
Here the orbifold group acts without fixed points.
The two gravitinos that would have been projected out combine with a
state
carrying $T^2$ momentun (or winding depending on the lattice shift)
and survive the orbifold projections.
They are massive, however, and their mass is an easily computable
function of the $T^2$ moduli.
In these orbifolds, $N=4$ supersymmetry is spontaneously broken
to $N=2$.

The partition function for the orbifold models under consideration can
be
written in the following way:
\begin{eqnarray}
Z^{\ \rm orb}_{\, \rm sp \ br}&=&{1 \over \im \vert\eta \vert^4}\,
 {1\over 2}\sum_{a,b=0}^1 (-1)^{a+b+ab}
 \left({\vartheta{a\atopwithdelims[]b}\over \eta}\right)^2 \cr
&& \mbox{}\times{1\over 2}\sum_{h,g=0}^1
 {\vartheta{a+h\atopwithdelims[]b+g}\over \eta}
{\vartheta{a-h\atopwithdelims[]b-g}\over \eta}\,
 Z_{4,20}^{\lambda}{h\atopwithdelims[]g}\,
 Z_{2,2}^{w}{h\atopwithdelims[]g} \, .
\label{hetsb}
\end{eqnarray}
The shifted lattice sum $\Gamma^{w}_{2,2}{h\atopwithdelims[]g}$
appearing in (\ref{25})
$$
Z_{2,2}^{w}{h\atopwithdelims[]g}
\equiv{\Gamma_{2,2}^{w}
{h\atopwithdelims[]g}\over
|\eta|^4}
$$
is given in
(\ref{44444}) and (\ref{4444l}). It depends on two integer-valued
two-vectors (see Appendix~A)
$\vec a$ and $\vec b$, whose components are defined modulo 2, and we
use the short-hand notation $w\equiv (\vec a, \vec b)$.
The modular properties are captured in a single $O(2,2,Z)$-invariant
parameter $\lambda
\equiv \vec a \vec b$, which allows us to distinguish two cases of
interest: $\lambda = 0$ and $\lambda = 1$ \footnote{It can be shown
(see
Appendix A) that
other values of $\l$ are related to the above
by lattice periodicity.}.

Modular invariance of the full partition function can be advocated for
determining how the $Z_2$-twisted contributions
$Z_{4,20}^{\l}{h\atopwithdelims[]g}$ should transform.
By using Eqs. (\ref{255b}) and (\ref{256b}) we find:
\be
\tau\to\tau+1 \ ,\ \ Z_{4,20}^{\l}{h\atopwithdelims[]g}\to
{\rm e}^{i\pi\left({4\over 3}+(1-\l){h^2 \over 2}\right)}\,
Z_{4,20}^{\l}{h\atopwithdelims[]h+g}
\label{255a}
\ee
\be
\tau\to-{1\over \tau} \ ,\ \ Z_{4,20}^{\l}{h\atopwithdelims[]g}\to
{\rm e}^{-i\pi(1-\l){hg}}\, Z_{4,20}^{\l}{g\atopwithdelims[]-h}\, .
\label{256a}
\ee
In the case $\l=0$, modular invariance allows us to use the same
twisted
partition functions $Z_{4,20}{h\atopwithdelims[]g}$ as those appearing
in $N=2$ ground states with a factorized two-torus (Eqs.
(\ref{het2helj}) and
(\ref{g2})). The case
$\l=1$, however, necessitates the introduction of slightly different
twists, since
$Z_{4,20}^{\l=1}{h\atopwithdelims[]g}$ must now transform with
different phases. Examples will be worked out in Section \ref{orb},
Eqs.
(\ref{420O1r})--(\ref{4201E8}).

Here we would like to pause and examine the possibility of
generalizing the construction presented so far to models where a $Z_2$
acts freely on the two-torus whereas {\it the compactification from ten
to
six dimensions is not necessarily an orbifold}. This can be achieved by
looking first at the helicity-generating function of the orbifold
models with spontaneously-broken $N=4$ supersymmetry (\ref{hetsb}).
This function
is actually given by (\ref{het2helj}) with
\be
C_{6,22}^{\rm orb}{1\atopwithdelims[]1}\left({v\over 2}\right)=
{1\over 2}\sum_{h,g=0}^1
C_{4,20}^{{\rm orb,}\, \l}\oao{h}{g}\left({v\over 2}\right)
\,
Z_{2,2}^{w}{h\atopwithdelims[]g} \, ,
\label{orbhelsb}
\ee
where
\be
C_{4,20}^{{\rm orb,}\, \l}\oao{h}{g}\left({v\over 2}\right)=
{\vartheta{1+h\atopwithdelims[]1+g}
\left({v\over 2}\right)\over \eta}
{\vartheta{1-h\atopwithdelims[]1-g}
\left({v\over 2}\right)\over \eta}\,
 Z_{4,20}^{\l} {h\atopwithdelims[]g}\, .
\label{corbsp}
\ee
Expression (\ref{orbhelsb}) is actually the most adequate for further
generalization. Indeed, instead of (\ref{corbsp}), we
can use more general blocks $C^{\l}_{4,20}\oao{h}{g}(v/2)$ such that
the internal $(4,20)$ theory has $N=4$
left-moving
superconformal symmetry. We can
thereby construct the most general
heterotic
four-dimensional ground states with $N=2$ supersymmetry, which can be
enhanced to
$N=4$. For these,
the helicity-generating function is (\ref{het2helj}) with
(\ref{helsb}), as advertised in Section~\ref{N2}.
Modular
covariance demands that
\be
\tau\to\tau+1 \ ,\ \ v\to v\ , \ \
C_{4,20}^{\l}{h\atopwithdelims[]g}\left({v\over 2}\right)\to
{\rm e}^{i\pi \left({5\over 3}-{\l}{h^2 \over 2} \right)}\,
C_{4,20}^{\l}{h\atopwithdelims[]h+g}\left({v\over 2}\right)
\label{gentr1}\ee
\be
\tau\to-{1\over \tau} \ ,\ \ v\to {v\over \tau}\ , \ \
C_{4,20}^{\l}{h\atopwithdelims[]g}\left({v\over 2}\right)\to
-{\rm e}^{i\pi\left({v^2 \over 2\tau}+{\l}{hg}\right)}\,
C_{4,20}^{\l}{g\atopwithdelims[]-h}\left({v\over 2}\right) .
\label{gentr2}\ee

In general, the model-dependent functions
$C_{4,20}^{\l}{h\atopwithdelims[]g}(v/2)$ depend on several (continuous
or discrete) moduli. The ($N=4$)-sector contribution
$C_{4,20}^{\l}{0\atopwithdelims[]0}(v/2)$ \textit{is in fact the one
given in
(\ref{corbsp}) for $(h,g)=(0,0)$} \cite{bd}, namely
\be
C_{4,20}^{\l}{0\atopwithdelims[]0}\left({v\over 2}\right)
 =\left({\vartheta{1\atopwithdelims[]1}
\left({v\over 2}\right)\over \eta}\right)^2 Z_{4,20}\,,
\label{corbsp00}
\ee
and does not depend on the choice of $\l=0,1$.
The other sectors, however,
might or might not be connected to some orbifold realization captured
in (\ref{corbsp}).
At $v=0$, $C_{4,20}^{\l}{0\atopwithdelims[]0}$ vanishes because of the
fermionic zero-modes and, as we will see later, for $(h,g)\neq (0,0)$,
$\left. C_{4,20}^{\l}{h\atopwithdelims[]g}\right|_{v=0}$ are purely
antiholomorphic functions. This last property puts severe constraints
on $\left. C_{4,20}^{\l}{h\atopwithdelims[]g}\right|_{v=0}$ and is
responsible for the absence of continuous-moduli dependence inside the
threshold corrections (see Section \ref{thrN4}).

\subsection{Decompactification limits and restoration
of $N=4$ supersymmetry}\label{spbr2}

As we mentioned above, there are two possible values for the parameter
$\l$, which lead to fundamentally different ground states. In the case
$\l =
0$, any model with spontaneously-broken $N=4$ supersymmetry of the type
(\ref{het2helj}) with
(\ref{helsb}) can be mapped onto a ground state with $N=2$
supersymmetry of
the type $K3 \times T^2$ studied in Section \ref{N2}. This mapping is
achieved
by defining the function $C_{4,20}(v/2)$, which appears in (\ref{g2})
in terms of
the blocks $C_{4,20}^{\l=0}{h\atopwithdelims[]g}(v/2)$ appearing in
(\ref{helsb}):
\be
C_{4,20}^{\vphantom{\l}}\left({v\over 2}\right)=
{1\over 2}\sum_{h,g=0}^1
C_{4,20}^{\l=0}{h\atopwithdelims[]g}\left({v\over 2}\right) ,
\label{map}
\ee
in agreement with all properties (modular transformations, \dots) that
these functions must satisfy.
For
$\l = 1$ it is not possible to establish such a kind of mapping.

This manipulation is actually deeper than
a formal construction, the two models being closely related in their
six-dimensional decompactification limit.
In fact, as we point out in Appendix A (see Eq. (\ref{l0lis}) for
the $\l=0$ shifted lattice sum I) any $\l=0$ shifted lattice sum
possesses a decompactification limit in which
$\Gamma^w_{2,2}\oao{h}{g}$ are equal for all $(h,g)$ (and in particular
equal to the limit of $\Gamma_{2,2}$). By comparing (\ref{g2})
and (\ref{helsb}), we thus conclude that,
in this six-dimensional limit\footnote{ For this lattice sum I, this
limit is $T_2\to0, U_2=1$.},
the $N=2$ ground state with spontaneously-broken supersymmetry
(i.e. with the shifted $(2,2)$
lattice) and the ordinary $N=2$ ground state (i.e. with the unshifted
$(2,2)$
lattice), which is mapped on the former through (\ref{map}), are in
fact
identical. It can also be argued that these two ground states, related
through (\ref{map}),
possess actually the same gauge group. Their matter content is,
however,
different.

On the other hand, any $\l=0$ or $\l=1$ model with
spontaneously-broken space-time supersymmetry of the type
(\ref{helsb})
can be mapped onto an $N=4$ model by keeping the $\oao{0}{0}$ sector
 only. Indeed, as we mentioned above,
this sector is the $N=4$ sector of the
original model whose conformal block
$C^\l_{4,20}\oao{0}{0}\left({v/2}
\right)$ is given in (\ref{corbsp00}). The corresponding
$N=4$ heterotic model (see  (\ref{het4helj})) is therefore
defined by a $(6,22)$ lattice factorized as:
\be
Z_{6,22}\to Z_{4,20}\, Z_{2,2}\, .
\label{mapp}
\ee

Again, this formal connection between an $N=2$ model with shifted
lattice and
an $N=4$ model can be made more concrete by observing that they do
have
a common six-dimensional limit. Indeed, either
in $\l=0$ or in $\l=1$ shifted sums, there is a
decompactification limit\footnote{
These limits are $T_2\to\infty$ and $U_2=1$ for both models I ($\l=0$)
and
X ($\l=1$) (see Eqs. (\ref{lolil}), (\ref{l1li}) and (\ref{l1li00})).}
where only $\Gamma_{2,2}^w\oao{0}{0}\equiv\Gamma_{2,2}$ survives,
thereby
selecting
the $N=4$ sector of the model (\ref{helsb}). Thus the
original $N=2$ ground state and the $N=4$ ground state obtained by
using the above
mapping are identical in that limit, provided the $(T,U)$ moduli are
appropriately rescaled in order to re-absorb the factor $1/2$
present in (\ref{helsb}) (for the lattices I and X, for example, we
must
perform $T\to2T$ in the $N=2$ model).

The previous observations show that $N=4$ supersymmetry is restored in
some appropriate six-dimensional decompactification limit of the
$N=2$
model built up with shifted lattices. This is a manifestation
of the underlying Scherk--Schwarz mechanism responsible for the
spontaneous breaking of the $N=4$
supersymmetry.
The same conclusion can be reached by analysing the behaviour of the
($T$- and $U$-dependent) mass of the two gravitinos (see \cite{tobe}
and \cite{decoa}).
For $\l=0$ ground states, there are two inequivalent limits in the
$(T,U)$
moduli
where the masses of both gravitinos either vanish or become infinite.
These two limits, when $N=4$ supersymmetry is and is not restored,
coincide respectively with the limits of some ordinary (i.e. with
factorized two-torus)
$N=4$
and
$N=2$ models. When the shift vector of the $(2,2)$
lattice
is of the type $\l=1$, the mass gap of two gravitinos always vanishes
at the
decompactification limit, and the $N=4$ supersymmetry is always
restored in
six dimensions.

We have summarized the above results in Figs. \ref{fig1} and
\ref{fig2}.
As
a final remark, we would like to mention another possibility that can
appear in $N=2$ ground states constructed with $(2,2)$ lattices such
that the shift vector satisfies
$\l=0$. As we will see in Section \ref{thrN4}, it can happen that the
ordinary
$N=2$ model obtained through the mapping (\ref{map}) possesses the
following
property:
\be
\sum_{h,g=0}^1C_{4,20}^{\l=0}{h\atopwithdelims[]g}\left({v\over
2}\right)
\stackrel{v\to0}{\longrightarrow}0\, .
\label{mapN4}
\ee
The $N=2$ supersymmetry of the latter model is actually promoted to
$N=4$
(see discussion at the end of Section \ref{thrN2}), therefore leading
to
the picture summarized in Fig. \ref{fig3}. Ground states that possess
the property (\ref{mapN4}) will be referred to as belonging to class
(\romannumeral2), whereas generic $\l = 0$ models (Fig. \ref{fig1})
will be of class (\romannumeral1).

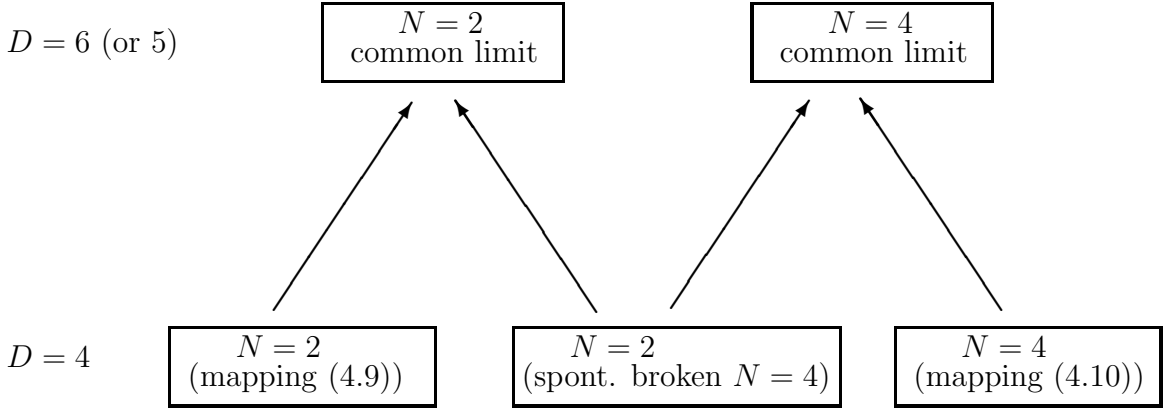
\begin{figure}
\setlength{\unitlength}{0.012500in}%
\setlength{\unitlength}{0.010000in}%
\begin{picture}(576,210)(40,525)
\thicklines
\put(431,695){\framebox(125,40){}}
\put(445,705){\makebox(0,0)[lb]{\smash{common limit}}}
\put(470,720){\makebox(0,0)[lb]{\smash{$N=4$}}}
\put(206,695){\framebox(125,40){}}
\put(245,720){\makebox(0,0)[lb]{\smash{$N=2$}}}
\put(220,705){\makebox(0,0)[lb]{\smash{common limit}}}
\put(126,525){\framebox(138,40){}}
\put(160,550){\makebox(0,0)[lb]{\smash{$N=2$}}}
\put(135,535){\makebox(0,0)[lb]{\smash{(mapping (\ref{map}))}}}
\put(306,525){\framebox(170,40){}}
\put(335,550){\makebox(0,0)[lb]{\smash{$N=2$}}}
\put(310,535){\makebox(0,0)[lb]{\smash{(spont. broken $N=4$)}}}
\put(506,525){\framebox(138,40){}}
\put(540,550){\makebox(0,0)[lb]{\smash{$N=4$}}}
\put(515,535){\makebox(0,0)[lb]{\smash{(mapping (\ref{mapp}))}}}
\put(180,575){\vector( 2, 3){ 72.308}}
\put(388,576){\vector( 2, 3){ 72.308}}
\put(349,574){\vector(-2, 3){ 73.846}}
\put(561,575){\vector(-2, 3){ 73.846}}
\put( 40,545){\makebox(0,0)[lb]{\smash{$D=4$}}}
\put( 40,710){\makebox(0,0)[lb]{\smash{$D=6$ (or 5)}}}
\end{picture}
\caption{\label{fig1}Decompactification scheme of generic models with
$\l=0$
shifted
$(2,2)$ lattice.}
\end{figure}
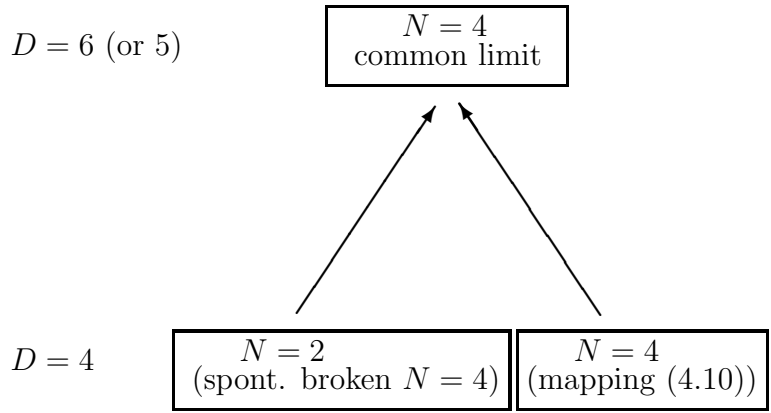
\begin{figure}
\setlength{\unitlength}{0.012500in}%
\setlength{\unitlength}{0.010000in}%
\begin{picture}(576,210)(40,525)
\thicklines
\put(306,695){\framebox(125,40){}}
\put(345,720){\makebox(0,0)[lb]{\smash{$N=4$}}}
\put(320,705){\makebox(0,0)[lb]{\smash{common limit}}}
\put(226,525){\framebox(175,40){}}
\put(260,550){\makebox(0,0)[lb]{\smash{$N=2$}}}
\put(235,535){\makebox(0,0)[lb]{\smash{(spont. broken $N=4$)}}}
\put(406,525){\framebox(130,40){}}
\put(435,550){\makebox(0,0)[lb]{\smash{$N=4$}}}
\put(410,535){\makebox(0,0)[lb]{\smash{(mapping (\ref{mapp}))}}}
\put(290,575){\vector( 2, 3){ 72.308}}
\put(449,574){\vector(-2, 3){ 73.846}}
\put(140,545){\makebox(0,0)[lb]{\smash{$D=4$}}}
\put(140,710){\makebox(0,0)[lb]{\smash{$D=6$ (or 5)}}}
\end{picture}
\caption{\label{fig2}Decompactification scheme of models with $\l=1$
shifted $(2,2)$ lattice.}
\end{figure}
\begin{figure}
\setlength{\unitlength}{0.012500in}%
\setlength{\unitlength}{0.010000in}%
\begin{picture}(576,210)(40,525)
\thicklines
\put(431,695){\framebox(125,40){}}
\put(445,705){\makebox(0,0)[lb]{\smash{common limit}}}
\put(470,720){\makebox(0,0)[lb]{\smash{$N=4$}}}
\put(206,695){\framebox(125,40){}}
\put(245,720){\makebox(0,0)[lb]{\smash{$N=4$}}}
\put(220,705){\makebox(0,0)[lb]{\smash{common limit}}}
\put(115,525){\framebox(170,40){}}
\put(120,550){\makebox(0,0)[lb]{\smash{$N=4$ (mapping (\ref{map})}}}
\put(130,535){\makebox(0,0)[lb]{\smash{with (\ref{mapN4}))}}}
\put(306,525){\framebox(170,40){}}
\put(335,550){\makebox(0,0)[lb]{\smash{$N=2$}}}
\put(310,535){\makebox(0,0)[lb]{\smash{(spont. broken $N=4$)}}}
\put(506,525){\framebox(140,40){}}
\put(540,550){\makebox(0,0)[lb]{\smash{$N=4$}}}
\put(515,535){\makebox(0,0)[lb]{\smash{(mapping (\ref{mapp}))}}}
\put(180,575){\vector( 2, 3){ 72.308}}
\put(388,576){\vector( 2, 3){ 72.308}}
\put(349,574){\vector(-2, 3){ 73.846}}
\put(561,575){\vector(-2, 3){ 73.846}}
\put( 40,545){\makebox(0,0)[lb]{\smash{$D=4$}}}
\put( 40,710){\makebox(0,0)[lb]{\smash{$D=6$ (or 5)}}}
\end{picture}
\caption{\label{fig3}Decompactification scheme for
models with $\l=0$ shifted
$(2,2)$ lattice of~class~(\romannumeral2).}
\end{figure}
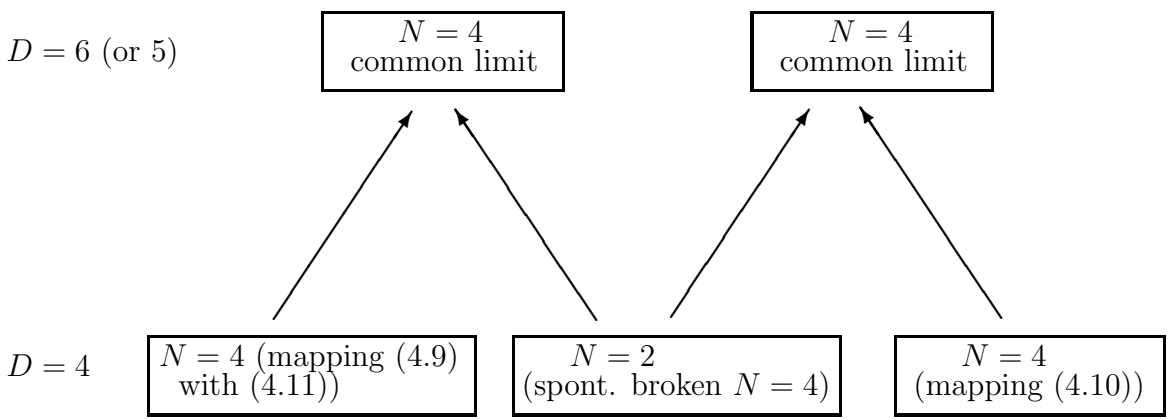

\boldmath
\section{Thresholds in models with spontaneously-broken $N=4$
supersymmetry}\label{thrN4}
\unboldmath

Our starting point is now (\ref{het2helj}), (\ref{helsb}). In that case
(\ref{2gau})  and (\ref{2grav})
read respectively\footnote{The prime summation
over $(h,g)$ stands for $(h,g) =
\{(0,1),(1,0),(1,1)\} $.}:
\be
F_{i}^{w}=\sump{\Gamma_{2,2}^{w}{h\atopwithdelims[]g}}\,
F_i^{\l}\oao{h}{g}
\label{gausb}
\ee
and
\be
F_{\rm grav}^{w}=\sump{\Gamma_{2,2}^{w} {h\atopwithdelims[]g}}\,
F_{\rm grav}^{\l}\oao{h}{g}\, ,
\label{au}
\ee
where we focused, as previously, on the corrections to gauge couplings
corresponding to the rank-20 factors of the gauge group. We have also
introduced
\be
F_i^{\l}\oao{h}{g}=-{1\over \bar \eta^{24}}
\left(\bP_{i}^2-{k_{i}\over 4\pi\im}
\right)
\overline{\Omega}^{\l}_{\vphantom 1}{h\atopwithdelims[]g}
\label{gausbhg}
\ee
and
\be
F_{\rm grav}^{\l}\oao{h}{g}
=-{1\over \bar \eta^{24}}
\left({\bE_2\over 12}-{1\over 4\pi\im}
\right)
\overline{\Omega}^{\l}_{\vphantom 1}{h\atopwithdelims[]g}
\, ,
\label{auhg}
\ee
where
\be
\bOmega^{\l}_{\vphantom 1}\oao{h}{g}=-{1\over 4}
\bar \eta^{20} \left. C_{4,20}^{\l}\oao{h}{g}\right\vert_{v=0}
\label{gb}
\ee
(notice that $\bOmega^{\l}\oao{0}{0}$ vanishes)\footnote{In the case of
orbifold models, the functions
$\overline{\Omega}^{\l}{h\atopwithdelims[]g}$
are given in (\ref{omorbsp}) in terms of the twisted lattices
$\Gamma_{4,20}^{\l}{h\atopwithdelims[]g}$.}.
By using (\ref{auhg}), one can recast (\ref{gausbhg}) in the form
\be
F_{i}^{\l}\oao{h}{g}=k_i^{\vphantom 1} \, F_{\rm grav}^{\l}\oao{h}{g}+
\bLambda_{i}^{\l}{h\atopwithdelims[]g}\, ,
\label{fi}
\ee
which involves the functions
\be
\bLambda_{i}^{\l}{h\atopwithdelims[]g}=
-{1 \over \bar \eta^{24}}
\left(\bP_{i}^2-k_{i}^{\vphantom{\bar H}}{{\bE_2\over 12}}
\right)
\overline{\Omega}^{\l}_{\vphantom i}{h\atopwithdelims[]g}
\, .
\label{lo}
\ee

The functions $\bOmega^{\l}\oao{h}{g}$ are antiholomorphic
for the same reason that $\overline{\Omega}$ in Eq. (\ref{58}) is
antiholomorphic in the case of $N=2$ models that are toroidal
compactifications of six-dimensional $N=1$ theories; the same holds
therefore for $\left. C_{4,20}^{\l}\oao{h}{g}\right\vert_{v=0}$, as we
advertised in  Section \ref{spbr}, as well as for
$\bLambda_{i}^{\l}{h\atopwithdelims[]g}$. The modular-transformation
properties of these functions are
(see (\ref{gentr1}) and (\ref{gentr2})):
\be
\tau\to\tau+1 \ ,\ \ \bOmega^{\l}{h\atopwithdelims[]g}\to
{\rm e}^{-i\pi{\l}{h^2 \over 2}}\,
\bOmega^{\l}{h\atopwithdelims[]h+g}\, , \ \
\bLambda^{\l}_i{h\atopwithdelims[]g}\to
{\rm e}^{-i\pi{\l}{h^2 \over 2}}\,
\bLambda^{\l}_i{h\atopwithdelims[]h+g}
\label{255c}
\ee
\be
\tau\to-{1\over \tau} \ ,\ \ \bOmega^{\l}{h\atopwithdelims[]g}\to
\bar \tau^{10}\,
{\rm e}^{i\pi{\l}{hg}}\, \bOmega^{\l}{g\atopwithdelims[]-h}\, ,\ \
\bLambda^{\l}_i{h\atopwithdelims[]g}\to
{\rm e}^{i\pi{\l}{hg}}\, \bLambda^{\l}_i{g\atopwithdelims[]-h}
\, .
\label{256c}
\ee
These transformation properties together with the singularity structure
inside the $\tau$-plane allow us to determine the most general
functions
$\bOmega^{\l}{h\atopwithdelims[]g}$
that could be obtained starting from any consistent ground state of the
type
(\ref{het2helj}), (\ref{helsb}). They turn out to depend on several
discrete Wilson lines
that appear
in the functions
$\left.C_{4,20}^{\l}{h\atopwithdelims[]g}\right\vert_{v=0}$, but most
of the model- and moduli-dependence present in
$C_{4,20}^{\l}{h\atopwithdelims[]g}(v/2)$ is lost\footnote{Similarly
to Section
\ref{thrN2}, the knowledge of
$C_{4,20}^{\l}{h\atopwithdelims[]g}$ at $v=0$ does not enable us
to reconstruct the full functions
$C_{4,20}^{\l}{h\atopwithdelims[]g}(v/2)$.}.
In the following, we will present this analysis for the relevant cases
($\l=0$ and $\l=1$). We will show how these functions
$\bOmega^{\l}{h\atopwithdelims[]g}$ can indeed be realized, and
eventually evaluate
$\bLambda^{\l}_i{h\atopwithdelims[]g}$.
We will therefore determine completely $F_i^{w}$ in terms
of several physical parameters of the model, among which the
beta-function
coefficients $b_i$,
much as we have reached (\ref{59}) for $N=2$
ground states with a factorized two-torus.

In order to make the subsequent analysis more transparent,
we introduce the functions
$$
F_i^{\l(\pm)}=F_i^{\l}\oao{1}{0}\pm F_i^{\l}\oao{1}{1}
$$
and similarly for $F_{\rm grav}^{\l(\pm)}$. The gauge and gravitational
functions
(\ref{gausb}) and (\ref{au}) now read:
\be
F_i^w=\Gamma_{2,2}^w\oao{0}{1}\,F_i^{\l}\oao{0}{1}
+\Gamma_{2,2}^{w(+)}\,F_i^{\l(+)}+
\Gamma_{2,2}^{w(-)}\,F_i^{\l(-)}
\label{gausbp}
\ee
and
\be
F_{\rm grav}^w=\Gamma_{2,2}^w\oao{0}{1}\,F_{\rm grav}^{\l}\oao{0}{1}+
\Gamma_{2,2}^{w(+)}\,F_{\rm grav}^{\l(+)}+
\Gamma_{2,2}^{w(-)}\,F_{\rm grav}^{\l(-)}\, ,
\label{aup}
\ee
respectively ($\Gamma_{2,2}^{w(\pm)}$ is given in (\ref{gpm})).

\subsection{The case $\l=0$}\label{l0}

The threshold corrections in this case have been calculated in
\cite{decoa} for a specific model that corresponds to the
Scherck--Schwarz version
of
the symmetric $Z_2$ orbifold. Here we will present the results
for the general case (see (\ref{helsb})), when $\l=0$.

The simplest way to derive the most general $\Omega^{\l}$'s for a given
value of $\lambda$ is to
extend the results of a particular model.
In the case $\lambda=0$, one could choose the symmetric $Z_2$  orbifold
of
\cite{decoa}. However, for simplicity, we shall consider the $E_8\times
E_8\times SO(8)\times {U(1)}^2$ model presented in Appendix
B, which leads to the functions
$\Omega^{\l=0}_{(0)}\oao{h}{g}$ given in Eq. (\ref{29b}).
A natural generalization of these functions is given by
\ba
\Omega^{\l=0}_{\vphantom{()}}\oao{0}{1} &=& g(1-x)\,
\Omega_{(0)}^{\l=0}\oao{0}{1}\nn\\
\Omega^{\l=0}_{\vphantom{()}}\oao{1}{0} &=& g(x)
\,\Omega_{(0)}^{\l=0}\oao{1}{0}\label{gom}\\
\Omega^{\l=0}_{\vphantom{()}}\oao{1}{1}&=& g\left(\frac{x}{x-1}\right)
\Omega_{(0)}^{\l=0}\oao{1}{1}\, ,\nn
\ea
where $x\equiv\left(\vartheta_2 / \vartheta_3\right)^4$. The function
$g(x)$ must
satisfy the constraints
$$
g(x)=g\left(\frac{1}{x}\right),
$$
required for modular covariance (see Eqs. (\ref{255c}) and
(\ref{256c})).
Unitarity now demands that $\Omega^{\l=0}\oao{h}{g}$ have a regular
expansion without poles
inside the fundamental domain,
except at $\tau = i\infty$. It follows
that
$g(x)$ can have poles at $x=0,1$ as well as at the roots of
$\Omega^{\l=0}\oao{1}{0}$. More details on the geometry and
singularities
on the three-punctured sphere
can be found in \cite{char}.
Putting everything together, we
obtain:
\be
g(x) = \xi_1\, \frac{x^2}{(x^2-x+1)^2} +\xi_2\, \frac{x}{x^2-x+1}
+\xi_3\, .
\label{glab}
\ee
In this representation, the $E_8\times E_8\times SO(8)\times {U(1)}^2$
ground state of
Appendix B corresponds therefore to $\xi_1=\xi_2=0$ and $\xi_3=1$.
It is clear from Eqs.
(\ref{gom})
and (\ref{glab}) that
\ba
\Omega^{\l=0}_{\vphantom 1}\oao{0}{1}&=&{\hphantom{-}}
\frac{1}{2}
\left(\th_3^4+\th_4^4\right)
\left(
\xi_1\, \th_3^8\, \th_4^8+
\xi_2\, \th_3^4\, \th_4^4 \, E_4^{\vphantom 3} +
\xi_3\, E_4^2
\right)
\nonumber\\
\Omega^{\l=0}_{\vphantom 1}\oao{1}{0}&=&-
\frac{1}{2}
\left(\th_2^4+\th_3^4\right)
\left(
\xi_1\,  \th_2^8\, \th_3^8+
\xi_2\,\th_2^4\, \th_3^4 \, E_4^{\vphantom 3} +
\xi_3\, E_4^2
\right)
\label{gom0xi}\\
\Omega^{\l=0}_{\vphantom 1}\oao{1}{1}&=&{\hphantom{-}}
\frac{1}{2}
\left(\th_2^4-\th_4^4\right)
\left(
\xi_1\,  \th_2^8\, \th_4^8-
\xi_2\,\th_2^4\, \th_4^4 \, E_4^{\vphantom 3}  +
\xi_3\, E_4^2
\right) ,\nonumber
\ea
which satisfy
\be
\sump\Omega^{\l=0}\oao{h}{g}
=\left(\xi_1+\xi_2
\right)E_4\, E_6
\, .\label{gom0id}
\ee

The above result deserves a discussion. The functions
$\Omega^{\l=0}\oao{h}{g}$
(the same actually
holds in the case $\l=1$) depend on three parameters only:
$(\xi_1,\xi_2,\xi_3)\equiv \vec \xi $, which, as we will see very soon,
are subject to several constraints and can take only some discrete
values.  These parameters exhaust all moduli dependence of
$\Omega^{\l}\oao{h}{g}$ and define a kind of universality classes.
Consequently, despite the model-dependence of
$C_{4,20}^{\l}{h\atopwithdelims[]g}(v/2)$, which is a priori a function
of a large number of moduli, $\left.
C_{4,20}^{\l}\oao{h}{g}\right\vert_{v=0}$ (see Eq.~(\ref{gb}))
is almost universal.

Our goal is to compute threshold corrections for the gauge couplings.
We must therefore determine the full gauge function $F_{i}^{w}$
(Eqs. (\ref{gausb}) and (\ref{fi})), which implies the computation of
the functions
$\Lambda_{i}^{\l}{h\atopwithdelims[]g}$, following (\ref{lo}). In the
general case, however, it is difficult to
proceed in this way, because the gauge group is
unknown and so is the action of the covariant derivative. Thus we shall
follow a different method. It consists in writing down the most
general functions $\Lambda_{i}^{\l}{h\atopwithdelims[]g}$
compatible with modular covariance, unitarity, etc., as we did for the
$\Omega$'s, and in determining the various free parameters that appear
in those functions in terms of
some low-energy physical quantities
(i.e. related to the massless spectrum)
such as the beta-function
coefficients
and the affine-Lie-algebra levels. This is exactly the method that
was used in order to reach (\ref{59}) in the case of $N=2$ models that
are toroidal compactifications from six to four dimensions.
As a corollary of our analysis, the universality-class vector $\vec\xi$
will
also be expressed in terms of physical parameters of the ground state.

We can find the most general functions
$\Lambda_{i}^{\l=0}{h\atopwithdelims[]g}$ by following
the same lines of thought as for the determination of the general
$\Omega$'s given in (\ref{gom}) and (\ref{glab}):
\ba
\Lambda_i^{\l=0}\oao{0}{1} &=&
f_i^{\lambda=0}(1-x)\, \Lambda_{(0)E_8}^{\l=0}\oao{0}{1}\nonumber\\
\Lambda_i^{\l=0}\oao{1}{0} &=&
f_i^{\lambda=0}(x)\, \Lambda_{(0)E_8}^{\l=0}\oao{1}{0}\label{lfe}\\
\Lambda_i^{\l=0}\oao{1}{1} &=&
f_i^{\lambda=0}\left(\frac{x}{x-1}\right)\Lambda_{(0)E_8}^{\l=0}
\oao{1}{1}\, ;\nonumber
\ea
here $f_i^{\lambda}(x)$ is the ratio of two polynomials of $x$,
satisfying
$f_i^{\lambda}(x) = f_i^{\lambda}({1/x})$, and
$\Lambda_{(0)E_8}^{\l=0}\oao{h}{g}$ are given in Eq. (\ref{L29b}).
Following arguments similar to the ones advocated
for $g(x)$, we obtain:
\be
f_i^{\lambda=0}(x)= \frac{A_i \, (x^6+1)+B_i\, x^2 (x^2+1) +C_i\, x
(x^4+1)+
D_i \, x^3} {(x^2-x+1)(x+1)^2(x-2)(2x-1)}\, .
\label{flab}
\ee

The determination of the constants  $A_i, B_i, C_i,D_i$ necessitates
the introduction of several constraints, which involve various physical
parameters.
The relevant quantities for this analysis are $F_{\rm
grav}^{\l=0}\oao{h}{g}$ and
$F_{i}^{\l=0}\oao{h}{g}$ (Eqs. (\ref{auhg}) and (\ref{fi})).
By using Eqs. (\ref{gom}), (\ref{glab}), (\ref{lfe}) and (\ref{flab})
in
Eqs. (\ref{auhg})
and (\ref{fi}), we obtain explicit expressions for
$F_{\rm
grav}^{\l=0}\oao{h}{g}$ and
$F_{i}^{\l=0}\oao{h}{g}$, which we can further expand in powers of
$\bar
q$.
The results for these expansions are summarized in Eqs.
(\ref{de1})--(\ref{f0m}).
In order to determine the various parameters
(namely $A_i, B_i, C_i,D_i$ and $\vec \xi $)
appearing in these
expressions in terms of low-energy quantities related to the model,
we proceed as follows.

We first observe that the tachyon,
 being the lowest-lying state, is not charged and, therefore,
cannot contribute to the gauge function (\ref{gausbp}). Taking into
account the structure of the shifted lattice,
namely the fact that the
lattice sum $\Gamma_{2,2}^w\oao{0}{1}$ is always of the form
$\Gamma_{2,2}^w\oao{0}{1}=1+\cdots$, whereas
$\Gamma_{2,2}^{w(\pm)}$ never contain the unity (see Appendix A,
Eqs. (\ref{l0t})--(\ref{l0h})),
we conclude that the coefficient of the $\frac{1}{\bar q}$-term in
$F_i^{\l=0}\oao{0}{1}$ of
(\ref{de1}) must be zero. Moreover, as
we notice in Appendix A, for any $w$, there is always a corner in
the $(T,U)$ moduli space where $\Gamma_{2,2}^w\oao{h}{g}$
are equal for all $(h,g)$ \footnote{For example, for a shift vector $w$
corresponding to model I in
Table A.1,
this happens indeed in the limit $T_2\to0,U_2=1,$ as is clear from
(\ref{l0lis}).} (up to exponentially suppressed terms).
In that limit, the $\frac{1}{\bar q}$-pole present in
$F_i^{\l=0(+)}$ of
(\ref{dep}) contributes as does the
$\frac{1}{\bar q}$-pole of $F_i^{\l=0}\oao{0}{1}$ of (\ref{de1}),
and its coefficient must then be set equal to zero.

We now turn to constraints originated from the identification of the
beta-function coefficients, according to (\ref{2a}). For {\it generic
values of the two-torus moduli}, the only contribution comes from the
constant
term of
$F_i^{\l=0}\oao{0}{1}$,
which has therefore to be identified with $b_i$. However, it
is important to observe
that there are
regions of the $(T,U)$ moduli space where {\it extra charged massless
states}
(vector multiplets and/or hypermultiplets) appear and contribute to the
beta-function coefficients, which therefore become $b_i \to b_i+\delta
b_i$;
those must in particular be considered in expressions such as (\ref{2})
in order to properly determine the thresholds.
This enlargement of the massless spectrum can occur at the
decompactification limit, where $\Gamma_{2,2}^w\oao{h}{g}$ become equal
for all $(h,g)$ (see Eq. (\ref{l0lis}) for the case
$\Gamma_{2,2}^{\I}\oao{h}{g}$). In this case only hypermultiplets might
become
massless (the gauge symmetry remains unchanged). The extra contribution
to the
beta-function
coefficients will be denoted $\delta_v b_i$, and will
be identified
with the constant term of
$F_i^{\l=0(+)}$ of (\ref{dep}). On the other hand, along the line
$T=f^w_h(U)$, extra vector multiplets and/or hypermultiplets become
massless.
This
enhancement of the massless spectrum is originated from the $(2,2)$
lattice,
as is
clear
from Eqs. (\ref{l0h}) and (\ref{dem});
we will have $\delta_h b_i$, which has to be identified
with twice the coefficient of the $\frac{1}{\sqrt{\bar q}}$-term of
$F_i^{\l=0(-)}$ in (\ref{dem})\footnote{Note that at some isolated
points
of this line, as explained in
Appendix A, the lattice multiplicity doubles and the beta-function
coefficients
become $b_i+2\delta_h b_i$ instead of $b_i+\delta_h
b_i$ which is
their value at a generic point along $T=f^w_h(U)$.}.

Finally, there are two constraints that are obtained by inspecting the
 gravitational function (\ref{aup}). The latter receives a tachyon
contribution.
 At generic values of the moduli $(T,U)$, this contribution is given by
the coefficient of the $\frac{1}{\bar q}$-term in
$ F_{\rm grav}^{\l=0}\oao{0}{1}$ of
(\ref{f01}), which must then be equal to $-1/12$ in our
normalizations.
 Furthermore, {\it the gravitational anomaly at generic values of the
two-torus
moduli}, $b_{\rm grav}$, has to be identified with the
constant term of
 $F_{\rm grav}^{\l=0}\oao{0}{1}$ (see Eq. (\ref{2ag})
and the structure of the shifted lattice sums). We will come back later
to the discontinuities $b_{\rm grav} \to b_{\rm grav}+ \delta b_{\rm
grav}$
that occur along special lines.

The seven constraints obtained so far are summarized in Appendix C,
where we
 solve them in order to express the parameters $A_i, B_i, C_i, D_i$ and
$\vec\xi$ in terms of $b_i, \delta_v b_i, \delta_h b_i$ and
$b_{\rm grav}$. By inspecting expressions (\ref{OX1}),
(\ref{OX2}) and (\ref{OX3}), which give $\vec \xi$, we can draw a
straightforward conclusion:
the parameters  $b_i$, $\delta_v b_i$  and $\delta_h b_i$ are
related in the sense that the combination
$2b_i-12\delta_h b_i - \delta_v b_i $ is necessarily
proportional to $k_i$ and the latter
captures the whole group-factor dependence. As we will see in the
following, in the framework of $\l=0$ models, the
constant of proportionality can be determined in terms of $b_{\rm
grav}$ only.

Before we turn to the determination of the threshold corrections,
several
comments related to the above analysis are in order.

We first observe that in the models under consideration, where the
two-torus
undergoes a shift leading to a spontaneous breaking of $N=4$
supersymmetry
down to $N=2$, there is room for discontinuities in the beta-function
coefficients. This phenomenon, as we pointed out in Section
\ref{N2orb},
does not occur in ground states where a two-torus is factorized. Here
it
occurs
along the line $T=f_h^w(U)$, where
extra massless states appear, charged under the rank-20 component of
the
gauge group.

The beta-function coefficients also suffer from discontinuity at the
decompactification limit where all
$\Gamma_{2,2}^w\oao{h}{g}$ become equal (see (\ref{l0lis})
and left-hand part of Fig. \ref{fig1})\footnote{We also have a trivial
discontinuity in the other decompactification
limit, namely (\ref{lolil}) depicted in the right-hand part of Fig.
\ref{fig1}.
In that limit $N=4$ supersymmetry is restored and the full functions
$F_i^w$
and $F_{\rm grav}^w$ vanish as a consequence of the exponential
suppression
of the lattice sums (see Eqs. (\ref{gausb}) and (\ref{au})). So do
the beta functions and the gravitational anomaly.
This discontinuity, however, does not
introduce any further physical parameter. The same phenomenon actually
appears
in the unique decompactification limit (\ref{l1li}) of $\l=1$ models
(see Fig. \ref{fig2}).}.
This is specific to $\l=0$ lattices and, as explained in Section
\ref{spbr2}, in this limit, the model at hand
becomes identical to an $N=2$ model with factorized two-torus obtained
through the mapping (\ref{map}) (see Fig. \ref{fig1}); therefore
$b_i+\delta_v b_i$ is to be identified with the beta-function
coefficient of the  latter model (called $\tilde b_i$ in Ref.
\cite{decoa}).

More information about this $N=2$ ground state with factorized
two-torus, sharing a common
limit with our
$N=4$ $\l=0$ model, can be obtained by analysing the
corresponding
functions $F_{\rm grav}$ and $F_i$.
According to the mapping (\ref{map}) these read:
$$
F_{\rm grav}^{\phantom{w}}=\Gamma_{2,2}^{\phantom{w}}
\sump F_{\rm grav}^{\l=0}\oao{h}{g}\, ,
$$
$$
F_i^{\phantom{w}}=k_i^{\phantom{w}}\,F_{\rm grav}^{\phantom{w}} +
\Gamma_{2,2}^{\phantom{w}}\sump
{\overline\Lambda}_i^{\l=0}\oao{h}{g}\, .
$$
By using the solutions (\ref{OA})--(\ref{OX2}) of Appendix C, we can
compute
$F_{\rm grav}^{\l=0}\oao{h}{g}$ as well as
$\Lambda_i^{\l=0}\oao{h}{g}$, and, thanks to (\ref{gom0id}), perform
the
summation with the result
(we keep here the explicit dependence with respect to $\xi_3$):
\be
F_{\rm grav}=-\frac{(1-\xi_3)}{12}\, \Gamma_{2,2}\, \frac{{\widehat
E_2}\, {\overline E}_4\,
{\overline E_6}}{\bar\eta^{24}}\label{liFg}
\ee
and
\be
F_{i}=-\frac{k_i(1-\xi_3)}{12}\, \Gamma_{2,2}\left(\frac{{\widehat
E_2}\, {\overline E}_4\,
{\overline E_6}}{\bar\eta^{24}}-\bar j+1008\right)+
\left(b_i+\delta_v b_i\right)\Gamma_{2,2}\, ,
\label{liFi}
\ee
which can be compared to the results for $N=2$ models with
factorized
$T^2$ (Eq.~(\ref{59})). Equation (\ref{liFg})
shows that in order for those limiting models to possess the correct
tachyon
contribution (normalization of the $\frac{1}{\bar q}$-pole in $F_{\rm
grav}$),
the original $\l=0$ models have either
\be
\xi_3=\cases{
0\ , &{\rm class \ (\romannumeral1)},\cr
1\ , &{\rm class \ (\romannumeral2);}\cr}
\label{b46}
\ee
we refer to the classes of $\l = 0$ models introduced in Section
\ref{spbr2}.

The models in the first class remain genuine $N=2$ in the limit under
consideration,
whereas those in class (\romannumeral2) actually become  $N=4$ models:
$F_{\rm grav}$
vanishes and $F_i$ must also vanish, implying $\delta_v b_i=-b_i$.
In this class the mapping (\ref{map}) actually satisfies
(\ref{mapN4}),
and
we are in the situation represented in Fig. \ref{fig3}. By using
Eq. (\ref{OX3}), we can recast (\ref{b46}) in terms of physical
parameters,
which therefore satisfy
\be
\cases{
4 b_i-24\, \delta_h b_i-2\delta_v b_i=
{9k_i}
\left(b_{\rm grav}-6\right)
\ ,\ \   {\rm   class \ (\romannumeral1),}\cr
\delta_v b_i=-b_i\, ,\ \ 2b_i-8\delta_h b_i=
{3k_i}
\left(b_{\rm grav}+2\right)
\ ,\ \   {\rm   class \ (\romannumeral2).}}
\label{b46p}
\ee
 These relations show that there always
exists in the string ground states considered here,
a combination of physical, gauge-group-dependent parameters
(such as
$b_i$, $\delta_v b_i$ and $\delta_h b_i$), which depends only
on the level
of the affine Lie algebras. As we will see in the following, this
implies
that there is no unambiguous way of defining a group-factor-independent
threshold
correction $Y$ as was the case in $N=2$ models with a factorized
two-torus (see Eq. (\ref{66a})).

Discontinuities like those discussed above
also occur in the gravitational anomaly. The expansions (\ref{f01}),
(\ref{f0p}) and (\ref{f0m}) of $F_{\rm grav}^{\l=0}\oao{h}{g}$,
together with the enhancement properties of the massless spectrum and
the
large- or
small-radius behaviours of the lattice sums
$\Gamma_{2,2}^{w}\oao{h}{g}$
(see Eqs. (\ref{l0t})--(\ref{l0lis})) show that there are several
possibilities.

In the limit that we have just analysed, where all
$\Gamma_{2,2}^w\oao{h}{g}$ become equal
(limit (\ref{l0lis}) for shifted lattice I), the gravitational anomaly
acquires an extra piece $\delta_v b_{\rm grav}$, which is the constant
term
of (\ref{f0p}). Equations (\ref{OX1}), (\ref{OX2}), (\ref{OX3}) and
(\ref{b46p}) then allow us
to recast this discontinuity as:
\be
\dv \bgrav=
\cases{
22-\bgrav\ , \ \ {\rm class\ (\romannumeral1),}\cr
-\bgrav\ , \ \ {\rm class \ (\romannumeral2),}
}\label{dvby}
\ee
where it appears, as expected, that the limiting ground states of class
(\romannumeral1)
are
 $N=2$ ground states with gravitational anomaly $22$, whereas for class
(\romannumeral2)
we
 reach $N=4$ models with vanishing gravitational anomaly.

Along the line $T=f_h^w(U)$, as can be seen from Eqs. (\ref{l0h})
and (\ref{f0m}), another discontinuity appears, $\delta_h\bgrav$,
which
has to be identified with twice the coefficient of the
$\frac{1}{\sqrt{\bar q}}$-term of $F_{\rm grav}^{\l=0(-)}$;
it
can be expressed as:
\be
\delta_h\bgrav=\frac{1}{24}
\cases{
2-3\bgrav\ ,\ \ {\rm class \ (\romannumeral1),}\cr
-30-3\bgrav\ ,\ \ {\rm class \ (\romannumeral2).}
}
\label{dhby}
\ee

Furthermore, in the case of the gravitational anomaly, extra
discontinuities
appear along the rational lines $T=U$ and $T=-1/U$ (see Eqs.
(\ref{l0t}) and (\ref{l0tt})),
which play no role in the beta-function coefficients of the rank-$20$
component
of the gauge group
($\delta_{t}b_i=\delta_{t} 'b_i=0$),
because of the absence of tachyonic contribution
in $F_i$. The same phenomenon also occurs in the ordinary $N=2$ models,
as
discussed at the end of Section \ref{thrN2},
although in that case the lines $T=U$ and $T=-1/U$
are equivalent as a consequence of the $SL(2,Z)_T$. Here the
$\frac{1}{\bar q}$-pole of
$F_{\rm grav}^{\l=0}\oao{0}{1}$ (\ref{f01}) leads to
\be
\delta_{t}\bgrav=-\frac{\delta_t^w}{6}\sp {\rm at \ }T=U
\label{dtby}
\ee
and
\be
\delta^{\prime}_{t}\bgrav=-\frac{\delta_t^{\prime w}}{6}\sp {\rm at \ }
T=-\frac{1}{U}\, ,
\label{dttby}
\ee
where
\be
\delta_t^w=(-)^{a_1-b_1}\, ,
\label{ddtby}
\ee
which becomes
$(-)^{a_1-b_1}+(-)^{a_2-b_2}$ when $T=U=i$, or
$(-)^{a_1-b_1}+(-)^{a_2-b_2}\left((-)^{a_1}+
(-)^{b_1}\right)$
if $T=U=\rho$ or $-1/\rho$, and
\be
\delta_t^{\prime w}=(-)^{a_2-b_2}\, ,
\label{ddttby}
\ee
which becomes
$(-)^{a_2-b_2}+(-)^{a_1-b_1}\left((-)^{a_2}+
(-)^{b_2}\right)$
if $T=-1/U=\rho$ or $-1/\rho$. As we will see in Section \ref{orb},
both  vector multiplets and
hypermultiplets (uncharged under the rank-20 component of the gauge
group)
can in general become massless along these lines, whereas in $N=2$
models with a
factorized two-torus, only vectors appear
(see Eq. (\ref{d69})). Therefore, {\it symmetry is not
necessarily enhanced. }

Finally, in ground states belonging to class (\romannumeral2), i.e.
when $\xi_3=1$, we
also
have
\be
\dv'\bgrav=\frac{1}{6}
\label{dvpby}
\ee
at the line $T=\fvw(U)$, as is clear from
Eqs. (\ref{f0p}) and (\ref{l0vp}). Notice, however, that along this
line
$\dv'b_i=0$ because of the absence of ${1\over \bar
q}$-pole in
$F_i^{\l=0(+)}$ of (\ref{dep}). This was one of the physical
constraints
imposed
above, namely the absence of charged tachyon anywhere in moduli
space.

We now come to the computation of the threshold corrections. Collecting
the results (\ref{gom})--(\ref{flab}) and (\ref{OA})--(\ref{OX3})
in Eqs.
(\ref{gausb})--(\ref{fi}), we can recast Eq. (\ref{2}),
{\it for generic values of $T$ and $U$,}
as was advertised in the introduction:
\be
\Delta_i^w= b_i^{\phantom{i}}\,  \Delta^w_{\phantom{i}}(T,U) +
\dhhp b_i^{\phantom{i}} \, H^w_{\phantom{i}}(T,U) +
\dvp b_i^{\phantom{i}}\, V^w_{\phantom{i}}(T,U) +
k_i^{\phantom{i}}\, Y^w_{\phantom{i}}(T,U) \, ,
\label{del}
\ee
where
\ba
 \Delta^w(T,U)&=&
\ifd \left(\sump\Gamma_{2,2}^w\oao{h}{g}\left(-\frac{1}{12}\,
\frac{\widehat E_2}{\bar\eta^{24}}\, {\overline
\Omega}_{(0)}^{\l}\oao{h}{g}\,
{\bar\delta}_g^{\l}\oao{h}{g}
+{\overline\Lambda}^{\l}_{(0)i}\oao{h}{g}\,
{\bar\delta}_f^{\l}\oao{h}{g}\right)-1\right)
\nonumber\\ &&+
\log\frac{2}{\pi}\frac{e^{1-\gamma}}{\sqrt{27}}
\label{d}
\\
H^w(T,U)&=&
\ifd \sump\Gamma_{2,2}^w\oao{h}{g}\left(-\frac{1}{12}\,
\frac{\widehat E_2}{\bar\eta^{24}}\, {\overline
\Omega}_{(0)}^{\l}\oao{h}{g}\,
{\bar h}_g^{\l}\oao{h}{g} +{\overline\Lambda}^{\l}_{(0)i}\oao{h}{g}\,
{\bar h}_f^{\l}\oao{h}{g}\right)
\label{h}
\\
V^w(T,U)&=&
\ifd \sump\Gamma_{2,2}^w\oao{h}{g}\left(-\frac{1}{12}\,
\frac{\widehat E_2}{\bar\eta^{24}}\, {\overline
\Omega}_{(0)}^{\l}\oao{h}{g}\,
{\bar v}_g^{\l}\oao{h}{g} +{\overline\Lambda}^{\l}_{(0)i}\oao{h}{g}\,
{\bar v}_f^{\l}\oao{h}{g}\right)
\label{v}
\\
Y^w(T,U)&=&
\ifd \sump\Gamma_{2,2}^w\oao{h}{g}\left(-\frac{1}{12}\,
\frac{\widehat E_2}{\bar\eta^{24}}\, {\overline
\Omega}_{(0)}^{\l}\oao{h}{g}\,
{\bar y}_g^{\l}\oao{h}{g} +{\overline\Lambda}^{\l}_{(0)i}\oao{h}{g}\,
{\bar y}_f^{\l}\oao{h}{g}\right)    \, .
\label{y}
\ea
The functions ${\delta}_{g,f}^{\l=0}$, ${h}_{g,f}^{\l=0}$,
 ${v}_{g,f}^{\l=0}$ and  ${y}_{g,f}^{\l=0}$
appearing in the above integrals are given in Eqs.~(\ref{0gf}).
The integrals themselves can be
evaluated by unfolding the fundamental domain and reducing the
summations over the
modular group orbits (the modular group now
being reduced as explained in Appendix A) in the spirit of
Refs. \cite{dkl,hm}. Some preliminary results were given in
\cite{ms,decoa}. More complete formulas are presented in Appendix D.

The
functions $\Delta^w(T,U)$ and $Y^w(T,U)$ defined in (\ref{d}) and
(\ref{y}) are finite all over the two-torus moduli space. However,
the function $H^w(T,U)$ becomes singular along the line
$T=f_h^w(U)$, because of the extra constant contribution of the
integrand in (\ref{h}), which originates from extra massless states
that
lead to a logarithmic divergence
(see Eqs. (\ref{Iths})). Along this line, we must therefore
substitute
in Eq. (\ref{del}):
\ba
H^w(T,U)&\to& H^w\left(f_h^w(U),U\right)
\nonumber\\
&=&
\ifd\left(
\sump\Gamma_{2,2}^w\oao{h}{g}\left(f_h^w(U),U\right) \right.\nonumber\\
&&\; \; \; \; \; \; \;\; \; \; \; \; \; \;  \left.\times
{\vphantom{\sump}}
\left(-\frac{1}{12}\,
\frac{\widehat E_2}{\bar\eta^{24}}\, {\overline
\Omega}_{(0)}^{\l}\oao{h}{g}\,
{\bar h}_g^{\l}\oao{h}{g} +{\overline\Lambda}^{\l}_{(0)i}\oao{h}{g}\,
{\bar h}_f^{\l}\oao{h}{g}\right)-1\right)
\nonumber\\
&&+\log\frac{2}{\pi}\frac{e^{1-\gamma}}{\sqrt{27}}
\, ,
\label{hp}
\ea
which accounts for the extra massless states by subtracting the
contribution
$\dhh b_i$. Equation~(\ref{del}) now leads to the correct thresholds.

The function $V^w(T,U)$ remains finite in the bulk of moduli space. On
the other
hand, it develops an extra logarithmic singularity in the
decompactification limit (\ref{l0lis}), where all
$\Gamma_{2,2}^w\oao{h}{g}$
become equal. Then, by formally substituting
\ba
V^w(T,U)&\to& V^w(T_{\rm lim},U_{\rm lim})
\nonumber\\
&=&
\ifd\left(
\sump\Gamma_{2,2}^w\oao{h}{g}(T_{\rm lim},U_{\rm
lim}) \right.\nonumber\\
&&\; \; \; \; \; \; \;\; \; \; \; \; \; \;  \left.\times
{\vphantom{\sump}}\left(-\frac{1}{12}\,
\frac{\widehat E_2}{\bar\eta^{24}}\, {\overline
\Omega}_{(0)}^{\l=0}\oao{h}{g}\,
{\bar v}_g^{\l=0}\oao{h}{g}
+{\overline\Lambda}^{\l=0}_{(0)i}\oao{h}{g}\,
{\bar v}_f^{\l=0}\oao{h}{g}\right)-1\right)
\nonumber\\
&&+\log\frac{2}{\pi}\frac{e^{1-\gamma}}{\sqrt{27}}\, ,
\label{vp}
\ea
which again regularizes the extra massless contributions, the threshold
(\ref{del}) matches in this limit (see Eqs. (\ref{liFg}) and
(\ref{liFi}))
the ordinary $N=2$ thresholds
\be
\Delta_i=(b_i+\dv b_i)\, \Delta -k_i \, (1-\xi_3)\,Y\,,
\label{66ap}
\ee
where $\Delta$ and $Y$ are given respectively in (\ref{67}) and
(\ref{68}).
For $T_2\to0$ and $U_2=1$ the latter behaves as (see Refs.
\cite{pr,kkpra}):
\be
\Delta_i\to (b_i+\dv b_i)\left(\frac{\pi}{3}\frac{1}{T_2}+
\log T_2 -\log 4\pi^2\, \vert\eta(i)\vert^4
\right)-k_i\, (1-\xi_3)\left(\frac{4\pi}{T_2}+20\,  \kappa T_2
\right) ,
\label{N2l}
\ee
up to exponentially suppressed terms ($\kappa$ is given in
(\ref{CON})).
For class-(\romannumeral1) models, the dominant behaviour is linear
with respect to
the volume of the decompactifying manifold.
If the model under consideration belongs instead to class
(\romannumeral2), the
matching
(\ref{66ap}) shows that $\Delta_i^w$ vanish, which reflects the
restoration of $N=4$ supersymmetry.

Actually, the substitution of (\ref{vp}) is formal in the sense that
the
function $V^w(T_{\rm lim},U_{\rm lim})$ defined in (\ref{vp}) is
divergent everywhere except for the limiting value $T_{\rm lim},U_{\rm
lim}$
(i.e.  $T_2\to0, U_2=1$ for the lattice I). Without
this substitution, $\Delta_i^w$ given in (\ref{del}) with $V^w$ given
in
(\ref{v}) possesses, for the models of class (\romannumeral2), a
logarithmic
divergence in
the limit considered here:
$\Delta_i^{\I}\sim-\dv b_i\log T_2=b_i\log T_2$
(more details about those limits, including subleading terms, can be
found
in Appendix D).
Although the
$N=4$
supersymmetry is restored, the accumulation of extra massless states
not properly regularized in the infra-red (one subtracts $b_i$
in the
integrals and not $b_i+\dv b_i$ as one would to follow the
formal
prescription (\ref{vp})) leads to that logarithmic behaviour. A proper
treatment of this infra-red divergence necessitates the introduction of
Wilson lines as explained in \cite{decoa}.

Finally, as we have already pointed out several times, in the
$\l=0$ models, there is another decompactification limit
($T_2\to\infty, U_2=1$ in models with shifted lattice I) where $N=4$
supersymmetry is always restored. In that limit, $\Delta_i^w$ given
in (\ref{del}) diverges
logarithmically (e.g. $\Delta_i^{\I}\sim-b_i\log T_2$), which
is the
same infra-red phenomenon as appeared in the previous case.

Our last comment about the gauge corrections $\Delta_i^w$ concerns the
group-factor-independent thresholds $Y^w(T,U)$ appearing in the
decomposition
(\ref{del}). In $N=2$ models with a factorized $T^2$ (see Section
\ref{thrN2}), the decomposition (\ref{66a}) is unique because there is
no
a priori relation between $b_i$ and $k_i$.
Moreover, $Y$ (see (\ref{68})) is absolutely model-independent, and
this is also a consequence of anomaly cancellations in six dimensions.
However, in ground states with
spontaneously-broken $N=4$ supersymmetry
under consideration, the various physical parameters appearing in the
decomposition (\ref{del}) are not independent. They are related through
(\ref{b46p}). We have therefore the freedom of adding to $Y^w(T,U)$,
defined in (\ref{y}), any function that is regular everywhere in the
moduli
space,
invariant under the relevant duality group, and properly behaved
in the decompactification limits; then, by using (\ref{b46p}),
we compensate the other functions $\Delta^w, H^w$ and $V^w$ without
disturbing
the decomposition in terms of $b_i, \dhh b_i, \dv b_i$ and
$k_i$. This arbitrariness cannot be reduced unless $Y^w(T,U)$ is related 
to some other
physical quantities such as the one-loop correction to the K\"ahler
potential in the spirit of
Ref. \cite{kkpra}, where this was done for ordinary $N=2$ models.
Furthermore, by using Eqs. (\ref{y}) and (\ref{0gf}), 
$Y^w(T,U)$ is recast as follows: 
\be
Y^w_{\vphantom 1}=Y^w_1 + \bgra\,  Y^w_2
\, ,
\label{ydeco}
\ee
which shows that
some {\it model-dependence}, captured in the parameter 
$\bgra$ and the shift vector $w$, is now left in the gauge-factor-independent
threshold. Notice that the 
freedom in the decomposition (\ref{del}) makes it possible for discarding 
either $Y^w_1$ or $Y^w_2$, which are both regular everywhere in the moduli space.

By repeating the above steps, we can proceed to the determination
of the gravitational corrections (\ref{gravth}). At {\it generic
points} of
the two-torus moduli space, they read:
\be
\Delta_{\rm grav}^w=\ifd\left(
-\frac{1}{12}\sump\Gamma_{2,2}^w\oao{h}{g} \, \frac{{\widehat
E}_2}{\bar\eta^{24}}\,
{\overline{\Omega}}_{(0)}^{\l}\oao{h}{g}\,  {\bar g}\oao{h}{g}
-\bgrav\right) ,
\label{delg}
\ee
where $g\oao{h}{g}$ are given in (\ref{glab}). This threshold is not
model-independent as it was in ground states with a factorized
two-torus
(see (\ref{69})), where anomaly cancellation in six dimensions was
advocated. Its model-dependence is captured in the parameters
$\vec\xi$, appearing in $g\oao{h}{g}$, that can be expressed in terms
of
$\bgrav$ by using (\ref{OX1})--(\ref{OX3}), together with
the result (\ref{b46p}):
\be
\cases{
\xi_1=\frac{31}{32}+\frac{3}{64}\bgrav \ ,\ \
\xi_2=\frac{1}{32}-\frac{3}{64}\bgrav \ ,\ \
\xi_3=0 \ ,\ \
{\rm class\ (\romannumeral1),}\cr
\cr
\xi_1=\frac{63}{32}+\frac{3}{64}\bgrav \ ,\ \
\xi_2=-\frac{63}{32}-\frac{3}{64}\bgrav \ ,\ \
\xi_3=1\ ,\ \
{\rm class\ (\romannumeral2).}\cr
}\label{I0}
\ee
Since the gravitational anomaly (see (\ref{bgrav}))
is related to the number of
massless vector multiplets and hypermultiplets, it is clear from the
above
expressions that the parameters $\vec \xi$ can only take discrete
rational
values, as advertised previously. This defines several universality
classes for the gravitational thresholds, which eventually read:
\be
\Delta_{\rm grav}^w = 
\Delta_{{\rm grav}, 1}^w
+ \bgrav\, 
\Delta_{{\rm grav}, 2}^w
\, .
\label{dgradec}
\ee
The actual expressions for  $\Delta_{{\rm grav}, 1}^w$
and  $\Delta_{{\rm grav}, 2}^w$
depend on whether the model belongs to the class (\romannumeral1) or (\romannumeral2), but are universal within each of the two classes where they turn out to depend only on the shift vector $w$.

The thresholds (\ref{delg}) diverge logarithmically along the lines
$T=U$, $T=-1/U$,
$T=f_h^w(U)$ and $T=f_v^w(U)$
(see Eqs. (\ref{Iths})), where the subtraction of $b_{\rm
grav}$
does not account for all massless contributions. The exact thresholds
 are obtained by replacing $\bgrav$ with the actual value of
the gravitational anomaly at the considered line
(see Eqs.~(\ref{dvby})--(\ref{dttby})). In the decompactification limit
(\ref{l0lis}), where an ordinary $N=2$ model is matched, the
gravitational thresholds diverge linearly for the class
(\romannumeral1) and
logarithmically for the class (\romannumeral2) since, then, $N=4$ is
actually
restored.
In the decompactification limit (\ref{lolil}), where $N=4$
supersymmetry
is systematically restored, the divergence is always logarithmic.
All these behaviours are summarized at the end of Appendix D.

\subsection{The case $\l=1$}\label{l1}

We now turn to the determination of threshold corrections in models
where
the shifted lattice
$\Gamma_{2,2}^w\oao{h}{g}$
is of the type $\lambda =1$. We must therefore determine the functions
${\Omega}^{\l=1}\oao{h}{g}$
and
$\Lambda^{\l=1}_i\oao{h}{g}$
appearing in
$F_{\rm
grav}^{\l=1}\oao{h}{g}$ and
$F_{i}^{\l=1}\oao{h}{g}$ (Eqs. (\ref{auhg}) and (\ref{fi})). Our
starting
point
will now be the model $E_8\times E_8 \times U(1)^2$ of
Appendix B.
We construct the generalized $\Omega^{\l=1}_{\vphantom i}\oao{h}{g}$
and
$\Lambda^{\l=1}_i\oao{h}{g}$ in a way similar to what was done in the
previous
section,
using the $\Omega_{(0)}^{\l=1}\oao{h}{g}$ and
$\Lambda_{(0)i}^{\l=1}\oao{h}{g}$
of
(\ref{129b}) and (\ref{L129b}), and Eqs. (\ref{gom}) and (\ref{lfe})
with $\lambda=1$ instead of $\lambda=0$.
Repeating the steps of the $\l=0$ calculation we find
that  $g(x)$  is again given by (\ref{glab}), which translates into
\ba
\Omega^{\l=1}_{\vphantom 1}\oao{0}{1}&=&{\hphantom{-}}
\th_3^2\, \th_4^2
\left(
\xi_1\, \th_3^8\, \th_4^8+
\xi_2\, \th_3^4\, \th_4^4 \, E_4^{\vphantom 3} +
\xi_3\, E_4^2
\right)
\nonumber\\
\Omega^{\l=1}_{\vphantom 1}\oao{1}{0}&=&-
\th_2^2\, \th_3^2
\left(
\xi_1\,  \th_2^8\, \th_3^8+
\xi_2\,\th_2^4\, \th_3^4 \, E_4^{\vphantom 3} +
\xi_3\, E_4^2
\right)
\label{gom1xi}\\
\Omega^{\l=1}_{\vphantom 1}\oao{1}{1}&=&-
\th_2^2\, \th_4^2
\left(
\xi_1\,  \th_2^8\, \th_4^8-
\xi_2\,\th_2^4\, \th_4^4 \, E_4^{\vphantom 3}  +
\xi_3\, E_4^2
\right)
;\nonumber
\ea
we also find that
\be
f_i^{\lambda =1}(x)= \frac{A_i\, (x^4-x^3+x^2-x+1)+B_i\,  x (x^2-x+1)
+C_i \, x^2}
{(x^2-x+1)(x-2)(2x-1)}\,.
\ee

The determination of the constants $A_i,B_i,C_i$ as well as of the
parameters
$\vec\xi$ of (\ref{glab}) in terms of physical parameters can be
carried
out as in the previous case. We first expand $F^{\l=1}_{i}\oao{h}{g}$
and $F^{\l=1}_{\rm grav}\oao{h}{g}$ (see Appendix C).
By taking into account the structure of the $\l=1$ shifted lattice as
explained in Appendix A (see Eqs. (\ref{l0t}), (\ref{l0tt}),
(\ref{l1v}) and
(\ref{l1h})), we
can identify the coefficients of the various negative or zero
powers of $\bar q$ in expressions (\ref{de11})--(\ref{f1m}) with
the physical parameters.

The absence of charged tachyon requires  the vanishing of the
$\frac{1}{\bar q}$-term in $F_i^{\l=1}\oao{0}{1}$.
The constant term of $F_i^{\l=1}\oao{0}{1}$ must  be identified with
the
beta-function
coefficient {\it at generic moduli} $b_i$. The $\frac{1}{{\bar
q}^{3/4}}$-term in $F_i^{\l=1(+)}$ plays a role along the line
$T=f_v^w(U)$
where extra states become massless. Twice its coefficient will be
therefore
identified with $\dv b_i$. Similarly, twice the coefficient of the
$\frac{1}{{\bar q}^{1/4}}$-term in $F_i^{\l=1(-)}$ will be called
$\dhh b_i$, which represents the discontinuity of the beta-function
coefficient along the line $T=f_h^w(U)$ \footnote{Note again that at
some isolated points of the lines
$T=\fv(U)$ or $T=\fh(U)$, the multiplicity of the relevant terms
in the lattice sums can double and the beta-function coefficients
become
$b_i+2\dv b_i$ or $b_i+2\dhh b_i$.}.
In contrast to what happens in $\l=0$ models, the two discontinuities
of the
beta-function coefficients arise at finite values of the moduli, where
in
general either vector multiplets and/or hypermultiplets appear.
Remember
that
for $\l=0$, this happens only for $\dhh b_i$, since $\dv b_i$ was the
discontinuity at the $N=2$ limit of the moduli space. Here  $N=4$
supersymmetry
is restored in both
limits
(see Fig. \ref{fig2}), with vanishing
of all beta-function coefficients and gravitational anomalies,
and without any new physical parameter.

Two more constraints are needed in order to determine all the above
parameters. These are obtained by inspecting the expansion
of $F_{\rm grav}^{\l=1}\oao{0}{1}$ (Eq. (\ref{f11})). Normalization of
the
tachyon contribution in the gravitational corrections imposes the
residue of the pole to be $-{1/12}$. Furthermore, the constant
term
is the gravitational anomaly.

The above constraints lead to six equations (\ref{e1}),
the solution of which
is given in (\ref{1A})--(\ref{1X3}). As we already mentioned in the
$\l=0$ case,
we observe that the combination $b_i-2\dhh b_i-8\dv b_i$ is
necessarily
proportional to $k_i$. In contrast to the $\l=0$ case
(see Eq.~(\ref{b46p})),
however, the proportionality constant cannot be expressed in terms of
$\bgrav$ only. It necessitates the introduction of a new
parameter,
although not independent, such as the discontinuity of the
gravitational
anomaly
along the line $T=\fv(U)$, $\dv\bgrav$, which is twice the
coefficient
of the $\frac{1}{{\bar q}^{3/4}}$-term in $F_{\rm grav}^{\l=1(+)}$
(\ref{f1p}).
We therefore find the relation:
\be
b_i-2\dhh b_i -8\dv b_i=3 k_i(16\, \dv\bgrav+\bgrav-2)\, .
\label{relv1}
\ee
Note that the gravitational anomaly possesses another discontinuity
$\dhh\bgrav$ along the $T=\fh(U)$, which  can be read off from
$F_{\rm grav}^{\l=1(-)}$ (\ref{f1m}) and
expressed in terms of other physical parameters by using
(\ref{1X1})--(\ref{1X3}) and
(\ref{relv1}):
\be
\dhh\bgrav=-28\, \dv\bgrav-\bgrav+\frac{10}{3}\, .
\label{relv11}
\ee

Further discontinuities $\delta_t\bgrav$ and $\delta_t'\bgrav$ arise
along $T=U$ and $T=-1/U$, respectively,  due to the
tachyon pole of $F_{\rm grav}^{\l=1}\oao{0}{1}$ combined with the
$\Gamma_{2,2}^w\oao{0}{1}$ lattice sum. These are the same as those
appearing in the $\l=0$ case, and are given in (\ref{dtby}) and
(\ref{dttby}).

The computation of the threshold corrections goes on as in the $\l=0$
situation.
The gauge corrections can be decomposed as in (\ref{del}) with all
$(T,U)$-dependent functions given in (\ref{d})--(\ref{y}) and
$\delta_{g,f}^{\l=1}$, \ldots, $y_{g,f}^{\l=1}$ displayed in
(\ref{1gf}). Again due to the relation (\ref{relv1}), the
definition of
the group-factor-independent contribution $Y^w(T,U)$ is not unique,
although it is taken to be
regular everywhere; it is also model-dependent through $\bgra$, and 
can be decomposed as in Eq. (\ref{ydeco}) (see
(\ref{1gf})).

Singularities appear at $T=\fh(U)$, where $H^w(T,U)$ exhibits
a logarithmic behaviour
(see Eqs. (\ref{Xths})).
This can be cured on the line $T=\fh(U)$
by properly subtracting the full contribution of the massless states,
i.e.
by performing the substitution (\ref{hp}). The same phenomenon occurs
across $T=\fv(U)$, where $V^w(T,U)$ diverges and where the substitution
\ba
V^w(T,U)&\to& V^w\left(f_v^w(U),U\right)
\nonumber\\
&=&
\ifd\left(
\sump\Gamma_{2,2}^w\oao{h}{g}\left(f_v^w(U),U\right) \right.\nonumber\\
&&\; \; \; \; \; \; \;\; \; \; \; \; \; \;  \left.\times
{\vphantom{\sump}}
\left(-\frac{1}{12}\,
\frac{\widehat E_2}{\bar\eta^{24}}\, {\overline
\Omega}_{(0)}^{\l=1}\oao{h}{g}\,
{\bar v}_g^{\l=1}\oao{h}{g}
+{\overline\Lambda}^{\l=1}_{(0)i}\oao{h}{g}\,
{\bar v}_f^{\l=1}\oao{h}{g}\right)-1\right)
\nonumber\\
&&+\log\frac{2}{\pi}\frac{e^{1-\gamma}}{\sqrt{27}}
\label{v1p}
\ea
is compulsory in order for the thresholds $\Delta_i^w$ to make sense.
Finally, at the limits (\ref{l1li}) (see Fig. \ref{fig2}), where the
$N=4$ supersymmetry is restored and
$F_i^{\l=1}$ vanishes, the thresholds (\ref{del}) diverge
logarithmically
(e.g. $\Delta_i^{\rm X}\sim\pm b_i\log T_2$, the minus sign
corresponding to
the large-$T_2$ limit, see Appendix D); this is,
as described previously, the consequence
of an incomplete infra-red regularization.

For the models at hand, gravitational corrections are still given in
(\ref{delg}), where the parameters $\vec\xi$ appearing in
$g\oao{h}{g}$ (see Eq. (\ref{glab})) are now taken in
(\ref{1X1})--(\ref{1X3}), which, thanks to
(\ref{relv1}), are expressed in the more convenient way:
\be
\xi_1=\frac{3}{4}\dv\bgrav+\frac{3}{64}\bgrav+\frac{27}{32}\
,\ \
\xi_2=-\frac{3}{2}\dv\bgrav-\frac{3}{64}\bgrav+\frac{5}{32}\
,\ \
\xi_3=\frac{3}{4}\dv\bgrav\, .
\label{I}
\ee
Here also we observe that only rational values are allowed for these
parameters. However, the decomposition (\ref{dgradec}) must now be replaced
with 
\be
\Delta_{\rm grav}^w = 
\Delta_{{\rm grav}, 1}^w
+ \bgrav\, 
\Delta_{{\rm grav}, 2}^w
+ \dv\bgrav\, 
\Delta_{{\rm grav}, 3}^w
\, ,
\label{iigradec}
\ee
where $\Delta_{{\rm grav}, i}^w$ $i=1,2,3$ are universal (only shift-vector-dependent) functions.
As far as the singularity of the  corrections is concerned, the same
comments as before apply here; details can be found in Appendix D.

\boldmath
\section{The case of orbifolds}\label{orb}
\unboldmath

\subsection{Some general results}

In Sections \ref{spbr} and \ref{thrN4} we described a general class of
heterotic
constructions with spontaneously-broken $N=4$ supersymmetry, for which
we
computed the gravitational and gauge threshold corrections. Those turn
out to
depend on moduli $(T,U)$ as well as on several low-energy parameters of
the
model, such as beta-function coefficients and the gravitational anomaly
and
their discontinuities across rational lines or on the border of the
moduli
space. The
gravitational thresholds, in particular, depend on a very specific
combination
of these low-energy data, namely on $\vec \xi$ (see (\ref{I0}) or
(\ref{I})).
The latter are discrete Wilson lines and are the only parameters
entering the
elliptic genus (see (\ref{gb}), (\ref{gom0xi}) and (\ref{gom1xi})),
which
therefore exhibits the universality properties
that we already discussed in Section \ref{thrN4}.

The above parameters (or equivalently the elliptic genus) do not
contain enough
information to reconstruct all properties of the massless spectrum,
such as
the
number of vector multiplets and hypermultiplets. Only the differences
$N_V - N_H$ or $\delta N_V - \delta N_H$ can be determined through
$b_{\rm grav}$ or $\delta b_{\rm grav}$. However, if we restrict
ourselves
to the subclass of the $Z_2$ orbifolds, more information can be
reached.
Indeed,  for these models the function
$C_{4,20}^{\l}\oao{h}{g}\left({v/ 2}\right)$
is given in (\ref{corbsp}), and it is possible to explicitly compute
the helicity supertrace $B_4$. We can then extract the massless part of
the
latter, and identify it with the low-energy formula, which reads:
\be
B_4^0=\frac{62 + 7 N_V - N_H}{4}\, ,
\label{B40}
\ee
for heterotic ground states. By using the low-energy expression for the
gravitational anomaly,
\be
b_{\rm grav}=\frac{22 - N_V + N_H}{12}\, ,
\label{bgrav}
\ee
we can determine $N_V$ and $N_H$ as well as the discontinuities of
these
numbers all over the moduli space.

The helicity supertrace
$B_4=\left\langle \left( \l+\bar \l \right)^4\right\rangle$ is obtained
at one loop by acting
on $Z(v,\bar v)$ (Eqs. (\ref{het2helj}),  (\ref{helsb}) and
(\ref{corbsp})) with
$\frac{1}{16\,  \pi^4}\left(\partial_v-\partial_{\bar v}\right)^4$ at
$v = \bar v = 0$. After some algebra (details can be found in Ref.
\cite{GKKOPP}), we find:
\be
B_4=\frac{3}{4}\frac{\Gamma_{2,2}\,\Gamma_{4,20}}{\bar\eta^{24}}
+\frac{1}{2}\sump {\Gamma_{2,2}^{w} {h\atopwithdelims[]g}}
\left( {H{h\atopwithdelims[]g}\over 2} + 2 -\bE_2 \right)
{\overline{\Omega}^{\l}{h\atopwithdelims[]g}\over \bar\eta^{24}}\, ,
\label{B4orb}
\ee
where
$$
H{h\atopwithdelims[]g}={12 \over  \pi i} \partial_\t \log {\th
{1-h\atopwithdelims[]1-g}\over
\eta}
= \cases{\hphantom{-} \th_3^4 + \th_4^4  \sp (h,g)=(0,1)\cr
          -  \th_2^4 - \th_3^4  \sp (h,g)=(1,0)\cr
\hphantom{-} \th_2^4 - \th_4^4  \sp (h,g)=(1,1)\, .\cr }
$$
Expression (\ref{B4orb}) is valid for $Z_2$-orbifold constructions with
spontaneously-broken $N=4$ to $N=2$ supersymmetry. The second term of
(\ref{B4orb}) results from $N=2$ sectors and possesses  the same
universality
properties as those  described previously for the gravitational
threshold
corrections: the model- and  moduli-dependence has shrunk to $(T,U)$
and $\vec
\xi$; put differently, this  term depends on the elliptic genus only.
However,
the first term of $B_4$, originated from the $N=4$ sector, spoils this
universality: as expected from  general considerations, it introduces
a full
dependence on the various moduli of $\Gamma_{4,20}$.

Let us now concentrate on the massless contributions to $B_4$. By using
the full
machinery introduced so far, we can compute $B_4^0$ in terms of $\vec
\xi$,
and use (\ref{I0}) or (\ref{I}) to trade the latter  for the parameter
$\bgra$. We find:
\be
B_4^0=\frac{3}{2}N_{\Gamma}+54-12\, \bgra \, .
\label{B40gen}
\ee
This formula is valid for any shift vector  $w$ ($\l = 0$ or 1), at any
{\it generic}  point of the $(T,U)$ moduli space. We also assumed that,
at
generic values of  the other moduli\footnote{For truly generic values
of the 80
moduli of
$\Gamma_{4,20}$, we would have $\Gamma_{4,20}=1+$ non-integer values of
$q$ and
$\bar q$. However, orbifold constructions often necessitate some of the
moduli
to be fixed at specific values.}, we have the behaviour:
$\Gamma_{4,20}=1+ 2N_{\Gamma}\bar q +\cdots$. This introduces a new
parameter of
the  orbifold, which captures all the extra moduli-dependence at the
level of
the  massless spectrum.  By using (\ref{B40}) and (\ref{bgrav}) we can
recast
(\ref{B40gen}) as: $N_V+N_H = 22 + 2N_{\Gamma}$, where $N_V$ and $N_H$
are the
number of vector multiplets and hypermultiplets {\it at generic $(T,U)$
moduli.} In turn, these are given by
\ba
N_V&=& 22 + N_{\Gamma}- 6 \bgra \, ,
\label{NVgen}\\
N_H&=& N_{\Gamma}+ 6 \bgra \, .
\label{NHgen}
\ea

We can go further and describe the behaviour of $B_4^0$ across rational
lines
in the $(T,U)$ moduli space. This can be achieved thanks to the various
expansions and limits introduced in Appendices A and C for lattices and
the
gravitational functions $F_{\rm grav}^{\l}\oao{h}{g}$. Notice that the
actual
value of $N_{\Gamma}$ plays no role in this analysis.
\newpage
\noindent {\sl a) The case $\l=0$}

\noindent {\sl -- The decompactification limit where ${\Gamma_{2,2}^{w}
{h\atopwithdelims[]g}}$ become equal for all $(h,g)$}

\noindent In this limit, present for any shift vector $w$, the massless
spectrum is
enhanced and discontinuities
$\delta_v b_i$ and $\delta_v \bgra$ appear. In the situation
(\romannumeral1) (see Section \ref{spbr2}), namely when
$\xi_3 =0$, this limit is also shared by an $N=2$ model with factorized
two-torus
(mapping (\ref{map}), see Fig. \ref{fig1}), which has beta-function
coefficients
$\tilde b_i=b_i + \delta_v b_i$, and gravitational anomaly
$\tilde \bgra= \bgra + \delta_v \bgra$. We find:
$$
\delta_v B^0_4=-3\delta_v\bgra\, ;
$$
this result can be recast in the form (see (\ref{dvby}), (\ref{B40})
and
(\ref{bgrav}))
$$
\delta_v N_V = 0 \ , \ \
\delta_v N_H = 264 -2 \bgra\, .
$$
The number of vector multiplets remains constant, as was already
stressed in the general case (see Section \ref{thrN4}); only extra
hypermultiplets
appear.

When $\xi_3=1$ (class (\romannumeral2)), in the limit under
consideration, the model
matches an $N=4$ orbifold with a factorized two-torus obtained through
the
mapping (\ref{map}) with (\ref{mapN4}) (see Fig. 3). We obtain now:
$$
\delta_v B^0_4=-18-3\delta_v\bgra \, .
$$
In the limit at hand, $N=4$ supersymmetry is restored and the number of
$N=2$ vector multiplets and hypermultiplets has no longer any meaning.
Instead
the number of $N=4$ vectors makes sense and turns out to be equal to
the number
of $N=2$ vector multiplets present before reaching the $N=4$ limit, as
can be
seen from the result
$$
\tilde B^0_4=B^0_4+ \delta_v B^0_4 = 3 +{3 \over 2} N_V \, ,
$$
$N_V$ given in (\ref{NVgen}). In other
words, the massless spectrum is reshuffled in such a way that the gauge
group
remains unchanged, as it should, when $N=4$
supersymmetry is restored.
\vskip 0.3cm
\noindent {\sl -- The line $T=f_h^w(U)$}

\noindent Here we find for both situations (\romannumeral1) and
(\romannumeral2)
$$
\delta_h B^0_4=-3\delta_h\bgra\, ,
$$
which leads to
(see (\ref{dhby}), (\ref{B40}) and
(\ref{bgrav}))
$$
\delta_h N_V = 0 \ , \ \
\delta_h N_H = \cases{
-\frac{3}{2}\bgra +1
\ ,  &{\rm class \ (\romannumeral1)},\cr
\cr
-\frac{3}{2}\bgra -15
\ ,  &{\rm class \ (\romannumeral2).}\cr}
$$
Now the absence of extra vectors is specific to orbifolds.
\vskip 0.3cm
\noindent {\sl -- The line $T=f_v^w(U)$}

\noindent For models of class (\romannumeral1) no extra massless states
appear along this line. In the
case (\romannumeral2), however, although the beta-function coefficients
(of the rank-20 factor of the
gauge group)
remain unchanged
and the gauge thresholds are regular, extra massless states appear
since the
gravitational anomaly has a discontinuity (see (\ref{dvpby})).
Similarly
$$
\delta_v' B^0_4=-\frac{1}{2}\, ,
$$
and consequently
$$
\delta_v' N_V = 0 \ , \ \
\delta_v' N_H = 2\, .
$$
The extra hypermultiplets are singlets under the rank-20 component of
the gauge group.
\vskip 0.3cm
\noindent {\sl b) The case $\l=1$}

\noindent
In this case $N=4$ supersymmetry is restored in all decompactification
limits,
and there is not much to say about them. The interesting phenomena
occur
along the lines $T=f_h^w(U)$ and $T=f_v^w(U)$, where we find:
$$
\delta_{h,v} B^0_4=-3\delta_{h,v}\bgra\, ,
$$
leading to
$$
\delta_{h,v} N_V = 0 \ , \ \
\delta_{h,v} N_H =12\, \delta_{h,v}\bgra
$$
(remember that in the situation $\l=1$, $\delta_{v}\bgra$ is considered
as an
input parameter together with $b_i$, $\delta_{v}b_i$, $\delta_{h}b_i$,
and
$\bgra$, whereas $\delta_{h}\bgra$ is related to the others through
(\ref{relv11})). Again, the absence of extra vectors is not to be
considered as
a generic feature of the class of models analysed in this paper, but
instead as
a property of the orbifold constructions.
\vskip 0.3cm
\noindent {\sl c) The lines $T=U$ and $T=-U^{-1}$}

\noindent
These deserve a special treatment, because they appear both in
models with
spontaneously-broken $N=4$ supersymmetry and in ordinary $N=2$ models
with a factorized two-torus (although in the latter they are
equivalent).
In all cases the beta-function coefficients (of the rank-20 factor of
the
gauge group) remain unchanged, $\delta_t b_i=\delta_t' b_i=0$, and the
corresponding
gauge
threshold corrections are regular. This is a consequence of the
absence of any
charged tachyon. The gravitational anomaly, however, receives extra
contributions. For the models with a factorized two-torus, this is
given in
(\ref{d69}), and accounts for the appearance of 2, 4 or 6 extra vector
multiplets:  the $U(1)^2$ factor of the two-torus becomes $U(1) \times
SU(2)$,
$SU(2) \times SU(2)$ or $SU(3)$ at $T=U$, $T=U=i$ or $T=U=\rho$.

In the case of models with spontaneously-broken $N=4$ supersymmetry we
are
considering here, the discontinuities of the gravitational anomaly are
given in
(\ref{dtby}) and (\ref{dttby}). Moreover, for orbifold constructions,
using (\ref{B4orb})
we find:
\be
\delta_{t} B^0_4=\frac{3}{2}\delta-12\, \delta_t\bgra\, ,
\label{dtB4}
\ee
which leads to
\be
\delta_t N_V = \delta +\delta^w_t \ , \ \
\delta_t N_H = \delta -\delta^w_t
\label{dtN}
\ee
for the line $T=U$, and similarly for the line $T=-1/U$ with
$\delta^w_t$ replaced with $\delta^{\prime w}_t $.
($\delta$, $\delta^w_t$ and $\delta^{\prime w}_t $ are defined in Eqs.
(\ref{d69}),
(\ref{ddtby}) and (\ref{ddttby}),
respectively).
We observe that, not only extra vector multiplets, but also extra
hypermultiplets, singlets under the rank-20 component of the gauge
group,
do appear when $T=U$ or $T=-1/U$ are reached.
Since $\delta^w_t\neq \delta^{\prime w}_t$, the spectrum of extra
massless states is different along the two lines under consideration,
which translates the breaking of the duality group.
Moreover, it is determined
exclusively by
the shift vector $w$ and does not depend on any low-energy parameter of
the
model (as, for instance, $\bgra$).

\subsection{Examples}

\noindent {\sl a) The case $\l =0$, class (\romannumeral1)}

\noindent These models fit to the scheme presented in Fig. \ref{fig1}.
They can
be seen as
deformations of the known six-dimensional $Z_2$ orbifolds. This
perspective has been adopted in \cite{decoa},
where a model was constructed as
a deformation
of the symmetric $Z_2$ orbifold. This will be our first example.
Following this procedure, more orbifold models can be constructed.
The number of possible models in this class is equal to the
number of ordinary $Z_2$ $N=2$ orbifolds. They are related by the
mapping
(\ref{map}).
\vskip 0.3cm
\noindent {\sl -- $E_8\times E_7\times SU(2) \times {U(1)}^2$}

\noindent In the notation used here this model can be recovered with
the lattice
\be
\Gamma_{4,20}^{\l=0}{h\atopwithdelims[]g}=
\Gamma_{4,4}^{\vphantom  y}{h\atopwithdelims[]g}\,
{1\over 2}\sum_{\bar a,\bar b=0}^1
\bar\vartheta^{6}_{\vphantom 1}{\bar a\atopwithdelims[]\bar b}\,
\bar\vartheta{\bar a+h\atopwithdelims[]\bar b+g}\,
\bar\vartheta{\bar a-h\atopwithdelims[]\bar b-g}\,
{\overline E_4}\, ,
\label{420O1r}
\ee
where
\be
Z_{4,4}{0\atopwithdelims[]0}\equiv Z_{4,4}\equiv{\Gamma_{4,4}\over
\vert\eta \vert^8}
\label{44}
\ee
is the partition function of four compactified bosons, which depends on
16 moduli while, for $(h,g)\neq (0,0)$,
\be
Z_{4,4}{h\atopwithdelims[]g}
\equiv {\Gamma_{4,4}{h\atopwithdelims[]g}\over \vert\eta \vert^8}
=
{16\,  \vert\eta \vert^4\over
\left\vert
\vartheta{1+h\atopwithdelims[] 1+g}\,
\vartheta{1-h\atopwithdelims[] 1-g}
\right\vert^2}
\label{44sym}
\ee
are the ordinary $Z_2$-twisted contributions.
Here $N_{\Gamma} = 240$.
One can use Eqs. (\ref{omorbsp}) to determine the corresponding
functions
$\Omega\oao{h}{g}$; after comparison with (\ref{gom0xi}), the latter
give
${\vec\xi}=(0,1,0)$. This determines the universality class of the
model,
and in particular $\bgra = -62/3$, $\delta_v\bgra = 128/3$ and
$\delta_h\bgra = 8/3$. We find $N_V=386$ and $N_H=116$, in agreement
with the
gauge group and
the matter massless spectrum, which is
$({\bf1},{\bf 56}, {\bf2}) + 4\times({\bf1},{\bf1,\bf1})$, with
$b_{E_8}=-60$, $b_{E_7}=-12$ and $b_{SU(2)}=52$. We also find
$\delta_v N_H = 512$ extra hypermultiplets in the $N=2$
decompactification
limit, originated from the twisted sector, and
falling in $8 \times({\bf1},{\bf 56}, {\bf1}) +
32\times({\bf1},{\bf1,\bf2})$.
They lead to $\delta_v b_{E_8}=0$, $\delta_v b_{E_7}= 96$ and $\delta_v
b_{SU(2)}=32$. Finally, we find $\delta_h N_H = 32$ extra
hypermultiplets
originated from the twisted sector in $16\times({\bf1},{\bf1,\bf2})$,
with
$\delta_h b_{E_8}=0$, $\delta_h b_{E_7}= 0$ and $\delta_h
b_{SU(2)}=16$. Equation (\ref{b46p}) is verified by the above
low-energy
parameters.
\newpage
\noindent {\sl -- $SO(12)\times SO(20)\times {U(1)}^2$}

\noindent In this case the $(4,20)$ lattice sum is
\be
\Gamma_{4,20}^{\l=0}{h\atopwithdelims[]g}=
\Gamma_{4,4}^{\vphantom  y}{h\atopwithdelims[]g}\,
{1\over 2}\sum_{\bar a,\bar b=0}^1
\bar\vartheta^6_{\vphantom 1}{\bar a\atopwithdelims[]\bar b}\,
\bar\vartheta^5_{\vphantom 1}{\bar a+h\atopwithdelims[]\bar b+g}\,
\bar\vartheta^5_{\vphantom 1}{\bar a-h\atopwithdelims[]\bar b-g}\,.
\label{420O5r}
\ee
Now $N_{\Gamma} = 240$,
${\vec\xi}=(1,0,0)$; $\bgra = 2/3$, $\delta_v\bgra = 64/3$ and
$\delta_h\bgra = 0$. We find $N_V=258$ and $N_H=244$, in agreement with
the
gauge group and
the matter massless spectrum, which is here
$({\bf 12},{\bf 20})+4\times({\bf1},{\bf1})$, with
$b_{SO(12)}=20$ and $b_{SO(20)}=-12$. We also find
$\delta_v N_H = 256$ extra hypermultiplets
falling in $8\times ({\bf32},{\bf1})$.
They lead to $\delta_v b_{SO(12)}=64$ and $\delta_v b_{SO(20)}=0$.
Finally, we
find $\delta_h N_H = 0$,
so that nothing happens along $T=f_h^w(U)$
($\delta_h N_V$ vanishes for all orbifold models).
All low-energy parameters are consistent with Eq. (\ref{b46p}).
\vskip 0.3cm
\noindent {\sl b) The case $\l =0$, class (\romannumeral2)}

\noindent
Let us now proceed to present two models where
$N=4$ supersymmetry is restored in both six-dimensional
limits (see Fig. \ref{fig3}). This
is related to the fact that they are constructed
by a pure shift in the $\Gamma_{2,2}$ lattice
and no twist action in the  $\Gamma_{4,4}$ lattice.
Since in both limits the shift action is effectively removed,
and no twist action is present,
supersymmetry
is always restored to $N=4$.
\vskip 0.3cm
\noindent {\sl -- $E_8\times E_8 \times SO(8) \times {U(1)}^2$}

\noindent The simplest model is the one  presented in Appendix B.
It corresponds
to the lattice (\ref{420E0E5}). Notice that it is quite remarkable to
find an $E_8\times E_8$ factor together with $N=2$ supersymmetry in a
four-dimensional construction. It has
$N_{\Gamma} = 252$,
${\vec\xi}=(0,0,1)$; $\bgra = -42$, $\delta_v\bgra = 42$ and
$\delta_h\bgra = 4$. We find $N_V=526$ and $N_H=0$. There is no matter
here and $b_{E_8}=-60$, $b_{SO(8)}= -12$. We have
$\delta_v b_{E_8}=60$, $\delta_v b_{SO(8)}= 12$ because of the $N=4$
restoration. We also find
$\delta_h N_H = 48$ extra hypermultiplets along the line $T=f^w_h(U)$
falling in $6\times ({\bf1},{\bf1},{\bf8})$.
They lead to $\delta_h b_{E_8}=0$ and $\delta_h b_{SO(8)}=12$, in
agreement
with Eq. (\ref{b46p}).
\vskip 0.3cm
\noindent {\sl -- $SO(40)\times {U(1)}^2$}

\noindent This model is obtained with
\be
\Gamma_{4,20}^{\l=0}{h\atopwithdelims[]g}=
{1\over 2}\sum_{\bar a,\bar b=0}^1
\vartheta^{2}_{\vphantom 1}{\bar a\atopwithdelims[]\bar b}\,
\vartheta{\bar a+h\atopwithdelims[]\bar b+g}\,
\vartheta{\bar a-h\atopwithdelims[]\bar b-g}\,
\bar\vartheta^{20}_{\vphantom 1}{\bar a+h\atopwithdelims[]\bar b+g}
\, .
\label{420linr}
\ee
It has the largest single group factor that can be obtained
in this construction. We find $N_{\Gamma} = 380$,
${\vec\xi}=(-1,1,1)$; $\bgra = -190/3$, $\delta_v\bgra = 190/3$ and
$\delta_h\bgra = 20/3$. There is no matter
here, $N_V=782$ and $b_{SO(40)}=-76$, $\delta_v b_{SO(40)}=76$. Now
$\delta_h
N_H = 80$ extra hypermultiplets appear along the line $T=f^w_h(U)$
falling in
$2\times ({\bf40})$. They lead to $\delta_h b_{SO(40)}=4$, in agreement
with Eq.
(\ref{b46p}).

The enlargement of the gauge group in  the last two models
is also a result of the absence of twist, which allows
the $\Gamma_{4,4}$ right-moving fermions to get gauged.
\newpage
\noindent {\sl c) The case $\l =1$}

\noindent
As explained in Section \ref{thrN4}, $\l=1$ models fall in the
situation
depicted
in Fig. \ref{fig2}. The two six-dimensional limits are equivalent
(which was
not true in the previous case) and therefore restore the $N=4$
supersymmetry
since the orbifold twist is removed. We will give three examples, one
of these
being the celebrated $E_8$ level 2 construction.
\vskip 0.3cm
\noindent {\sl -- $E_8\times E_8\times {U(1)}^2$}

\noindent This is the
simplest  model of this category and is the one presented
in Appendix B. The $(4,20)$ lattice is
given in (\ref{42013}). It has
$N_{\Gamma} = 240$,
${\vec\xi}=(0,0,1)$; $\bgra = -118/3$, $\delta_v\bgra = 4/3$ and
$\delta_h\bgra = 16/3$. We find $N_V=498$, and $N_H=4$ hypermultiplets,
which
are singlets of the non-Abelian gauge group factor. The beta-function
coefficient is $b_{E_8}=-60$. Moreover, along the rational lines
$T=f^w_v(U)$ and $T=f^w_h(U)$, we find
$\delta_v N_H = 16$ and $\delta_h N_H = 64$ extra hypermultiplets,
singlets of
the non-Abelian gauge group factor, leading to
$\delta_v b_{E_8}=0$ and $\delta_h b_{E_8}=0$. This is in agreement
with Eq. (\ref{relv1}).
\vskip 0.3cm
\noindent {\sl -- $SO(16)\times SO(16)\times {U(1)}^2$}

\noindent Another model can be obtained by using the
lattice
\be
\Gamma_{4,20}^{\l=1}{h\atopwithdelims[]g}=
\Gamma_{4,4}^{\vphantom  y}{h\atopwithdelims[]g}\,
{1\over 2}\sum_{\bar a,\bar b=0}^1
\bar\vartheta^8_{\vphantom 1}{\bar a\atopwithdelims[]\bar b}\,
\bar\vartheta^8_{\vphantom 1}{\bar a+h\atopwithdelims[]\bar b+g}\, .
\label{42011r}
\ee
We now have
$N_{\Gamma} = 240$,
${\vec\xi}=(1,0,0)$, $\bgra = 10/3$, $\delta_v\bgra = 0$ and
$\delta_h\bgra = 0$. We find $N_V=242$ and $N_H=260$, in agreement with
the
gauge group and
the matter massless spectrum, which is here
$ ({\bf 16},{\bf 16})
+
4\times({\bf1,\bf1})$, with
$b_{SO(16)}=4$. Finally $\delta_{v,h} N_H=0$ and
$\delta_{v,h} b_{SO(16)}=0$, in agreement
with Eq. (\ref{relv1}).
\vskip 0.3cm
\noindent {\sl -- $E_8 \times {U(1)}^2$}

\noindent This model has recently
attracted much attention in the framework of heterotic/type II
dual-pair
construction \cite{FHSV}. On the $\Gamma_{0,16}$ lattice, the
$Z_2$ permutes the two $E_8$'s. A single $E_8$ current algebra
survives, which
is realized at level 2. Eventually, the $(4,20)$ lattice sum reads:
\be
\Gamma_{4,20}^{\l=1}{h\atopwithdelims[]g}=
\Gamma_{4,4}^{\vphantom{l}}{h\atopwithdelims[]g}\,
\Gamma_{E_{8 \vert 2}}^{\vphantom{l}}{h\atopwithdelims[]g}
\, ,
\label{4201E82}
\ee
where
\be
\Gamma_{E_{8 \vert 2}}{0\atopwithdelims[]0} =
\bE_4^2(\bar \tau)\ ,  \ \
\Gamma_{E_{8 \vert 2}}{0\atopwithdelims[]1} =
\bE_4(2\bar \tau)\frac{\bar \eta^{16}(\bar \tau)}{\bar \eta^{8}(2 \bar
\tau)}\,  ,
\label{4201E8}
\ee
and $\Gamma_{E_{8 \vert 2}}{1\atopwithdelims[]0 {\rm \ or \ } 1}$ are
obtained
by the modular transformations $\tau \to \tau +1$ and
$\tau \to -\tau^{-1}$.

In the model at hand,  $N_{\Gamma} = 240$,
${\vec\xi}=(15/16,1/16,0)$; $\bgra = 2$, $\delta_v\bgra = 0$ and
$\delta_h\bgra = 4/3$. We find $N_V=250$, and $N_H=252$
hypermultiplets, which
are in $({\bf 248})+4\times({\bf1})$. The beta-function
coefficient is $b_{E_8}=0$. Moreover, along the rational line
$T=f^w_v(U)$, $\delta_v N_H = 0$  and consequently
$\delta_v b_{E_8} = 0$. On the other hand,
$\delta_h N_H = 16$ hypermultiplets appear at  $T=f^w_h(U)$
and, being singlets of $E_8$, give $\delta_h b_{E_8}=0$. Again,
beta-function
coefficients fulfil  Eq. (\ref{relv1}).

\section{Conclusions}

In this paper we have analysed threshold corrections to gauge and
gravitational couplings in four-dimensional heterotic models where
$N=4$  space-time supersymmetry is spontaneously
broken to $N=2$.

Such ground states can be viewed as obtained by compactifying the
ten-dimensional heterotic string on a six-dimensional compact manifold
of $SU(2)$ holonomy. This manifold is locally but not globally
of the product form $K3\times T^2$.
In these models there are two massive gravitinos, whose masses
are calculable functions of the torus moduli.
These masses become vanishing, and thus supersymmetry is restored to
$N=4$,
in an appropriate  decompactification limit. The analysis of the
decompactification limits exhibits three subclasses of models ($\l = 0$
(\romannumeral1) and
(\romannumeral2), and $\l = 1$).

The properties mentioned above are expected to significantly affect
the high-energy running of effective coupling constants; this was
shown
to be true in some sample ground states in \cite{decoa}.

Here we have derived explicit expressions for generic models of the
above type
without knowledge of their detailed structure. The important
ingredients
that appear in the expressions for the one-loop gauge and gravitational
thresholds are properties of the massless and BPS spectrum;
more precisely, beta-function coefficients and affine-Lie-algebra
levels, as well as jumps
of the beta-functions along submanifolds of the torus moduli space
where extra BPS multiplets become massless. In fact, in contrast to
what happens in models with a factorized two-torus, several rational
lines appear, where the gauge threshold corrections are singular
(singularities of the gravitational thresholds appear independently of
the factorization of the
two-torus). However, these lines do not necessarily correspond to an
enhancement of gauge symmetry: $\delta N_V$ and $\delta N_H$ are not a
priori determined.

We have thus found that the universality properties, observed in
$K3\times
T^2$-like compactifications \cite{kkpr, kkpra}
as a consequence of six-dimensional anomaly cancellations,
are  slightly modified here,
although they
can still be traced to modular invariance and unitarity, and
to the fact that the couplings studied are of the
BPS-saturated
type \cite{bk}. For the gravitational thresholds, the
explicit expression exhibits a model-dependence, which is captured in
the shift vector $w$ and the rational parameters $\vec \xi$
namely $\bgra$ (and $\dv \bgra$ for $\l = 1$).
The latter can be interpreted as discrete Wilson lines
(or instanton numbers of the $Z_2$-shift embedding), and define the
various universality classes where all models under consideration fall.
These are genuine classes in the sense that they contain more than a
single representative.
As far as the gauge threshold corrections are concerned, the ususal
decomposition in two terms no longer holds. A gauge-factor-independent
term
can still be defined. However, there is some arbitrariness in its
definition due to a relation between the various low-energy parameters
involved. Moreover,
this term depends explicitly on the value of the gravitational anomaly
of the ground state.

By using our expressions for the threshold corrections (for which we
have also explicitly performed the integrals over the fundamental
domain), we have analysed the behaviour at large radii of
compactification. In agreement with the expected
supersymmetry-restoration properties, the thresholds are linearly or
logarithmically divergent. In the second case, the $N=4$ supersymmetry
is restored, and the logarithmic divergence is actually an infra-red
artefact due to an accumulation of massless states, which can be lifted
by switching on appropriate Wilson lines. Indeed, the thresholds should
vanish as expected when supersymmetry is extended to $N=4$.

For generic orbifold constructions falling in our general class of
heterotic ground states, the enhancement of the massless spectrum along
specific submanifolds of the moduli space can be unambiguously
determined. Except for the lines $T=U$ and $T=-1/U$, only
hypermultiplets become
massless. In the framework of orbifolds, we have also presented several
specific constructions, where the gauge group contains factors such as
$E_8 \times E_8$, $SO(40)$ or even $E_{8\vert 2}$ (in four dimensions).

The results presented here are a priori applicable to $N=2$
supersymmetric
theories. In fact,
they can serve for realistic $N=1$ models that are orbifolds of the
ground states studied in this paper.
The internal moduli-dependence of the couplings would be coming from
$N=2$
sectors and will thus be given by the expressions we have derived
above.

The formalism we developed so far can also be useful for analysing the
issue of  non-perturbative phenomena in $N=2$ type II dual models. The
extra $\delta N_V$
and $\delta N_H$ massless states that appear on the rational lines will
then correspond to monopoles or dyons {\it \`a la} Seiberg--Witten.
Work in this direction will appear soon \cite{GKP}.

\vskip 0.56cm
\centerline{\bf Acknowledgements}
\vskip 0.25cm
\noindent
We thank A. Gregori, J.-P. Derendinger and S.
Stieberger for useful remarks on this work. We
also thank N. Obers for helpful
discussions on the computation of the fundamental-domain integrals.
E. Kiritsis thanks the Centre de Physique Th\'eorique de l'Ecole
Polytechnique for hospitality.
E.~Kiritsis and C. Kounnas were  supported in part by EEC
contracts TMR-ERB-4061-PL95-0789 and TMR-ERBFMRX-CT96-0045.
P.M. Petropoulos was  supported in part by EEC contract
TMR-ERBFMRX-CT96-0090.
J. Rizos thanks the CERN Theory Division  for
hospitality
and  acknowledges financial
support from the EEC contracts ERBCHBG-CT94-0634 and
TMR-ERBFMRX-CT96-0090.

\vskip 0.3cm
\setcounter{section}{0}
\setcounter{equation}{0}
\renewcommand{\theequation}{A.\arabic{equation}}
\section*{\normalsize{\centerline{\bf Appendix A: Two-torus lattice
sums}}}

In this appendix we give our notation and conventions for the usual
and $Z_2$-shifted ($2,2$) lattice sums. We also analyse the behaviours
of those sums all over the moduli space
as well as in various decompactification limits.
\vskip 0.3cm
\noindent {\sl A.1 $Z_2$-shifted lattice sums}

\noindent
The ($2,2$) lattice sum is given by
\ba
\Gamma_{2,2}\left(T,U,\bT,\bU\right)=
\sum_{{\vec m},{\vec n}\in Z}
\exp
\bigg(\!\!\!\!\!\!\!\!\!\!&&
2\pi i\bar\t
{\vec m} {\vec n}\cr &&-
{\pi\im\over T_2 U_2}
\left|T n_{1} + TUn_{2}+Um_1-m_2\right|^2
\bigg)\, .
\label{z22}
\ea
It is invariant under the full
target-space duality group $SL(2,Z)_T^{\vphantom U} \times
SL(2,Z)^{\vphantom T}_U \times Z_2^{T \leftrightarrow U}$.

The $Z_2$-shifted lattice sum of the two-torus $\Gamma_{2,2}^{w}
{h\atopwithdelims[]g}$
depends on two integer-valued two-vectors
$(\vec a, \vec b)\equiv w$.
Independently of the shift vector $w$,
\be
\Gamma_{2,2}^{w}{0\atopwithdelims[]0}\equiv\Gamma_{2,2}\,,
\ee
given in (\ref{z22}); for $(h,g)\ne(0,0)$,
$\Gamma^{w}_{2,2}{h\atopwithdelims[]g}$
is obtained from
$\Gamma_{2,2}^{\vphantom A}$ by inserting
$(-1)^{g  \left( \vec n \vec a  +  \vec m \vec b\right)}$ and shifting
$\vec m\to \vec m + \vec a h/2$ and $\vec n\to \vec n + \vec b h/2$.
There are many choices for the $Z_{2}$ translation on the $T^2$. The
choice of the
vectors $\vec a$ and $\vec b$ determines the kind of states
(winding and/or momentum) that are projected out by the orbifold.
We find:
\ba
\!\!\!\!\!\!\!\!\!\!\!\!\!\!\!\!\!\!&&\Gamma_{2,2}^{w}
{h\atopwithdelims[]g}=
\sum_{\vec m,\vec n \in Z}
(-1)^{g  \left( \vec n \vec a  +  \vec m \vec b\right)}
\exp
\Bigg(
2\pi i\bar\t\left( \vec m + \vec a \, {h \over 2}\right)
\left(\vec n + \vec b \, {h \over 2}\right)\cr
\!\!\!\!\!\!\!\!\!\!\!\!\!\!\!\!\!\!&&-
{\pi\im\over T_2 U_2}
\left|T \left( n_{1}+b_1\, {h \over 2}\right) + TU\left( n_{2}+b_2 \,
{h \over
2}\right)+U\left( m_1+a_1\, {h \over 2}\right)-\left( m_2+a_2\, {h
\over 2}\right)\right|^2
\Bigg)
\label{44444}
\ea
in the Hamiltonian representation, or
\ba
\!\!\!\!\!\!
\Gamma_{2,2}^{w}{h\atopwithdelims[]g}&=&{ T_2 \over\im}
\sum_{\vec m,\vec n \in Z}
{\rm e}^{i \pi\vec a  \left( \vec n g  -  \vec m h -\vec b \, {gh\over
2}\right)}
\exp-
{\pi \over \im}\sum_{i,j}
\left(m_i + b_i\, {g\over 2}+\left(n_i + b_i\, {h\over
2}\right)\t\right)
\cr
&&\ \ \ \ \ \ \ \ \ \ \ \ \ \ \ \ \ \ \ \ \ \ \ \ \ \ \ \ \ \ \ \ \ \ \
\ \ \ \ \left(G_{ij}+B_{ij}\right)
\left(m_j + b_j\, {g\over 2}+\left(n_j + b_j\, {h\over 2}\right)\bar
\tau\right)
\label{4444l}
\ea
in the Lagrangian representation, where as usual
\be
G={\iT \over \iU}
\left(\matrix{1 & U_2 \cr U_2 & |U|^2 \cr}\right)
\ ,\ \ B=T_2 \left(\matrix{ 0 & -1 \cr
1 & {\hphantom -}0 \cr}\right)\, .
\label{I2}
\ee

It is easy to check the periodicity properties ($h,g$ integers)
\be
\Gamma_{2,2}^{w}{h\atopwithdelims[]g}=
\Gamma_{2,2}^{w}{h+2\atopwithdelims[]g}=
\Gamma_{2,2}^{w}{h\atopwithdelims[]g+2}=
\Gamma_{2,2}^{w}{-h\atopwithdelims[]-g}\, ,
\label{25b}
\ee
as well as the modular transformations that expression
\be
Z_{2,2}^{w}{h\atopwithdelims[]g}
\equiv{\Gamma_{2,2}^{w}
{h\atopwithdelims[]g}\over
|\eta|^4}
\label{25}
\ee
obeys:
\be
\tau\to\tau+1 \ ,\ \ Z_{2,2}^{w}{h\atopwithdelims[]g}\to
{\rm e}^{i\pi{\vec a \vec b}{h^2 \over 2}}\, Z_{2,2}^{w}
{h\atopwithdelims[]h+g}
\label{255b}\ee
\be
\tau\to-{1\over \tau} \ ,\ \ Z_{2,2}^{w}
{h\atopwithdelims[]g}\to
{\rm e}^{-i\pi{\vec a \vec b}{hg}}\, Z_{2,2}^{w}
{g\atopwithdelims[]-h}\, .
\label{256b}\ee
The relevant parameter for these transformations is $\l \equiv \vec a
\vec b$.

We would now like to give a few properties of the shifted lattice sums.
It is clear from expression (\ref{44444}) or (\ref{4444l}) that the
integers $a_i$ and $b_i$ are defined modulo 2, in the sense that adding
2 to anyone of them amounts at most to a change of sign in
$\Gamma_{2,2}^{w}{1\atopwithdelims[]1}$. Such a
modification is necessarily compensated by an appropriate one in
$C_{4,20}^{\l}{1\atopwithdelims[]1}$ (see Eq. (\ref{helsb})) in order
to ensure modular
invariance, and thus we are left with the same string ground state. On
the
other
hand, adding 2 to $a_i$ or $b_i$ translates into adding a multiple of 2
to $\l$. Therefore,
although $\l$ can be any integer, {\it only $\l=0$ and $\l=1$
correspond to
truly different situations.}

In Tables A.1 and A.2, we list all physically distinct models
with $\l=0$ and $\l=1$, respectively. In each of these classes, all the
models are related to one another by transformations 
that belong to $SL(2,Z)_T^{\vphantom U} \times
SL(2,Z)^{\vphantom T}_U \times Z_2^{T \leftrightarrow U}$.
\begin{center}
\begin{tabular}{| c | c | c | c | c | }
\hline
{\rm Case}&$\vec{a}$ & $\vec{b}$ \\ \hline
I   &$(0,0) $  & $(1,0)$  \\ \hline
II  &$(0,0) $  & $(0,1)$  \\ \hline
III &$(0,0) $  & $(1,1)$  \\ \hline
IV  &$(1,0) $  & $(0,0)$  \\ \hline
V   &$(0,1) $  & $(0,0)$  \\ \hline
VI  &$(1,1) $  & $(0,0)$  \\ \hline
VII &$(1,0) $  & $(0,1)$  \\ \hline
VIII&$(0,1) $  & $(1,0)$  \\ \hline
IX  &$(1,-1)$  & $(1,1)$  \\ \hline
\end{tabular}
\end{center}
\centerline{Table A.1: The nine  models with
$\l=0$.}
\begin{center}
\begin{tabular}{| c | c | c | c | c | }
\hline
{\rm Case}&$\vec{a}$ & $\vec{b}$ \\ \hline
X   &$(1,0) $  & $(1,0)$  \\ \hline
XI  &$(1,0) $  & $(1,1)$  \\ \hline
XII &$(1,1) $  & $(1,0)$  \\ \hline
XIII&$(0,1) $  & $(0,1)$  \\ \hline
XIV &$(0,1) $  & $(1,1)$  \\ \hline
XV  &$(1,1) $  & $(0,1)$  \\ \hline
\end{tabular}
\end{center}
\centerline{Table A.2: The six  models with
$\l=1$.}

Another issue that we
would like to discuss here is that of target-space duality in the
presence of a $Z_2$
translation. The moduli dependence of the two-torus shifted sectors
(see Eq. (\ref{44444}) or (\ref{4444l})) reduces in general  the
duality group
to some
subgroup\footnote{The subgroups
of $SL(2,Z)$ that will actually appear in the following are
$\Gamma^{\pm}(2)$ and $\Gamma(2)$. If
$\left(\matrix{a&b \cr c&d}\right)$
represents an element of the modular group,
$\Gamma^+(2)$ is defined by $a,d$ odd and $b$ even, while for
$\Gamma^-(2)$ we have $a,d$ odd and $c$ even. Their intersection is
$\Gamma(2)$.}
of
$SL(2,Z)_T^{\vphantom U} \times SL(2,Z)^{\vphantom T}_U \times Z_2^{T
\leftrightarrow U}$. Transformations that do not belong to this
subgroup map a model $w=(\vec a, \vec b)$ to some other model
$w'=(\vec a
', \vec b ')$, {\it leaving however $\l=\vec a \vec b=\vec a
'
\vec b '$ invariant.} This plays an important role in string
constructions such as
those described in (\ref{het2helj}) with
(\ref{helsb}), where a $Z_2$ translation appears, giving a
moduli-dependent mass to half of the gravitinos. Indeed, for such a
model,
decompactification limits that are related
by transformations that do not belong to the actual duality group are
no longer equivalent. Therefore, the spontaneously-broken $N=4$
supersymmetry might or might not be restored (see Section \ref{spbr}).

To be more specific, by using expression (\ref{44444}), we can
determine the transformation properties of $\Gamma_{2,2}^{w}
{h\atopwithdelims[]g}$ under the full group
$SL(2,Z)_T^{\vphantom U} \times SL(2,Z)^{\vphantom T}_U \times Z_2^{T
\leftrightarrow U}$:
$$
SL(2,Z)_T \ :\ \
\left(\matrix{a_1 \cr a_2 \cr b_1 \cr b_2 \cr}\right)
\to
\left(
\matrix{
d&{\hphantom{-}}0&0&b\cr 0&{\hphantom{-}}d&-b{\hphantom{-}}&0\cr
0&-c&a&0\cr c&{\hphantom{-}}0&0&a\cr}
\right)
\left(
\matrix{a_1 \cr a_2 \cr  b_1 \cr b_2\cr}
\right)\ ,\ \ ad-bc=1\, ,
$$
$$
SL(2,Z)_U \ :\ \
\left(\matrix{a_1 \cr a_2 \cr b_1 \cr b_2 \cr}\right)
\to
\left(
\matrix{
{\hphantom{-}}a'&-c'{\hphantom{-}}&0&0\cr -b'&d'&0&0\cr
{\hphantom{-}}0&0&d'&b'\cr {\hphantom{-}}0&0&c'&a'\cr}
\right)
\left(
\matrix{a_1 \cr a_2 \cr  b_1 \cr b_2\cr}
\right)\ ,\ \ a'd'-b'c'=1
$$
and
$$
Z_2^{T \leftrightarrow U} \ :\ \
\left(
\matrix{a_1 \cr a_2 \cr b_1 \cr b_2 \cr}
\right)
\to
\left(
\matrix{0&0&1&0\cr 0&1&0&0\cr 1&0&0&0\cr 0&0&0&1\cr}
\right)
\left(\matrix{a_1 \cr a_2 \cr b_1 \cr b_2 \cr}\right)\, .
$$
Thus, we can determine the duality group for a given model by demanding
that the components of the vectors $\vec a$ and $\vec b$ remain
invariant modulo 2. For example, in the situation I ($\l = 0$) defined
by $\vec a=(0,0)$ and $\vec b=(1,0)$, the target-space duality group
turns
out to be
$\Gamma ^+ (2)_T \times \Gamma ^- (2)_U$, whereas for the case X with
$\l
= 1$ and $\vec a=(1,0)$, $\vec b=(1,0)$, we find
$\Gamma  (2)_T \times \Gamma  (2)_U \times Z_2^{T \leftrightarrow U}$.
\vskip 0.3cm
\noindent {\sl A.2 Rational lines and asymptotic behaviours}

\noindent
Finally, we would like to analyse the behaviour of the shifted
lattices over the moduli space. This includes the identification
of special lines in the $(T,U)$-plane, where extra massless states
can appear in the spectrum, as well as some large-radius properties.
Notice that these special lines are not necessarily lines of enhanced
symmetry, since
in some situations only extra hypermultiplets appear.
For this analysis we indroduce the combinations
\be
\Gamma_{2,2}^{w(\pm)}=\frac{1}{2}
\left(\Gamma_{2,2}^w\oao{1}{0}\pm\Gamma_{2,2}^w\oao{1}{1}
\right)\,,
\label{gpm}
\ee
which turn out to be convenient in the computation of the threshold
corrections (see Section~\ref{thrN4}).
\vskip 0.3cm
\noindent {\sl a) The case $\l=0$}

\noindent We focus here on the appearance of $O({\bar q})$ terms in
$\Gamma_{2,2}^w\oao{0}{1}$ or
$\Gamma_{2,2}^{w(+)}$,
and $O(\sqrt{\bar q})$ terms in
$\Gamma_{2,2}^{w(-)}$.
These situations are indeed possible along some specific lines in the
moduli space, although they are not simultaneously realized. The
results
are summarized as follows:
\ba
\Gamma_{2,2}^w\oao{0}{1}&=&1+(-)^{a_1-b_1}2\,{\bar q}+\cdots\ , \ \
{\rm for\ } T=U \label{l0t}\\
                        &=&1+(-)^{a_2-b_2}2\,{\bar q}+\cdots\ , \ \
{\rm for\ } T=-\frac{1}{U} \label{l0tt}\\
\Gamma_{2,2}^{w(+)}&=&\cdots+2\,{\bar q}+\cdots\ , \ \  {\rm for} \
T=f^w_v(U)\label{l0vp}\\
\Gamma_{2,2}^{w(-)}&=&\cdots+2\,\sqrt{\bar q}+\cdots\ , \ \  {\rm for}
\
T=f^w_h(U)\, .\label{l0h}
\ea
The lines $T=U$ and $T=-1/U$ are no longer equivalent.
In these expressions, the multiplicities are valid for generic
points along the indicated lines. They can, however, be modified at
some
particular values of the moduli. For instance,
$\Gamma_{2,2}^w\oao{0}{1}=1+\Big((-)^{a_1-b_1}+
(-)^{a_2-b_2}\Big)2\,{\bar q}
+\cdots$
 for $T=U=i$,
$\Gamma_{2,2}^w\oao{0}{1}=1+\Big((-)^{a_1-b_1}
+(-)^{a_2-b_2}\Big((-)^{a_1}
+(-)^{b_1}\Big)\Big)2\,{\bar q}+\cdots$
 for $T=U=\rho$ or $-1/\rho$,  and
$\Gamma_{2,2}^w\oao{0}{1}=1+\Big((-)^{a_2-b_2}
+(-)^{a_1-b_1}\Big((-)^{a_2}
+(-)^{b_2}\Big)\Big)2\,{\bar q}+\cdots$
 for $T=-1/U=\rho$ or $-1/\rho$.
On the other hand, the functions $f_v^w(U)$ and $f_h^w(U)$ depend on
the particular shift vector $w$. For concreteness, we concentrate
on the particular case I (see Table A.1); any other situation is
obtained by
duality transformation. In this case, $f^{\I}_v=4U$ and $f_h^{\I}=2U$.
Moreover, for $\Gamma_{2,2}^{\I (+)}$, the multiplicity is doubled at
$T=4U=1+i\sqrt{3}$, whereas it is doubled at $T=2U=1+i$ for
$\Gamma_{2,2}^{\I (-)}$.

Whatever the value of $\lambda$, the existence of the above lines
$T=f^w_{v,h}(U)$ translates an underlying $Z_2$ symmetry of the shifted
lattice or of a sublattice of the latter. For example,
$\Gamma_{2,2}^{\I (-)}$ is invariant under
$T \leftrightarrow 2U$, whereas only a sublattice of $\Gamma_{2,2}^{\I
(+)}$
is invariant under $T \leftrightarrow 4U$; similarly, a sublattice of
$\Gamma_{2,2}^w\oao{0}{1}$ is invariant under $T \leftrightarrow U$
or $T \leftrightarrow -1/U$.

In contrast to what happens in the case of ordinary lattice sums,
the behaviour of the $\l=0$ shifted lattice sums in the
decompactification
limit depends on whether one considers large or small moduli. This is
due to
the partial breaking of the duality group.
For definiteness, let us focus on model I and consider two
six-dimensional
limits: $T_2\to\infty, U_2=1$ (i.e. $R_1\to\infty,
R_2\to\infty$ \footnote{Remember that when $T_1=U_1=0$, $T_2$ and $U_2$
are parametrized
as follows: $U_2=R_2/R_1$ and $T_2=R_1\,R_2$, where $R_1$ and $R_2$ are
the
radii of compactification.}) on the one hand, and $T_2\to0,U_2=1$ (i.e.
$R_1\to0, R_2\to0$) on the other. These two limits are mapped onto
each
other under the combined transformation $T\to-1/T$
and $U\to-1/U$, which does not leave model I invariant
(it actually gives model IV). Therefore, they are not expected to be
equivalent and it is easy to verify that
\be
\Gamma_{2,2}^{\I}
\oao{h}{g}
\underarrow{T_2\to\infty,U_2=1}\cases{
                                  T_2/\tau_2\ , \ \  {\rm for} \
h=g=0\, ,\cr
                                  0\ ,\ \ {\rm otherwise},
          }
\label{lolil}
\ee
whereas
\be
\Gamma_{2,2}^{\I}\oao{h}{g}
{\underarrow{T_2\to0,U_2=1}}\frac{1}{T_2\tau_2}
\ \ \forall h,g\ ,
\label{l0lis}
\ee
up to exponentially suppressed terms.

Similar conclusions can be reached for other $\l=0$ models by
considering
the relevant $SL(2,Z)_T\times SU(2,Z)_U\times Z_2^{T\leftrightarrow U}$
 transformations: {\it there are always two distinct decompactification
limits where either all
$\Gamma_{2,2}^w\oao{h}{g}$ survive and are equal, or only
$\Gamma_{2,2}^w\oao{0}{0}$ survives.}

We would like to emphasize again that the nature of the extra massless
states (vector
multiplets and hypermultiplets) appearing across the lines $T=U$,
$T=-1/U$,
$T=f^w_{v,h}(U)$ as well as
in the two distinct decompactification limits, is not determined by the
structure of the
shifted lattice only: it depends on the full structure of the
string ground state.
\vskip 0.3cm
\noindent{\sl b) The case $\l=1$}

\noindent In this case, we are interested in terms of order ${\bar q}$
in $\Gamma_{2,2}^w\oao{0}{1}$. These are given in (\ref{l0t}) and
(\ref{l0tt}), with the
same modifications of their multiplicity at $T=U=i$ and at other
special points, as
explained above. Moreover, terms of order ${\bar q}^{3/4}$
and ${\bar q}^{1/4}$ are generated in
$\Gamma_{2,2}^{w(+)}$ and $\Gamma_{2,2}^{w(-)}$, respectively:
\ba
\Gamma_{2,2}^{w(+)} &=&\cdots+2\,{\bar q}^{\frac{3}{4}}+\cdots\ , \ \
{\rm for}\ T=f_v^w(U)\label{l1v}\\
\Gamma_{2,2}^{w(-)} &=&\cdots+2\,{\bar q}^{\frac{1}{4}}+\cdots\ , \ \
{\rm for}\ T=f_h^w(U)\, ,\label{l1h}
\ea
where for the model X (see Table A.2) $f_v^{\rm X}=3U$
or $U/3$
and
$f_h^{\rm X}=U$. Again, the generic multiplicity is 2 and can
be promoted to 4 at some particular points on the lines
$T=f_{v,h}^w(U)$. Results for other models
in Table A.2 are obtained by performing appropriate
$SL(2,Z)_T\times SL(2,Z)_U\times Z_2^{T\leftrightarrow U}$
transformations.

We now turn to the decompactification limit of $\l=1$ models. Let us
consider again a specific model, namely model X, in the limits
$T_2\to\infty,
U_2=1$ and  $T_2\to0,U_2=1$. There is a major difference with respect
to
the $\l=0$ case studied above: {\it the duality transformation that
maps
the
limits at hand onto each other now leaves the model invariant.}
These two limits are therefore equivalent and the $\l=1$ shifted
lattice
under consideration possesses a unique behaviour, which is
\be
\Gamma_{2,2}^{\rm X}\oao{h}{g}\to0\ \ \forall \ (h,g)\ne(0,0)
\label{l1li}
\ee
in both $T_2\to\infty, U_2=1$ and $T_2\to0,U_2=1$ limits, whereas
\be
\Gamma_{2,2}^{\rm X}\oao{0}{0}\equiv\Gamma_{2,2}\to
\cases{
{T_2/\tau_2}\ , \ \ {\rm for}  \ \ {T_2\to\infty,U_2=1}\, ,\cr
{1/T_2\tau_2}\ , \ \ {\rm for}  \ \ {T_2\to0,U_2=1}\, .}
\label{l1li00}
\ee
The same holds for more general $\l=1$ models. There is essentially a
{\it unique decompactification limit where only
$\Gamma_{2,2}^{w}\oao{0}{0}$
survives.}

\vskip 0.3cm
\setcounter{section}{0}
\setcounter{equation}{0}
\renewcommand{\theequation}{B.\arabic{equation}}
\boldmath
\section*{\normalsize{\centerline{\bf Appendix B: Two four-dimensional
$E_8\times E_8$ orbifold models}}}
\unboldmath

We present here two typical $Z_2$-orbifold models with $N=4$
supersymmetry broken to $N=2$ and
determine some quantities relevant in Section  \ref{thrN4} to the
general
analysis of the threshold corrections.
The partition function for the $Z_2$-orbifold constructions is given in
(\ref{hetsb}), which we recall here:
\begin{eqnarray}
Z^{\ \rm orb}_{\, \rm sp \ br}&=&{1 \over \im \, \eta^{12} \, \bar
\eta^{24}}\,
 {1\over 2}\sum_{a,b=0}^1 (-1)^{a+b+ab}\,
 \vartheta^2{a\atopwithdelims[]b} \cr
&& \times {1\over 2}\sum_{h,g=0}^1
 \vartheta{a+h\atopwithdelims[]b+g}\,
 \vartheta{a-h\atopwithdelims[]b-g}\,
 \Gamma_{4,20}^{\lambda}{h\atopwithdelims[]g}\,
 \Gamma_{2,2}^{w}{h\atopwithdelims[]g} \, ,
\nn
\end{eqnarray}
where  $\Gamma_{2,2}^{w}{h\atopwithdelims[]g}$
is the shifted two-torus lattice sum (see Eq. (\ref{44444}) or
(\ref{4444l})). For these constructions, we can recast the threshold
functions
$\bOmega^{\l}\oao{h}{g}$
defined in (\ref{gb})
by using
(\ref{corbsp}). We find:
\ba
\bOmega^{\l}_{\vphantom 1}\oao{0}{1}&=&{\hphantom{-}}
{1\over \vartheta_3^2 \, \vartheta_4^2}
\Gamma_{4,20}^{\l}{0\atopwithdelims[]1}
\nonumber \\
\bOmega^{\l}_{\vphantom 1}\oao{1}{0}&=&-
{1\over \vartheta_2^2 \, \vartheta_3^2}
\Gamma_{4,20}^{\l}{1\atopwithdelims[]0}
\label{omorbsp}\\
\bOmega^{\l}_{\vphantom 1}\oao{1}{1}&=&-
{1\over \vartheta_2^2 \, \vartheta_4^2}
\Gamma_{4,20}^{\l}{1\atopwithdelims[]1}\, .
\nonumber
\ea

We also recall the Eisenstein series, which will appear in the
following considerations:
\be
E_{2}^{\vphantom 1}=
{12\over i \pi}\partial_{\t}\log \eta
=1-24\sum_{n=1}^{\infty}{n\, q^n\over 1-q^n}
\label{61b}
\ee
\be
E_{4}^{\vphantom 1}=
{1 \over 2}\left(
{\vartheta}_2^8+
{\vartheta}_3^8+
{\vartheta}_4^8
\right)
=1+240\sum_{n=1}^{\infty}{n^3q^n\over 1-q^n}
\ee
\be
E_{6}^{\vphantom 1}=
\frac{1}{2}
\left({\vartheta}_2^4 + {\vartheta}_3^4\right)
\left({\vartheta}_3^4 + {\vartheta}_4^4\right)
\left({\vartheta}_4^4 - {\vartheta}_2^4\right)
=1-504\sum_{n=1}^{\infty}{n^5q^n\over 1-q^n}
\, .
\ee
\vskip 0.3cm
\noindent {\sl a) The case $\l=0$}

\noindent
We can choose the following $(4,20)$ twisted lattice:
\be
\Gamma_{4,20}^{\l=0}{h\atopwithdelims[]g}=
{1\over 2}\sum_{\bar a,\bar b=0}^1
\vartheta^{2}_{\vphantom 1}{\bar a\atopwithdelims[]\bar b}\,
\vartheta{\bar a+h\atopwithdelims[]\bar b+g}\,
\vartheta{\bar a-h\atopwithdelims[]\bar b-g}\,
\bar\vartheta^4_{\vphantom 1}{\bar a+h\atopwithdelims[]\bar b+g}\,
\bE_4^2
\, ,
\label{420E0E5}
\ee
which leads to an $N=2$  four-dimensional model with gauge group
$E_8\times E_8\times
SO(8)\times U(1)^2$ with $N_V=526, N_H=0$.
Using  (\ref{omorbsp}) we can explicitly  determine the $\Omega$'s,
which  now read:
\ba
\Omega_{(0)}^{\l=0}\oao{0}{1}&=&{\hphantom{-}}
\frac{1}{2}\, E_4^2
\left(\th_3^4+\th_4^4\right)\ = \ -{1\over2} \vartheta_3^{20}
(x^2-x+1)^2 (x-2)
\nonumber\\
\Omega_{(0)}^{\l=0}\oao{1}{0}&=&-
\frac{1}{2}\, E_4^2
\left(\th_2^4+\th_3^4\right)\ = \  -{1\over2} \vartheta_3^{20}
(x^2-x+1)^2 (x+1)
\label{29b}\\
\Omega_{(0)}^{\l=0}\oao{1}{1}&=&{\hphantom{-}}
\frac{1}{2}\, E_4^2
\left(\th_2^4-\th_4^4\right)\ = \ {\hphantom{-}}   {1\over2}
\vartheta_3^{20}
(x^2-x+1)^2 (2x-1)
\, .
\nonumber
\ea
We introduced, as previously, the variable $x=\left(\vartheta_2 /
\vartheta_3\right)^4$, which allows us in particular to recast
$E_4=\th_3^8 \, (x^2-x+1)$.

The $\Lambda$'s corresponding to the $E_8$ factors of the gauge group
are determined in a straightforward way, by using Eq. (\ref{lo}) as
well as
the
identity \cite{pr}:
$$
 -\frac{E_4}{{\eta}^{24}}\left(P^2_{E_8} -
\frac{E_2}{12}\right) E_4=  \frac{E_4}{E_6}\, \frac{j - j(i)}{12}\, .
$$
We find:
\bea
\Lambda_{(0)E_8}^{\l=0}{0\atopwithdelims[]1} &=&{\hphantom{-}}
\frac{1}{24}\frac{E_4\,E_6}{\eta^{24}} (\vartheta_3^4+\vartheta_4^4)=
-\frac{16}{3}\frac{(x^2-x+1)(x+1)(x-2)^2(2x-1)}{x^2(x-1)^2}\nn\\
\Lambda_{(0)E_8}^{\l=0}{1\atopwithdelims[]0} &=&
-\frac{1}{24}\frac{E_4\,E_6}{\eta^{24}} (\vartheta_2^4+\vartheta_3^4)=
-\frac{16}{3}\frac{(x^2-x+1)(x+1)^2(x-2)(2x-1)}{x^2(x-1)^2}
\label{L29b}\\
\Lambda_{(0)E_8}^{\l=0}{1\atopwithdelims[]1} &=&{\hphantom{-}}
\frac{1}{24}\frac{E_4\,E_6}{\eta^{24}} (\vartheta_2^4-\vartheta_4^4)=
\hphantom{-}\frac{16}{3}
\frac{(x^2-x+1)(x+1)(x-2)(2x-1)^2}{x^2(x-1)^2}\, .\nn
\eea
\vskip 0.3cm
\noindent {\sl b) The case $\l=1$}

\noindent Similarly a $\l=1$ model can be obtained  with
\be
\Gamma_{4,20}^{\l=1}{h\atopwithdelims[]g}=
\Gamma_{4,4}^{\vphantom{l}}{h\atopwithdelims[]g}\,
\bE_4^2
\, .
\label{42013}
\ee
The gauge group (in a generic point of the
$\Gamma_{4,4}{h\atopwithdelims[]g}$ lattice) is now
$E_8\times E_8\times{U(1)}^2$ and $N_V=498$,
$N_H=4$.
Following the same procedure as in the previous case, we obtain:
\ba
\Omega_{(0)}^{\l=1}\oao{0}{1}&=&
{\hphantom{-}}
E_4^2\,
\th_3^2\, \th_4^2={\hphantom{-}}
\vartheta_3^{20}(x^2-x+1)^2\sqrt{1-x}\nonumber\\
\Omega_{(0)}^{\l=1}\oao{1}{0}&=&-
E_4^2\,
\th_2^2\, \th_3^2=
-\vartheta_3^{20}(x^2-x+1)^2\sqrt{x}
\label{129b}\\
\Omega_{(0)}^{\l=1}\oao{1}{1}&=&-
E_4^2\,
\th_2^2\, \th_4^2 =
-\vartheta_3^{20}(x^2-x+1)^2\sqrt{x(1-x)}\, ,\nonumber
\ea
and, for the $E_8$ factors,
\bea
\Lambda_{(0)E_8}^{\l=1}{0\atopwithdelims[]1} &=&{\hphantom{-}}
\frac{1}{12}\frac{E_4\,E_6}{\eta^{24}}\,  \vartheta_3^2\,
\vartheta_4^2=
\hphantom{-}
 \frac{32}{3}\frac{(x^2-x+1)(x+1)(x-2)(2x-1)\sqrt{1-x}}
{x^2(x-1)^2}\nn\\
\Lambda_{(0)E_8}^{\l=1}{1\atopwithdelims[]0} &=&
-\frac{1}{12}\frac{E_4\,E_6}{\eta^{24}} \, \vartheta_2^2\,
\vartheta_3^2=
-\frac{32}{3}\frac{(x^2-x+1)(x+1)(x-2)(2x-1)\sqrt{x}}
{x^2(x-1)^2}
\label{L129b}\\
\Lambda_{(0)E_8}^{\l=1}{1\atopwithdelims[]1} &=&
-\frac{1}{12}\frac{E_4\,E_6}{\eta^{24}} \, \vartheta_2^2\,
\vartheta_4^2=
-\frac{32}{3}\frac{(x^2-x+1)(x+1)(x-2)(2x-1)\sqrt{x(1-x)}}
{x^2(x-1)^2}\, .\nn
\eea

\vskip 0.3cm
\setcounter{section}{0}
\setcounter{equation}{0}
\renewcommand{\theequation}{C.\arabic{equation}}
\section*{\normalsize{\centerline{\bf Appendix C:
Some details on the threshold calculation}}}

In this appendix we collect technicalities that appear in the
determination
of the threshold corrections (see Section \ref{thrN4}) for models
with spontaneously-broken $N=4$ supersymmetry, for
which the helicity-generating function is given in Eqs.
(\ref{het2helj}) and (\ref{helsb}).
\vskip 0.3cm
\noindent {\sl C.1 Models with $\l=0$ shifted lattice}

\noindent In order to express the constants $A_i, B_i, C_i, D_i$ and
$\vec\xi$, which appear in the functions $F_i^{\l=0}$ and
$F_{\rm grav}^{\l=0}$, in terms of the physical
parameters of the model, namely $b_i, \dhh b_i, \dv b_i$ and
$\bgrav$, we
must identify the latter with the various coefficients that appear
in the large-$\tau_2$ expansions of $F_{\rm grav}^{\l=0}\oao{h}{g}$ and
$F_i^{\l=0}\oao{h}{g}$ (Eqs. (\ref{auhg}) and (\ref{fi})). Neglecting
the $\frac{1}{\im}$-suppressed
contributions, which
play no role in our argument, these expansions read:
\ba
F_i^{\l=0}\oao{0}{1}&=&\frac{1}{\bar
q}\left(-\frac{2\,A_i+2\,B_i+2\,C_i+D_i}{48}-
\frac{\xi_1+\xi_2+\xi_3}{12}\,k_i\right)\nonumber\\
&&+\frac{1}{6}\left(-282\,A_i-26\,B_i-122\,C_i+3\,D_i
+(4\,\xi_1-124\,\xi_2-252\,\xi_3)\,k_i\right)\nonumber\\
&&
+
O(\bar q)
\label{de1}\\
F_i^{\l=0(+)}
&=&\frac{1}{\bar q}\left(-\frac{A_i}{24}+\frac{\xi_3}{12}\,k_i\right)
\nonumber\\
&&
\mbox{}+\left(-47\,A_i-\frac{32}{3}\,B_i-\frac{80}{3}\,C_i+
\frac{1}{3}\left(64\,\xi_1+128\,\xi_2+
126\,\xi_3\right)k_i\right)+O(\bar q)
\label{dep}\\
F_i^{\l=0(-)}&=&
\frac{1}{\sqrt{\bar
q}}\,\left(-\frac{6\,A_i+2\,C_i}{3}+\frac{4\,\xi_2+6\,\xi_3}{3}
\,k_i\right) +
O\left(\sqrt{\bar q}\right)
\label{dem}
\ea
and
\ba
F_{\rm grav}^{\l=0}\oao{0}{1}&=&
\frac{1}{\bar q}
\left(-\frac{\xi_1+\xi_2+\xi_3}{12}\right)
+\frac{2}{3}\left(\xi_1-31\,\xi_2-63\,\xi_3\right)
+O(\bar q)
\label{f01}\\
F_{\rm grav}^{\l=0(+)}&=&
\frac{1}{\bar q}\,\frac{\xi_3}{12}+
\frac{64\,\xi_1+128\,\xi_2+126\,\xi_3}{3}+
O(\bar q)
\label{f0p}
\\
F_{\rm grav}^{\l=0(-)}&=&
\frac{1}{\sqrt{\bar q}}\,\left(\frac{4\,\xi_2+6\,\xi_3}{3}\right)
+O\left(\sqrt{\bar q}\right)
\label{f0m}\,.
\ea

The various constraints and identifications explained in the text lead
to
the following equations:
\ba
2\,  A_i +2\,  B_i+2\,  C_i +D_i+4\, (\xi_1+\xi_2+\xi_3)\,  k_i&=&0
\nn \\
\frac{1}{6}\left(-282\, A_i-26\,  B_i-122 \, C_i +3\,  D_i +
\frac{1}{6}(4\, \xi_1-124\, \xi_2-252\, \xi_3)\right)\, k_i&=&b_i
\nn \\
-A_i+2\, \xi_3\,  k_i&=&0
\nn \\
-\frac{3\, A_i+C_i}{3}+\frac{2\, \xi_2+3\, \xi_3}{3}k_i &=& \frac{\dhh
b_i}{4}
\label{e0} \\
-47\,  A_i-\frac{32}{3} B_i-\frac{80}{3} C_i+
\frac{2}{3}(32\, \xi_1+64\, \xi_2+63\, \xi_3)k_i &=& \dv b_i\nn \\
\xi_1+\xi_2+\xi_3 &=&1
\nn \\
\frac{2}{3}(\xi_1-31\, \xi_2-63\, \xi_3) &=&\bgrav\, .\nn
\ea

The solutions read:
\ba
A_i&=&\frac{1}{36}(4 \, b_i-24 \, \dhh b_i-2 \, \dv b_i +54\,  k_i-9\,
\bgrav \, k_i)
\label{OA}\\
B_i&=&\frac{1}{144}(53\,  b_i-48\,  \dhh b_i-40 \, \dv b_i +990\,
k_i-99\,
\bgrav\, k_i)
\label{OB}\\
C_i&=&\frac{1}{288}(-112\,  b_i+456\,  \dhh b_i + 56\,  \dv b_i -1494\,
k_i+225\, \bgrav \, k_i)
\label{OC}\\
D_i&=&\frac{1}{44}(-26 \, b_i - 168\,  \dhh b_i + 40 \, \dv b_i -
1494\,
k_i + 45\,  \bgrav\,  k_i)
\label{OD}\\
\xi_1&=&\frac{1}{576\,  k_i}(32 \, b_i-192 \, \dhh b_i-16 \, \dv b_i +
990\,  k_i-45\,  \bgrav k_i)
\label{OX1}\\
\xi_2&=&\frac{1}{576\,  k_i}(-64 \, b_i +384\,  \dhh b_i +32\,  \dv b_i
-846\,  k_i + 117 \, \bgrav\, k_i)
\label{OX2}\\
\xi_3&=&\frac{1}{72\,  k_i}(4\,  b_i-24 \, \dhh b_i-2\,  \dv b_i + 54\,
k_i-9\, \bgrav\, k_i)\, .
\label{OX3}
\ea

Let us now introduce several ``elementary" functions, which will
enable us to express the quantities appearing in (\ref{d})--(\ref{y})
in a compact way. As usual, $f(x)=f\oao{1}{0}$ and consequently
$f\oao{0}{1}=f(1-x)$, $f\oao{1}{1}=f\left({x/(x-1)}\right)$. We
have:
\ba
\sigma(x) &=&-\frac{(x-1)^2}{3\,x}\nn\\
\phi(x) &=&{\hphantom{-}}\frac{(x-1)^6}{(x^2-x+1)(x+1)^2(x-2)(2x-1)}
\nn\\
\chi(x) &=&{\hphantom{-}}\frac{(x-1)^4}{(x^2-x+1)(x-2)(2x-1)}
\nn\\
\psi(x) &=&{\hphantom{-}}\frac{(x-1)^2}{2\, (x^2-x+1)}\, .\nn
\ea
With these conventions:
\ba
\delta_g^{\l=0} &=&{\hphantom{-}}\frac{2}{9} \psi^2 \nn \\
h_g^{\l=0} &=& -\frac{4}{3} \psi^2 \nn \\
v_g^{\l=0} &=& -\frac{1}{9} \psi^2 \nn \\
y_g^{\l=0} &=& {\hphantom{-}}
\Bigg(
3-\frac{1}{24}\frac{1}{\sigma}
+\frac{4}{9}\frac{1}{\sigma^2}-
\frac{\bgrav}
{16}\left(8-\frac{1}{\sigma}\right)
\Bigg)\psi^2
 \nn \\
\delta_f^{\l=0} &=&  {\hphantom{-}}
\left(\frac{1}{9}-\frac{5}{48}\frac{1}{\sigma}
-\frac{1}{48}\frac{1}{\sigma^2}\right)\phi
\label{0gf}
\ea
\ba
h_f^{\l=0} &=& {\hphantom{-}}
\left(-\frac{2}{3}+\frac{29}{36}\frac{1}{\sigma}
\right)\phi
 \nn \\
v_f^{\l=0} &=&  {\hphantom{-}}
\left(-\frac{1}{18}+\frac{5}{108}\frac{1}{\sigma}
\right)\phi
 \nn \\
y_f^{\l=0} &=&  {\hphantom{-}}
\Bigg(\left(\frac{3}{2}-\frac{61}{48}\frac{1}{\sigma}
-\frac{1}{24}\frac{1}{\sigma^2}+
\frac{4}{27}\frac{1}{\sigma^3}\right)
-\bgrav\left(\frac{1}{4}-\frac{23}{96}\frac{1}{\sigma}
-\frac{1}{48}\frac{1}{\sigma^2}\right)\Bigg)\phi\, .\nn
\ea
\vskip 0.3cm
\noindent {\sl C.2 Models with $\l=1$ shifted lattice}

\noindent We now express the constants $A_i, B_i, C_i$ and $\vec\xi$ of
$F_i^{\l=1}$ and  $F_{\rm grav}^{\l=1}$ in terms of the various
physical parameters.
Neglecting the $\frac{1}{\im}$-suppressed contributions,
the expansions of $F^{\l=1}_{i}\oao{h}{g}$
and $F^{\l=1}_{\rm grav}\oao{h}{g}$ are given by:
\ba
F_i^{\l=1}\oao{0}{1}&=&\frac{1}{\bar
q}\left(-\frac{A_i+B_i+C_i}{12}-\frac{\xi_1+\xi_2+\xi_3}{12}\, k_i
\right)
\nonumber\\
 &~&\mbox{}+
\frac{2}{3}\left(-97\,A_i-33\,B_i-C_i+
(5\, \xi_1-27\, \xi_2-59\, \xi_3)\, k_i\right)+
O(\bar q)
\label{de11}\\
F_i^{\l=1(+)}
&=&\frac{1}{{\bar q}^{\frac{3}{4}}}
\left(-\frac{A_i}{3}+\frac{2\,\xi_3}{3}\,k_i\right)+
O\left({\bar q}^{\frac{1}{4}}\right)
\label{de1p}\\
F_i^{\l=1(-)}
&=&\frac{1}{{\bar q}^{\frac{1}{4}}}
\left(-\frac{44\,A_i+16\,B_i}{3}+\frac{32\,\xi_2+8\,\xi_3}{3}\right)+
O\left({\bar q}^{\frac{3}{4}}\right)\, ,
\label{de1m}
\ea
and
\ba
F_{\rm grav}^{\l=1}\oao{0}{1}&=&
\frac{1}{\bar q}
\left(-\frac{\xi_1+\xi_2+\xi_3}{12}\right)
+\frac{2}{3}\left(5\,\xi_1-27\,\xi_2-59\,\xi_3\right)
+O(\bar q)
\label{f11}\\
F_{\rm grav}^{\l=1(+)}&=&
\frac{1}{{\bar q}^{\frac{3}{4}}}\,\frac{2\,\xi_3}{3}+
O\left({\bar q}^{\frac{1}{4}}\right)
\label{f1p}\\
F_{\rm grav}^{\l=1(-)}&=&
\frac{1}{{\bar q}^{\frac{1}{4}}}\,\frac{32\,\xi_2+8\,\xi_3}{3}+
O\left({\bar q}^{\frac{3}{4}}\right)\, .
\label{f1m}
\ea
The equations now are
\be
\eqalign{
A_i + B_i + C_i +(\xi_1 + \xi_2 + \xi_3) k_i &= 0
\cr
\frac{2}{3}\Big(-97\, A_i -33\, B_i - C_i + (5\, \xi_1 -27\, \xi_2 -59
\, \xi_3)
\,  k_i\Big)
&= b_i
\cr
-\frac{A_i}{6}+\frac{\xi_3}{3}\, k_i&= \frac{\delta_v b_i}{4}
\cr
\frac{-22\, A_i-8\, B_i}{3}+\frac{16\, \xi_2+4\, \xi_3}{3} k_i &=
\frac{\dhh
b_i}{4}
\cr
\xi_1+\xi_2+\xi_3 &= 1\cr
\frac{2}{3}(5\, \xi_1-27\, \xi_2-59\, \xi_3) &= \bgrav \, ,
\cr
}\label{e1}
\ee
which we can solve as:
\ba
A_i &=& \frac{1}{32}
\left(b_i - 2 \, \dhh b_i - 56  \,\dv b_i + 6 \, k_i -3  \,\bgrav\,
k_i\right)
\label{1A}\\
B_i &=& \frac{1}{64}
\left(9  \,b_i + 12 \, \dhh b_i + 336 \, \dv b_i - 34 \, k_i + 21 \,
\bgrav \, k_i\right)
\label{1B}\\
C_i &=& \frac{1}{64}
\left(7 \, b_i - 8  \,\dhh b_i - 224  \,\dv b_i - 42 \, k_i - 15 \,
\bgrav  \,k_i\right)
\label{1C}\\
\xi_1 &=& \frac{1}{64 \,k_i}
\left(b_i - 2  \,\dhh b_i - 8 \, \dv b_i + 60 \, k_i \right)
\label{1X1}\\
\xi_2 &=& \frac{1}{64 \, k_i}
\left(-2  \,b_i + 4  \,\dhh b_i + 16 \, \dv b_i - 2  \,k_i  + 3 \,
\bgrav \, k_i\right)
\label{1X2}\\
\xi_3 &=& \frac{1}{64 \, k_i}
\left(b_i - 2 \, \dhh b_i - 8 \, \dv b_i + 6  \,k_i  - 3  \,\bgrav \,
k_i\right)\, .
\label{1X3}
\ea

Finally, we have:
\ba
\delta_g^{\l=1} &=& {\hphantom{-}}\frac{1}{16}\psi^2\nn \\
h_g^{\l=1} &=& -\frac{1}{8}\psi^2\nn \\
v_g^{\l=1} &=& -\frac{1}{2}\psi^2\nn \\
y_g^{\l=1} &=& {\hphantom{-}}\left(\frac{3}{8}-\frac{3}{32}\bgrav
-\frac{5}{24}
\frac{1}{\sigma}+\frac{4}{9}\frac{1}{\sigma^2}\right)\psi^2\label{1gf} \\
\delta_f^{\l=1} &=& {\hphantom{-}}\left(\frac{1}{32}+
\frac{1}{64}\frac{1}{\sigma}
\right)\chi\nn \\
h_f^{\l=1} &=& -\frac{1}{16}\chi\nn \\
v_f^{\l=1} &=& -\frac{7}{4}\chi\nn \\
y_f^{\l=1} &=& {\hphantom{-}}\Bigg(\frac{3}{16}-\frac{1}{96}
\frac{1}{\sigma}-\frac{1}{9}\frac{1}{\sigma^2}
-\bgrav
\left(\frac{3}{32}+\frac{1}{64}\frac{1}{\sigma^2}\right)\Bigg)\chi\, .
\nn
\ea

\vskip 0.3cm
\setcounter{section}{0}
\setcounter{equation}{0}
\renewcommand{\theequation}{D.\arabic{equation}}
\section*{\normalsize{\centerline{\bf Appendix D:
Fundamental-domain integrals}}}
\vskip 0.3cm
\noindent {\sl D.1 General evaluation of the integrals}

\noindent
In this appendix, we evaluate the following integrals:
\be
I(T,U) = \ifd\left(\sump\Gamma_{2,2}^w\oao{h}{g}\,
\bLambda^\lambda\oao{h}{g}-c_0\right)
\label{i}
\ee
and
\be
\tilde{I}(T,U)=\ifd\left(\sump\Gamma_{2,2}^w
\oao{h}{g}\,{\widehat{E}}_2\,
{\overline{\Phi}}^\lambda\oao{h}{g}-\hat{c}_0\right)\,,
\label{it}
\ee
and present some relevant asymptotic behaviours. 
Integrals invariant
under $\Gamma(2)$ such as (\ref{i}) were first evaluated in
\cite{ms} and later in \cite{decoa} in special cases, and then more
generally in \cite{GKKOPP} and the last of \cite{st}.

The functions
$\Lambda^\lambda\oao{h}{g}$ and $\Phi^\lambda\oao{h}{g}$ possess the
following properties:
(\romannumeral1) they transform in a way that ensures modular
invariance
of the first term of the integrand in both $\l=0$ and $\l=1$ cases (see
Eqs. (\ref{255c}) and (\ref{256c}) with $\Phi^\lambda\oao{h}{g}
\sim\frac{1}{\eta^{24}}\Omega^{\l}\oao{h}{g}$);
(\romannumeral2) they are holomorphic with Fourier
expansion\footnote{As usual, $f^{(\pm)}=f\oao{1}{0}\pm f\oao{1}{1}$.}
in terms of $q$,
\ba
\Lambda^\lambda\oao{0}{1}&=&\sum_{n\ge-1}c_n\,q^n\nonumber\\
\Lambda^{\l(+)}&=&\sum_{n\ge-1}a_n\,q^{n+\frac{\l}{4}}\nonumber\\
\Lambda^{\l(-)}&=&\sum_{n\ge-1}b_n\,q^{n+\frac{1}{2}
+\frac{\l}{4}}\nonumber\\
\Phi^\l\oao{0}{1}&=&\sum_{n\ge-1}c_n\,q^n\nonumber\\
\Phi^{\l(+)}&=&\sum_{n\ge-1}a_n\,q^{n+\frac{\l}{4}}\label{fex}\\
\Phi^{\l(-)}&=&\sum_{n\ge-1}b_n\,q^{n+\frac{1}{2}
+\frac{\l}{4}}\nonumber\\
E_2\,\Phi^\lambda\oao{0}{1}&=&\sum_{n\ge-1}\hat{c}_n\,q^n\nonumber\\
E_2\,\Phi^{\l(+)}&=&\sum_{n\ge-1}\hat{a}_n\,
q^{n+\frac{\l}{4}}\nonumber\\
E_2\,\Phi^{\l(-)}&=&\sum_{n\ge-1}\hat{b}_n\,
q^{n+\frac{1}{2}+\frac{\l}{4}}\, .
\nonumber
\ea
As corollary of these properties, in the $\l=0$ case, we have:
$$
\sump\Lambda^{\l=0}\oao{h}{g} = \alpha+\beta\,j\, ,
$$
which implies that the coefficients $a_n+c_n$ are in this case closely
related to
the Fourier coefficients of the $j$-function. Similarly,
$$
\sump\Phi^{\l=0}\oao{h}{g} = \gamma \, \frac{E_4\,E_6}{\eta^{24}}+
\delta\,\frac{E_4}{E_6}
$$
($\alpha$, $\beta$, $\gamma$ and $\delta$ are constants) in general,
although in our computations
of gravitational corrections $\delta$ turns out to vanish
systematically.

The above integrals are expected to converge in the $(T,U)$ plane, with
logarithmic singularities on the lines $T=U$, $T=-1/U$,
$T=f_v^w(U)$ and $T=f_h^w(U)$ due to the presence of $c_{-1}$, $a_{-1}$
and
$b_{-1}$
terms, respectively, in (\ref{fex}) (see Appendix A).

The starting point is the Hamiltonian representation of the lattice
sums,
which reads:
\be
\tau_2\,\Gamma_{2,2}^w\oao{h}{g} = \sum_A T^w[A]\oao{h}{g}\, ,
\label{lpr}
\ee
where, in some specific Poisson-resummed form,
\be
T^w[A]\oao{h}{g} = T_2\,
e^{-i\pi\l \frac{h g}{2}}\,
e^{i\pi\left({{\matrix{a_1 &
a_2}}}\right)A\left({{\matrix{g \cr
-h}}}\right)}\,
e^{2\pi i\bT\det A}\,
e^{-\frac{\pi T_2}{\tau_2 U_2}
{\left|\left(\matrix{1 & U}\right)A\left(\matrix{\tau \cr
1}\right)\right|}^2}
\label{cpr}
\ee
and the summation is performed over a set of matrices of the form
(remember that $w=(\vec a,\vec b)$ with
$\vec a=(a_1,a_2)$ and $\vec b=(b_1,b_2)$)
$$
A=\left(\matrix{
n_1+b_1\frac{h}{2}&m_1+b_1\frac{g}{2}\cr
n_2+b_2\frac{h}{2}&m_2+b_2\frac{g}{2}
}\right)\, .
$$

In order to evaluate the integrals (\ref{i}) and (\ref{it}), we
generalize the method of modular
orbits, which was first introduced in \cite{dkl} and later applied to
various situations.
The idea is to reduce the set of matrices to a fundamental one and
simultaneously
unfold the integration domain by performing $PSL(2,Z)$ transformations
on the
$\tau$ variable. In this way, each term of the resulting series can be
integrated separately. This operation assumes the exchange of
summations and
integrations,
which can be invalid because of tachyon-like divergences. Depending on
the values of
the moduli $T$ and $U$, we must therefore utilize other Poisson
resummations than the
 one presented in (\ref{cpr}).

The set of fundamental matrices depends on the vector $\vec{b}$. For
concreteness,
we will analyse two situations only, in which the shift vectors are
$w_{\I}$
and
$w_{\X}$, corresponding respectively to $\l=0$ and $\l=1$ lattices.
Any other
case  in Tables A.1 and A.2 can be  obtained by duality
transformations.
\vskip 0.3cm
\noindent {\sl a) Evaluation of $I$ for shift vectors $w_{\I}$ and
$w_{\X}$}

\noindent In the case at hand, $\vec{b}=(1,0)$ and there is no null
orbit:
\be
I = I_{\rm nd} + I_{\rm dg}\, , \label{Ig}
\ee
where ``nd" and ``dg" stand for non-degenerate and degenerate orbits,
respectively, and
\ba
I_{\rm nd} &=& \sump I_{\rm nd} \oao{h}{g}\label{Ind}\\
I_{\rm dg} &=& I_{\rm dg} \oao{0}{1}\label{Idg}\, .
\ea
After the identification of the set of fundamental matrices, we obtain:
\ba
I_{\rm nd}\oao{0}{1}&=&2\ihd\sum_{k>0}\sum_{k>j\ge0}\sum_{p\ne0}
\, T^{\rm I \, or \, X}
\left[A=\left(
\matrix{k&j+\frac{1}{2}\cr 0&p}
\right)\right]
\oao{0}{1}\,\bLambda\oao{0}{1}
\label{Ind01}\\
I_{\rm nd}\oao{1}{0}&=&2\ihd\sum_{k\ge0}\sum_{k\ge j\ge0}\sum_{p\ne0}
\, T^{\rm I \, or \, X}
\left[A=\left(
\matrix{k+\frac{1}{2}&j\cr 0&p}
\right)\right]
\oao{1}{0}\,\bLambda\oao{1}{0}
\label{Ind10}\\
I_{\rm nd}\oao{1}{1}&=&2\ihd\sum_{k>0}\sum_{k>j\ge0}\sum_{p\ne0}
\, T^{\rm I \, or \, X}
\left[A=\left(
\matrix{k+\frac{1}{2}&j+\frac{1}{2}\cr 0&p}
\right)\right]
\oao{1}{1}\, \bLambda\oao{1}{1}
\label{Ind11}\\
I_{\rm dg}\oao{0}{1}&=&\lim_{N\to\infty}\Bigg\{\ijd{2}\sum_{j,p}
\, T^{\rm I \, or \, X}
\left[A=\left(
\matrix{0&j+\frac{1}{2}\cr 0&p}
\right)\right]
\oao{0}{1}\, \bLambda\oao{0}{1}
\left(1-e^{-\frac{N}{\tau_2}}\right)\nonumber\\
&&\ \ \ \ \ \ \ \ \ \ \ \ \ \ \ \ \ \ \ \ \
-c_0\left(\log N +\gamma+1+\log\frac{2}{3\sqrt{3}}\right)
\Bigg\}\, ,
\label{Igd01}
\ea
where ${\cal H}$ is the upper half-plane and ${\cal S}$ is the strip
$\left\{\tau\in{\cal H},|\tau_1|<{1/2}\right\}$.
By using the standard machinery and the appropriate Poisson resummation
of (\ref{cpr}) to cover the whole moduli space, we obtain the following
results:
\ba
I^{\I}(T,U) &=& -c_0\left(\log\left|\vartheta_4(T)\right|^4
\left|\vartheta_2(U)\right|^4\,T_2\,U_2-\gamma+1+
\log\frac{\pi}{6\sqrt{3}}\right)+
\left(c_0-\frac{a_0}{2}\right)
\log\left|\frac{\vartheta_4(T)}{\eta(T)}\right|^4
\nonumber\\
&&+\frac{\pi}{9}\Big(a_0-2\,c_0-48\left(a_{-1}+c_{-1}\right)\Big)
\left(\frac{T_2}{2}\,\Theta\left(\frac{T_2}{2}-2\,U_2\right)+
2\,U_2\,\Theta\left(2\,U_2-\frac{T_2}{2}\right)\right)\nonumber\\
&&+\frac{\pi}{9}\Big(a_0-2\,c_0+24\left(a_{-1}+c_{-1}\right)\Big)
\Big(T_2\,\Theta\left(T_2-U_2\right)+
     U_2\,\Theta\left(U_2-T_2\right)\Big)\nonumber\\
&&+4\,\Re\Bigg\{
-c_{-1}\,\Li_1
\left(e^{2\pi i
\left(T_1-U_1+i\left|T_2-U_2\right|\right)}\right)
+a_{-1}\,\Li_1
\left(e^{2\pi i
\left(\frac{T_1}{2}-2U_1+i\left|\frac{T_2}{2}-
2U_2\right|\right)}\right)
\nonumber\\
&&\ \ \ \ \ \ \ \ \ \ \ \,
+b_{-1}\,\Li_1
\left(e^{2\pi i
\left(\frac{T_1}{2}-U_1+i
\left|\frac{T_2}{2}-U_2\right|\right)}\right)\nonumber\\
&&\ \ \ \ \ \ \ \ \ \ \ \,
+\sum_{k,\ell>0}
\Bigg(
-c_{k\ell}\,\Li_1
\left(e^{2\pi i
(T k +U \ell)}\right)
+a_{k\ell}\,\Li_1
\left(e^{2\pi i
\left(\frac{T}{2} k +2 U \ell\right)}\right)
\nonumber\\
&&\ \ \ \ \ \ \ \ \ \ \ \ \ \ \ \ \ \ \ \ \ \ \
+\left(2\, c_{2k\ell}-a_{2k\ell}\right)
\Li_1\left(e^{2\pi i(T k +2 U \ell)}\right)\nonumber\\
&&\ \ \ \ \ \ \ \ \ \ \ \ \ \ \ \ \ \ \ \ \ \ \
+b_{2k\ell-k-\ell}\,\Li_1
\left(e^{2\pi i
\left(\frac{T}{2} (2k-1) + U (2 \ell-1)\right)}\right)\Bigg)
\Bigg\}\label{II}
\ea
and
\ba
I^{\rm X}(T,U) &=& -c_0\left(
\log\left|\vartheta_2(T)\right|^4
\left|\vartheta_2(U)\right|^4\, T_2 \,  U_2
-\gamma+1+\log\frac{\pi}{96\, \sqrt{3}}
\right)\nonumber\\
&&-\pi\,c_0\,
\Big(T_2\,\Theta\left(T_2-U_2\right)+U_2\,
\Theta\left(U_2-T_2\right)\Big)
\nonumber\\
&&
+4\,\Re\Bigg\{
c_{-1}\,\Li_1\left(
e^{2\pi i\left(T_1-U_1+i\left|T_2-U_2\right|\right)}
\right)
\nonumber\\
&&\ \ \ \ \ \ \ \ \,
+a_{-1}\,\Li_1\left(
e^{2\pi i\left(\frac{3T_1}{2}-\frac{U_1}{2}+
i\left|\frac{3T_2}{2}-\frac{U_2}{2}\right|\right)}
\right)
\nonumber\\
&&\ \ \ \ \ \ \ \ \,
+a_{-1}\,\Li_1\left(
e^{2\pi i\left(\frac{T_1}{2}-\frac{3U_1}{2}+
i\left|\frac{T_2}{2}-\frac{3U_2}{2}\right|\right)}
\right)
\nonumber\\
&&\ \ \ \ \ \ \ \ \,
+b_{-1}\,\Li_1\left(
e^{2\pi i\left(\frac{T_1}{2}-\frac{U_1}{2}+
i\left|\frac{T_2}{2}-\frac{U_2}{2}\right|\right)}
\right)
\nonumber\\
&&\ \ \ \ \ \ \ \ \,
+\sum_{k,\ell>0}\Bigg(
-c_{k\ell}\,\Lii{T k+U \ell}+2\,c_{4k\ell}\,\Lii{2 T k +2 U \ell}
\nonumber\\
&&\ \ \ \ \ \ \ \ \ \ \ \ \ \ \ \ \ \ \;
+2\,c_{4k\ell-2k-2\ell+1}\,\Lii{T(2k-1)+U(2\ell-1)}\nonumber\\
&&\ \ \ \ \ \ \ \ \ \ \ \ \ \ \ \ \ \ \;
+a_{4k\ell-3k-3\ell+2}\,\Lii{\frac{T}{2}(4k-3)+
\frac{U}{2}(4\ell-3)}\nonumber\\
&&\ \ \ \ \ \ \ \ \ \ \ \ \ \ \ \ \ \ \;
+a_{4k\ell-k-\ell}\,\Lii{\frac{T}{2}(4k-1)+
\frac{U}{2}(4\ell-1)}\nonumber\\
&&\ \ \ \ \ \ \ \ \ \ \ \ \ \ \ \ \ \ \;
+b_{4k\ell-k-3\ell}\,\Lii{\frac{T}{2}(4k-3)+
\frac{U}{2}(4\ell-1)}\nonumber\\
&&\ \ \ \ \ \ \ \ \ \ \ \ \ \ \ \ \ \ \;
+b_{4k\ell-3k-\ell}\,\Lii{\frac{T}{2}(4k-1)+
\frac{U}{2}(4\ell-3)}\Bigg)
\Bigg\}\, .\label{IX}
\ea
The polylogarithms are defined as usual:
\ba
\Li_1(x) &=& -\log(1-x)\nonumber\\
\Li_2(x) &=& \sum_{j>0}\frac{x^j}{j^2}\nonumber\\
\Li_3(x) &=& \sum_{j>0}\frac{x^j}{j^3}\, .\nonumber
\ea

Several comments are in order here. We first notice
the partial breaking of the duality group
$SL(2,Z)_T\times SL(2,Z)_U\times Z_2^{T\leftrightarrow U}$ as
explained in Appendix A (the $Z_2^{T\leftrightarrow U}$ symmetry
survives
in the second case). We also
observe the appearance
of logarithmic singularities,
as expected
from Eqs. (\ref{l0t})--(\ref{l0h}) and (\ref{l1v}), (\ref{l1h}).
In $I^{\I}$, these take place
at $T=U$, $T=-1/U$, $T=4U$ and $T=2U$. For the situation $I^{\X}$
($\l=1$), the
divergences occur at $T=U$, $T=-1/U$, $T=3U$ and $T=U/3$. The leading
behaviours
are
\ba
I^{\I}&\sim & \hphantom{-}4\, c_{-1}\, \log  \vert T-U\vert
\sp {\rm at \ }T=U \nn\\
I^{\I}&\sim & -4\, a_{-1}\, \log  \vert T-4\, U\vert
\sp {\rm at \ }T=4 U \label{Isb}\\
I^{\I}&\sim & -4\, b_{-1}\, \log  \vert T-2\, U\vert
\sp {\rm at \ }T=2 U\, , \nn
\ea
and
\ba
I^{\X}&\sim & -4\left(c_{-1}+b_{-1}\right)  \log  \vert T-U\vert
\sp {\rm at \ }T=U \nn\\
I^{\X}&\sim & -4\, a_{-1}\, \log  \vert T-3 U\vert
\sp {\rm at \ }T=3\, U \label{Xsb}\\
I^{\X}&\sim & -4\, a_{-1}\, \log  \left\vert T-{U\over 3}\right\vert
\sp {\rm at \ }T={U\over 3}\, . \nn
\ea
The residues at  $T=-1/U$ can be determined in both cases I and X by
performing appropriate Poisson resummations.

The most obvious example for the situation with $\l=0$ is the
constant function. In that case only the first term of (\ref{II})
survives,
in agreement with \cite{decoa}. A somewhat less trivial situation is
provided by
the function $\Lambda\oao{h}{g}=j$, $\forall (h,g)\neq (0,0)$, for
which
\ba
I^{\I}[j]&=&-744 \left(\log\left|\vartheta_4(T)\right|^4
\left|\vartheta_2(U)\right|^4\, T_2 \, U_2
-\gamma+1+\log\frac{\pi}{6\sqrt{3}}\right)\nonumber\\
&&-8\,\log\left|j\left(\frac{T}{2}\right)-j(2\, U)\right|+
4\,\log\left|j(T)-j(U)\right|\, .\nonumber
\ea
This is precisely what is obtained by using the results of \cite{hm}
together with the identity
\be
\sump\Gamma_{2,2}^{\I}\oao{h}{g}=2\, \Gamma_{2,2}\left(\frac{T}{2},2\,
U\right)-
\Gamma_{2,2}(T,U)\, .
\label{Iid}
\ee

Finally, we would like to analyse the behaviour of the above integrals
in the
two limits
that were considered in Appendix A, and which play a role in our
analysis
of the decompactification problem. Up to exponentially suppressed
terms,
\be
I^{\I}(T,U)\, ,\; I^{\X}(T,U)\ \underarrow{T_2\to\infty,U_2=1}
-c_0\,\log T_2 - c_0\,
\mu\label{IXll}
\ee
\ba
I^{\I}(T,U) \ \underarrow{T_2\to 0,U_2=1}&&\frac{\pi}{3}
\Big(a_0+c_0-24\left(a_{-1}+c_{-1}\right)\Big)\frac{1}{T_2}+c_0\,\log
T_2 \nn\\
&&-c_0\,\mu-\left(3c_0+\frac{a_0}{2}\right)\log 2\label{Ils}
\ea
and
\be
I^{\X}(T,U)\ \underarrow{T_2\to0,U_2=1}\ c_0\,\log T_2 - c_0\,
\mu\, ;\label{Xls}
\ee
we have assumed $T_1=U_1=0$ and $\mu$ is a constant:
\be
\mu=4\, \log|\eta(i)|-\gamma+1+\log\frac{\pi}{3\sqrt{3}}\, .
\label{CON1}
\ee
\vskip 0.3cm
\noindent {\sl b) Evaluation of $\tilde{I}$ for shift vectors $w_{\I}$
and $w_{\X}$}

\noindent  The insertion of
$\widehat E_2=\bE_{2}-{3\over \pi\im}$, for the cases at hand (i.e.
without
null orbit), leads to the following result:
\be
\tilde{I}=\hat{I}-\frac{3}{\pi}\,I'\, ,
\label{Igt}
\ee
where $\hat{I}$ is the integral (\ref{Ig}) evaluated above, with
all
coefficients $c_n,a_n,b_n$ substituted with
$\hat{c}_n,\hat{a}_n,\hat{b}_n$;
on the other hand,
\be
I'=\sump I'_{\rm nd}\oao{h}{g}+I'_{\rm dg}\oao{0}{1}  ,\label{Ip}
\ee
where $I'_{\rm nd}\oao{h}{g}$ are given in (\ref{Ind01})--(\ref{Ind11})
with
all $\bLambda\oao{h}{g}$ substituted with
$\tau_2^{-1} \overline{\Phi}\oao{h}{g}$
and
\be
I'_{\rm dg}\oao{0}{1}  = \ijd{3}\sum_{j,p}\,
T^{\rm I \, or \, X}
\left[A=\left(
\matrix{0&j+\frac{1}{2}\cr 0&p}
\right)\right]
\oao{0}{1}\, \overline{\Phi}\oao{0}{1}
\label{Idgp}
\ee
(this integral is infra-red-finite;
the cut-off present in (\ref{it}) plays a role in $\hat{I}$ only).
After some
algebra we find:
\ba
{I^{\I}}'(T,U) \ = \ \frac{4}{T_2\, U_2}
\Re &&\Bigg\{
            \sum_{k>0}\left((c_0-a_0)\,\PP{T k}+
                             a_0\, \PP{\frac{T}{2} k}\right)\nonumber\\
&&+\sum_{\ell>0}\Big(-c_0\,\PP{U \ell}+2c_0\, \PP{2 U \ell}\Big)
  +\frac{c_0}{4\pi}\,\zeta(3)\nonumber\\
&& +\frac{\pi^2}{108}\Bigg[
    \frac{1}{30}\left(16\, c_0+a_0\right)
    \left(16\, U_2^3\,\TTH{\frac{T_2}{2}-2U_2}
    +\frac{T_2^3}{4}\,\TTH{2U_2-\frac{T_2}{2}}\right)\nonumber\\
&&\ \ \ \ \ \ \  +
\frac{1}{15}\left(7c_0-8a_0\right)\left(U_2^3\,\TTH{T_2-U_2}
+{T_2^3}\,\TTH{U_2-T_2}\right)\nonumber\\
&&\ \ \ \ \ \ \ +
\left(a_0+c_0-48(a_{-1}+c_{-1})\right)T_2\,U_2\,
\left(\frac{T_2}{2}\,\TTH{\frac{T_2}{2}-2U_2}\right.\nonumber\\
&&\ \ \ \ \ \ \ +
 \left.2U_2\,\TTH{2U_2-\frac{T_2}{2}}
-\frac{T_2}{2}\,\TTH{T_2-U_2}-\frac{U_2}{2}\,
\TTH{U_2-T_2}\right)\Bigg]\nonumber\\
&& -c_{-1}\, \PP{T_1-U_1+i\ABS{T_2-U_2}} \nonumber\\
&&
+a_{-1}\PP{\frac{T_1}{2}-2U_1+i\ABS{\frac{T_2}{2}-2U_2}}\nonumber\\
&&+b_{-1}\PP{\frac{T_1}{2}-U_1+i\ABS{\frac{T_2}{2}-U_2}}\nonumber\\
&&
+\sum_{k,\ell>0}\left(-c_{k\ell}\, \PP{T k +U \ell}+
a_{k\ell}\, \PP{\frac{T}{2}k+2 U \ell}\right.\nonumber\\
&&\ \ \ \ \ \ \ \ \ \, \, \,+\left(2c_{2k\ell}-a_{2k\ell}\right)\PP{T k
+
2 U \ell}\nonumber\\
&&\ \ \ \ \ \ \ \ \ \, \, \,+\left.b_{2 k \ell - k
-\ell}\,
\PP{\frac{T}{2}(2k-1)+U(2\ell-1)}\right)\Bigg\}
\label{IIp}
\ea
and
\ba
{I^{\X}}'(T,U)  \ = \ \frac{4}{T_2\, U_2}\Re
&&
  \Bigg\{
             c_0\,\sum_{k>0}\Big(2\,\PP{2 T k}-\PP{T k}\Big)
  +c_0\,\sum_{\ell>0}\Big(2\,\PP{2 U \ell}-\PP{U \ell}\Big)\nonumber\\
&&
  +\frac{c_0}{4\pi}\,\zeta(3)
  +\frac{\pi^2}{12}\, c_0\,\left(U_2^3\,\TTH{T_2-U_2}
  +T_2^3\,\TTH{U_2-T_2}\right)\nonumber\\
&&
  +c_{-1}\, \PP{T_1-U_1+i\ABS{T_2-U_2}} \nonumber\\
&&
  +a_{-1}\,
\PP{\frac{3T_1}{2}-\frac{U_1}{2}+i\ABS{\frac{3T_2}{2}-\frac{U_2}{2}}}
\nonumber\\
&&
  +a_{-1}\,
\PP{\frac{T_1}{2}-\frac{3U_1}{2}+i\ABS{\frac{T_2}{2}-\frac{3U_2}{2}}}
\nonumber\\
&&
  +b_{-1}\,
\PP{\frac{T_1}{2}-\frac{U_1}{2}+i\ABS{\frac{T_2}{2}-
\frac{U_2}{2}}}\nonumber\\
&& +\sum_{k,\ell>0}\left({\vrule height .6cm width 0pt}
   -c_{k\ell}\, \PP{T k +U \ell}+2c_{4k\ell}\, \PP{2 T k+2 U
\ell}\right.\nonumber\\
&& \ \ \ \ \ \ \ \ \ \ \ \,
   +2 c_{4k\ell-2k-2\ell+1}\, \PP{T (2k-1) + U(2\ell-1)}\nonumber\\
&& \ \ \ \ \ \ \ \ \ \ \ \,
		 +a_{4k\ell-3k-3\ell+2}\, \PP{\frac{T}{2}(4k-3) +
\frac{U}{2}(4\ell-3)}\nonumber\\
&& \ \ \ \ \ \ \ \ \ \ \ \,
   +a_{4k\ell-k-\ell}\, \PP{\frac{T}{2}(4k-1) +
\frac{U}{2}(4\ell-1)}\nonumber\\
&& \ \ \ \ \ \ \ \ \ \ \ \,
   +b_{4k\ell-k-3\ell}\, \PP{\frac{T}{2}(4k-3) +
\frac{U}{2}(4\ell-1)}\nonumber\\
&& \ \ \ \ \ \ \ \ \ \ \ \,
   +b_{4k\ell-3k-\ell}\, \PP{\frac{T}{2}(4k-1) + \frac{U}{2}(4\ell-3)}
\Bigg)\Bigg\}\label{IXp}\, ,
\ea
where we have introduced \cite{hm}
$$
\PP{x}=\Im x\, \Li_2\left(e^{2\pi i x}\right)+
\frac{1}{2\pi}\, \Li_3\left(e^{2\pi i x}\right).
$$

Regarding the singularities and the breaking of the duality group,
the same observations can be made, as in the cases without insertion of
$\widehat E_2$. In particular, Eqs. (\ref{Isb}) and (\ref{Xsb}) hold
also for the functions $\tilde I^{\I}(T,U)$ and $\tilde I^{\X}(T,U)$.
When the shift vector is $w_{\I}$, the simplest situations
arise with $\Phi\oao{h}{g}=E_4\,E_6/\eta^{24}$ or
$E_4/E_6$, for all $(h,g)\neq (0,0)$.
We can compute these integrals by using the identity (\ref{Iid}) and
the results of \cite{hm}; they turn out to be in agreement with our
general
formulas (\ref{Igt}), (\ref{IIp}).

The asymptotic behaviours read here:
\be
\tilde{I}^{\I}(T,U)\, ,\; \tilde{I}^{\X}(T,U)\
\underarrow{T_2\to\infty,U_2=1}\
-\left(c_0-24\,c_{-1}\right)\log T_2 -\left(c_0-24\,c_{-1}\right)\mu
-\frac{c_0\,\rho}{T_2}
\label{IXtll}
\ee
\ba
\tilde{I}^{\I}(T,U)\ {\underarrow{T_2\to0,U_2=1}}&&
-8 \pi\left((a_{-1}+c_{-1})-\frac{a_0+c_0}{48}\right)\frac{1}{T_2}
+\left(c_0-24\,c_{-1}\right)\log T_2
\nonumber\\
&&
-\left(3\left(c_0-24\,c_{-1}\right)+
\frac{1}{2}\left(a_0-24\,a_{-1}\right)\right)\log2
\nonumber\\
&&
-\left(c_0-24\,c_{-1}\right)\mu
-\left(c_0\, \kappa+a_0\, \nu\right)T_2
\label{Itls}
\ea
\be
\tilde{I}^{\X}(T,U)\ \underarrow{T_2\to0,U_2=1}\
\left(c_0-24\,c_{-1}\right)\log T_2 -\left(c_0-24\,c_{-1}\right)\mu
-c_0\,\rho\, T_2\, ,
\label{Xtls}
\ee
up to exponentially suppressed terms. Again, we assumed $T_1=U_1=0$
and introduced the constants
\ba
\rho &=&\frac{12}{\pi}\sum_{j>0}\left(\frac{1}{j^2}
\left(\frac{1}{2}+\frac{1}{{\sinh}^2 2\pi j}
-\frac{1}{4\sinh^2 \pi j}\right)
+\frac{1}{j^3}\,\frac{1}{4\pi}\tanh \pi j\right)
\nonumber\\
\kappa &=&\frac{12}{\pi}\sum_{j>0}\left(\frac{1}{j^2}
\left(\frac{1}{30}+\frac{1}{4\sinh^2 \pi j}
\right)
+\frac{1}{j^3}\,\frac{1}{4\pi} \coth\pi j\right)
\label{CON}\\
\nu &=&\frac{12}{\pi}\sum_{j>0}\left(\frac{1}{j^2}
\left(-\frac{7}{240}+\frac{1}{8\sinh^2 \frac{\pi j}{2}}\right)
+\frac{1}{j^3}\,\frac{1}{4\pi}\frac{1}{\sinh \pi j}\right)\, .\nonumber
\ea
\newpage
\noindent {\sl D.2 Application to threshold corrections}

\noindent
One can use the results obtained so far to further investigate the
threshold
corrections,
Eqs.~(\ref{del}) and (\ref{delg}), of the models with
spontaneously-broken
supersymmetry described in Section~\ref{spbr}.

For lattices with shift vectors $w_{\I}$ and $w_{\X}$ ($\l=0$ and $1$
respectively)
we obtain the following singularity properties (see (\ref{Isb}),
(\ref{Xsb})):
$$
\Delta^{\I}_{\rm grav} \sim -\frac{1}{3} \log|T-U| \ ,\ \
\Delta_i^{\I}  {\rm \ finite, \ at \ }  T=U
$$
$$
\Delta^{\I}_{\rm grav} \sim \frac{1}{3} \log\left|T+\frac{1}{U}\right|
\ ,\ \
\Delta_i^{\I}  {\rm \ finite, \ at \ }  T=-\frac{1}{U}
$$
\be
\Delta^{\I}_{\rm grav} \sim \left.
\cases{ {\rm finite \ in \ class \ (\romannumeral1)}, \cr
-\frac{1}{3} \log  |T-4 U| {\rm \ in \ class \ (\romannumeral2)}
\cr}\right\}  ,\ \
\Delta_i^{\I}{\rm \ finite, \ at \ }  T=4U
\label{Iths}
\ee
$$
\Delta^{\I}_{\rm grav} \sim -2\dhh\bgrav\log|T-2U| \ ,\ \
\Delta_i^{\I} \sim    -2\dhh b_i\log|T-2U|\, ,
{\rm \ at \ }  T=2U \, ,
$$
and
$$
\Delta^{\X}_{\rm grav} \sim
\left(\frac{1}{3}-2\dhh\bgrav\right)\log|T-U| \ \ ,\
\Delta_i^{\X} \sim -2\dhh b_i\log|T-U|\, ,
{\rm \ at \ }  T=U
$$
$$
\Delta^{\X}_{\rm grav} \sim
\frac{1}{3}\log\left|T+\frac{1}{U}\right| \ \ ,\
\Delta_i^{\X}   {\rm \ finite, \ at \ }  T=-\frac{1}{U}
$$
\be
\Delta^{\X}_{\rm grav} \sim
-2\dv\bgrav\log|T-3 U| \ ,\ \
\Delta_i^{\X}\sim   -2\dv b_i\log|T-3U|\, ,
{\rm \ at \ }  T=3U
\label{Xths}
\ee
$$
\Delta^{\X}_{\rm grav} \sim
-2\dv\bgrav\log|U-3 T| \ ,\ \
\Delta_i^{\X}\sim  -2\dv b_i\log|U-3T|\, ,
{\rm \ at \ }  T=\frac{U}{3}\, .
$$

Finally, we can analyse the behaviour of the corrections in the
various
decompactification
limits. We will give the results containing leading terms and
subleading
corrections, up to
exponentially suppressed ones. We assume again $T_1=U_1=0$,
and use Eqs.
(\ref{IXll})--(\ref{Xls})
and
(\ref{IXtll})--(\ref{Xtls}).
\vskip 0.3cm
\noindent {\sl a) The limit $T_2\to\infty\,,\ U_2=1$}

\noindent In this limit, $N=4$ supersymmetry is restored in both $\l=0$
and
$\l=1$ lattices. The behaviours are
$$
\Delta^{\I}_{\rm grav}\ ,\ \ \Delta^{\X}_{\rm grav}\to
- b_{\rm grav}
\left(\log T_2 + \mu \right) - \left(b_{\rm
grav}-2\right)\frac{\rho}{T_2}\, ,
$$
and
$$
\Delta^{\I}_{i}\ , \ \ \Delta^{\X}_{i}\to - b_{i}
\left(\log T_2+ \mu - \log {2\over \pi}{e^{1-\gamma}\over 3\sqrt{3}}
\right)
- k_i\left(b_{\rm grav}-2\right)\frac{\rho}{T_2}\, .
$$
\vskip 0.3cm
\noindent {\sl b) The limit $T_2\to0\,,\ U_2=1$}

\noindent  This limit is ($N=4$)-supersymmetric for $\l=0$ models of
class (\romannumeral2) and models with $\l=1$. For $\l=0$ models
belonging to class (\romannumeral1),
the supersymmetry remains $N=2$:
\ba
\Delta^{\I}_{\rm grav}\ {\rm (class \ (\romannumeral1))} &\to &
\frac{4\pi}{T_2}
+b_{\rm grav} \left(\log T_2-\mu-\frac{5}{2} \log 2\right) -11 \log 2
\nonumber\\
&&-\left(\kappa\left(b_{\rm grav}-2\right)-
\nu\left(b_{\rm grav}-22\right)\right) T_2
\nonumber
\ea
\ba
\Delta^{\I}_{i}\ {\rm (class \ (\romannumeral1))} &\to& \left(b_i+\dv
b_i-12 \, k_i\right)
\frac{\pi}{3T_2}
+b_{i} \left(\log T_2
-\mu  + \log {1\over \pi}{e^{1-\gamma}\over 12\sqrt{3}}\right)
\nonumber\\
&&
-\frac{\dv b_i}{2}\log 2 - k_i\left(\kappa \left(b_{\rm
grav}-2\right)-\nu\left(b_{\rm
grav}-22\right)\right) T_2
\nn
\ea
$$
\Delta^{\I}_{\rm grav}\ {\rm (class \ (\romannumeral2))}\ \to
b_{\rm grav} \left(\log T_2-\mu-\frac{5}{2}\log 2\right)
-\left(b_{\rm grav}-2\right)\left(\kappa-\nu\right) T_2
$$
$$
\Delta^{\I}_{i}\ {\rm (class \ (\romannumeral2))}\ \to
b_{i} \left(\log T_2-\mu + \log {1\over \pi}{e^{1-\gamma}\over
6\sqrt{6}} \right)
-k_i\left(b_{\rm grav}-2\right)\left(\kappa-\nu\right) T_2
$$
$$
\Delta^{\X}_{\rm grav}\to
-b_{\rm grav} \left(-\log T_2+\mu\right)
-\left(b_{\rm grav}-2\right) \rho  T_2
$$
$$
\Delta^{\X}_{i}\to
-b_{i} \left(-\log T_2+\mu - \log {2\over \pi}{e^{1-\gamma}\over
3\sqrt{3}} \right)
-k_i\left(b_{\rm grav}-2\right) \rho T_2 \ .
$$

\end{document}